\documentclass[a4paper,11pt,english]{article}
\pdfoutput=1

\usepackage{jheppub}
\usepackage{amsmath,amssymb,amsfonts,amsthm}
\usepackage[framemethod=tikz]{mdframed}
\usepackage{placeins}
\usepackage{xcolor}
\usepackage{framed}
\definecolor{shadecolor}{gray}{0.95}
\usepackage{booktabs}
\usepackage{mathtools}
\definecolor{darkblue}{rgb}{0.1,0.1,.7}
\usepackage[]{amsmath}
\usepackage[]{graphicx}
\usepackage[]{latexsym}
\usepackage{amscd}
\usepackage[all,cmtip]{xy}
\usepackage{mathrsfs}
\usepackage{bbold}
\usepackage{ytableau}
\usepackage{youngtab}
\usepackage{simplewick}
\usepackage{changepage}
\usepackage[]{algorithm2e}
\usepackage{multirow}
\usepackage{slashed}
\usepackage[OT2,OT1]{fontenc}
\usepackage{braket}
\usepackage{tikz}
\usetikzlibrary{decorations.pathmorphing,circuits.logic.US,circuits.logic.IEC,fit,positioning,arrows,patterns,decorations.markings}
\tikzset{
	Witten diagram/.style={
		execute at begin picture={
			\draw[blue, line width=1.5pt] circle[radius=\pgfkeysvalueof{/tikz/Witten/radius}];
			\path node (X){\phantom{X}};
		},
		baseline={(X.base)}
	},
	vertex/.style={circle,fill,inner sep=1.5pt,node contents={}},
	Witten/.cd,
	radius/.initial=3cm
}

\newcommand{\ML}[1]{\textcolor{orange}{ML: #1}}

\theoremstyle{remark}

\makeatletter
\def\@fpheader{\ }
\makeatother

\title{What is a photon in de Sitter spacetime?}
\author[a]{Manuel Loparco,} 
\author[a]{João Penedones,}
\author[a,b,c]{Yannis Ulrich}

\affiliation[a]{Fields and Strings Laboratory, Institute of Physics\\
École Polytechnique Fédéral de Lausanne (EPFL)\\
Route de la Sorge, CH-1015 Lausanne, Switzerland}
\affiliation[b]{Max Planck Institute for Solid State Research\\
D-70569 Stuttgart, Germany}
\affiliation[c]{School of Natural Sciences\\
Technische Universität München\\
D-85748 Garching, Germany}
\emailAdd{manuel.loparco@gmail.com, joao.penedones@epfl.ch, y.ulrich@fkf.mpg.de}
\abstract{The states of a single photon in four-dimensional de Sitter (dS) spacetime form a Unitary Irreducible Representation (UIR) of SO(1,4), which we call the photon UIR. While in flat spacetime photons are intimately tied to gauge symmetry, we demonstrate that in de Sitter, photon states emerge generically in any quantum field theory, even without an underlying U(1) gauge field.
We derive a Källén-Lehmann representation for antisymmetric tensor two-point functions and show that numerous composite operators constructed from massive free fields can create states in the photon UIR. Remarkably, we find that some of these operators exhibit two-point functions with slower late-time and large-distance decay than the electromagnetic field strength itself, challenging the conventional notion that photons dominate the infrared regime.
Using our spectral representation, we establish non-perturbative bounds on the late-time behavior of electric and magnetic fields in de Sitter, with potential implications for primordial magnetogenesis. Through one-loop calculations, we demonstrate that both the creation of photon states and the enhanced late-time large-distance behavior persist in weakly interacting theories.}
\begin{document}
\maketitle

\newpage
\section{Introduction}
\label{sec:introduction}
Quantum Field Theory (QFT) in de Sitter (dS) spacetime is both a powerful laboratory for testing our understanding of QFT in curved spacetimes and a phenomenologically relevant framework. Since inflationary cosmology is well described by a quasi-de Sitter phase, developing the tools to study QFT in this context is essential in order to understand the quantum origin of structure in our universe \cite{Baumann:2009ds}.

Many of the familiar features of QFT in flat space undergo a significant modification in de Sitter. Most notably, the absence of a globally timelike Killing vector prevents the construction of a non-perturbative S-matrix, obscuring our usual notions of asymptotic states and scattering amplitudes \cite{Bousso:2004tv}\footnote{For recent perturbative definitions of S-matrices in de Sitter see \cite{Melville:2023kgd,Melville:2024ove,Donath:2024utn}.}. In this setting, a useful guiding principle is to take seriously the structure of the Hilbert space itself. In particular, the Hilbert space of a QFT in four-dimensional de Sitter decomposes into Unitary Irreducible Representations (UIRs) of the isometry group $SO(1,4)$. This decomposition leads to powerful non-perturbative constraints on correlation functions of local operators, such as spectral representations of two- and four-point functions \cite{Hogervorst:2021uvp,DiPietro:2021sjt,Loparco:2023rug,Schaub:2023scu}, as well as non-trivial constraints on renormalization group flows \cite{Loparco:2024ibp}. 

In this work, we focus on the photon UIR — the representation labeled by $\Delta=2$ and $SO(3)$ spin 1, which is furnished by the states of a single photon \cite{Angelopoulos:1980wg,Garidi:2006ey,RiosFukelman:2023mgq}.

 In flat spacetime, constructing a QFT whose Hilbert space contains states in the UIR of the photon typically requires a $U(1)$ gauge symmetry. Moreover, among all spin-1 excitations, these states are the ones that dominate at large distances. In de Sitter spacetime, we find a more surprising picture: \emph{photon states appear generically in the Hilbert space of any QFT in dS}, even in the absence of an underlying gauge symmetry. This feature had been anticipated from group-theoretic arguments in \cite{MartindS,Penedones:2023uqc}, where it was shown that photon states arise in the tensor product of massive representations.


To understand these observations in the language of QFT, we derive the Källén-Lehmann representation for two-point functions of antisymmetric tensor operators with two indices, motivated by the fact that such is the index structure of the Maxwell field strength. Then, computing the spectral decomposition of a variety of composite operators $B_{\mu\nu}$ made of massive fields in free and weakly coupled theories, we find that many of them interpolate between the Bunch-Davies vacuum and states in the photon UIR
\begin{equation}
    \langle0|B_{\mu\nu}(x)|\text{photon}\rangle\neq0\,.
\end{equation}
This is not entirely surprising if we consider the fact that in the coordinate patch where we \textit{do} have a Hamiltonian, namely the static patch, the Bunch-Davies vacuum is a thermal state \cite{PhysRevD.15.2738}. 

Perhaps more surprisingly, some of these operators have two-point functions which decay more slowly at late times and large spatial separations than those of the electromagnetic field strength. In other words, among all spin 1 states, \emph{photons are \textbf{not} the ones that dominate the IR regime of QFT in de Sitter}. For example, we find that the field strength of a massive vector has a two-point function which dominates over that of the field strength of the photon at late times\footnote{In particular, a Proca field with intrinsic parity even has a dominant ``electric field" while a Proca field with parity odd has a dominant ``magnetic field".}. This fact can also be understood more intuitively if we think in terms of energy densities: radiation dilutes faster than matter in an expanding universe. 
\begin{table}
\centering
\begin{tabular}{c|c|c|c}
    Operator  & Creates photons  & Late time $\langle\mathbf{E}\mathbf{E}\rangle$ & Late time $\langle\mathbf{B}\mathbf{B}\rangle$\\
      \hline
     Photon $F_{\mu\nu}$ & \checkmark & $(\eta/\mathbf{y})^4$ & $(\eta/\mathbf{y})^4$ \\
     \hline
     Massless $\partial_{[\mu}\phi_1\partial_{\nu]}\phi_2$ & \checkmark &  & $(\eta/\mathbf{y})^4$\\
     \hline
     Massive $\partial_{[\mu}\phi_1\partial_{\nu]}\phi_2$ & \checkmark & $(\eta/\mathbf{y})^{2\Delta}$ (*) & \\
     \hline
     Massive $V_{1[\mu}V_{2\nu]}$ & \checkmark & & \\
     \hline
     Massive $V_{[\mu}\partial_{\nu]}\phi$ & \checkmark & $(\eta/\mathbf{y})^{2\Delta}$ (*)& \\
     \hline
     CFT $\mathcal{O}^{\mu\nu}$ & \checkmark & & \\
     \hline
     Proca $F_{\mu\nu}$ & & $m^2(\eta/\mathbf{y})^{2}$ & $(\eta/\mathbf{y})^4$\\
     \hline
     Pseudo-Proca $F_{\mu\nu}$ &  & $(\eta/\mathbf{y})^4$ & $m^2(\eta/\mathbf{y})^{2}$\\
     \hline
\end{tabular}
\caption{A variety of normal ordered composite operators in free or weakly coupled theories create states in the UIR of the photon. In the rightmost columns we report the leading late time and large distance scaling of the two-point functions of their ``electric" and ``magnetic" components when they dominate or equate the free Maxwell EM fields. $\Delta$ is a dimensionless number constrained to be $\Delta>1$. The asterisks (*) indicate that such behavior dominates over free Maxwell theory only if the elementary fields are light enough. We use coordinates $ds^2=\frac{-\mathrm{d}\eta^2+\mathrm{d}\mathbf{y}^2}{\eta^2}$ and report the two-point functions in the locally inertial frame of a free falling observer, with the late time limit being $\eta\to0$. For the Proca fields, the result reported is for $m^2\ll H^2$.}
\label{tab:summary}
\end{table}\\
\noindent The late time behavior of a two-point function is also encoded in its K\"allén-Lehmann representation, more precisely in the analytic structure of its spectral densities \cite{Hogervorst:2021uvp,Loparco:2023rug}. Under a set of fairly general assumptions, we can exploit this fact to prove a non-perturbative bound on the late time and large distance scaling of antisymmetric two-point functions in the Bunch-Davies vacuum, which in particular applies to electric and magnetic fields \footnote{See \eqref{eq:EBdefinitions} for the precise definition of electric and magnetic fields.}$^{,}$\footnote{This bound applies to parity invariant theories. It would be  interesting to study what happens when one relaxes this assumption.}
\begin{equation}
\begin{aligned}
    \langle0|\mathbf{E}^a(\eta,\mathbf{y})\mathbf{E}^b(\eta,0)|0\rangle&\stackrel{\eta\to0}{\longrightarrow} c\left(\frac{\eta^2}{\mathbf{y}^2}\right)^{\Delta}\left(\delta^{ab}-2\frac{\mathbf{y}^a\mathbf{y}^b}{\mathbf{y}^2}\right)\\
    \langle0|\mathbf{B}^a(\eta,\mathbf{y})\mathbf{B}^b(\eta,0)|0\rangle&\stackrel{\eta\to0}{\longrightarrow} c\left(\frac{\eta^2}{\mathbf{y}^2}\right)^{\Delta}\left(\delta^{ab}-2\frac{\mathbf{y}^a\mathbf{y}^b}{\mathbf{y}^2}\right)
\end{aligned}
\qquad\quad
\vcenter{\hbox{$\Delta>1$}}
\end{equation}
It would be extremely interesting to understand whether these bounds are of interest in the context of cosmological magnetogenesis \cite{Giovannini:2003yn,Subramanian:2015lua}. In particular, we find that they are saturated, respectively, by the field strengths of a free vector and a free pseudo-vector with small but nonzero mass, as reported in Table \ref{tab:summary}.

\vspace{1em}
\noindent The results we present in this paper highlight the utility of a representation-theoretic approach to QFT in de Sitter. Hilbert space methods — based on symmetry, unitarity, and representation theory — can yield new physical insights that go against our intuitions from flat space QFT. As we make progress towards understanding QFT and quantum gravitational aspects of cosmology, this perspective may prove to be indispensable.

\paragraph{Outline} This paper is structured as follows. We begin with a review of some preliminaries on QFT in dS in Section \ref{sec:preliminaries}, including a discussion on free Maxwell theory in \ref{subsec:freemaxwell}. 

Then, in Section \ref{sec:kallanlehmann} we derive the K\"allén-Lehmann decomposition for two-point functions of antisymmetric operators in flat space and in dS, and present inversion formulas for both cases.  At the end of the section, in \ref{subsec:latetimes}, we show how to derive the late-time behavior of a two-point function from its K\"allén-Lehmann decomposition and identify what conditions must be satisfied in order for the two-point function to have a late time behavior that dominates over that of the Maxwell field strength. Finally, in \ref{subsec:pmf}, we highlight how the K\"allén-Lehmann representation implies a bound on the late time behavior of electromagnetic fields, and sketch possible connections to the question of primordial magnetic fields in cosmology.

Section \ref{sec:photonsfrommass0} is devoted to using the K\"allén-Lehmann decomposition to show that many operators in free massive theories in dS create states in the photon UIR when they act on the vacuum, and that some of these operators show a large distance behavior at late times that dominates over that of the Maxwell field strength.

In Section \ref{sec:photonsfrommass1} we study the effect of weak interactions on the statements made in Section \ref{sec:photonsfrommass0}. We start by reviewing the treatment of mass renormalization in the theory of two scalars coupled cubically at one loop \ref{sec:cubicscalar}. We phrase the results in a slightly new way in the language of spectral densities. Then, we move to certain classes of interactions involving the operators of Section \ref{sec:photonsfrommass0} and photons and we find that in every case the creation of states in the photon UIR and the peculiar late time behavior persist at one loop.

In Section \ref{sec:conclusion} we discuss possible future directions.
\section{Preliminaries}
\label{sec:preliminaries}
In this section we review the basics of QFT in four-dimensional de Sitter spacetime. We start by describing the geometry of dS in the embedding space formalism \ref{subsec:geometry} and how tensor fields \ref{subsec:fields} and states \ref{subsec:states} can be uplifted to the embedding. Finally, we review some crucial aspects of electromagnetism in de Sitter \ref{subsec:freemaxwell}. For a review of the classification of UIRs of de Sitter, see Appendix \ref{sec:UIRs}.
\subsection{Geometry}
\label{subsec:geometry}
de Sitter can be embedded in a Minkowski spacetime with one more spatial dimension. We use $Y^A$ with $A=0,1,\ldots,4$ for vectors in the embedding space when they satisfy the constraint
\begin{equation}
    -(Y^0)^2+(Y^1)^2+(Y^2)^2+(Y^3)^2+(Y^4)^2=R^2\,,
\end{equation}
where $R$ is the characteristic length scale of de Sitter, related to the Hubble constant $H$ through $R=H^{-1}$. We will now set $R=1$ and restore it when useful. 

A coordinate system which is frequently used in the literature is the \textbf{conformally flat} or \textbf{Poincaré} coordinates, defined through the embedding condition
\begin{equation}
Y^0 =\, \frac{\eta^2 - |\mathbf{y}|^2 -1}{2\eta}~,\qquad
Y^{i} = -\,\frac{\mathbf{y}^i}{\eta}~, \qquad
Y^{4} = \,\frac{\eta^2 - |\mathbf{y}|^2 +1}{2\eta}
\label{Xetax}
\end{equation}
where $\eta\in(-\infty,0)$, $\mathbf{y}\in\mathbb{R}^3$, $i$ runs over $1,2,3$ and we use the notation $y^\mu=(\eta,\mathbf{y})$. Then, the metric becomes
\begin{equation}
    ds^2=\frac{-\mathrm{d}\eta^2+\mathrm{d}\mathbf{y}^2}{\eta^2}\,.
    \label{eq:coordinates}
\end{equation}
This coordinate system covers only half of global de Sitter spacetime. 

It is common to refer to the timeslice at $\eta\to 0^-$ as the \textbf{late-time surface}. Physically, it is associated with the Universe at the time of reheating. In the inflationary picture, this surface is glued to a period of radiation domination, and the rest of the evolution is governed by the classic Big Bang model. Fluctuations of quantum fields imprinted at this surface are thus the initial conditions for the rest of the evolution of the Universe.

Every point at the late time boundary of de Sitter
is in one-to-one correspondence with a lightray in embedding space. Light-rays $P^A$ are defined through
\begin{equation}
    -(P^0)^2+(P^1)^2+(P^2)^2+(P^3)^2+(P^4)^2=0\,, \qquad P^A\sim\lambda P^A\,, \qquad P^0>0\,.
\end{equation}
In conformally flat coordinates, a possible relation for this correspondence is
\begin{equation}
    P^0=\frac{1+|\mathbf{y}|^2}{2}\,, \qquad P^i=\mathbf{y}^i\,, \qquad P^{4}=\frac{|\mathbf{y}|^2-1}{2}\,.
\end{equation}
The embedding space realization of covariant derivatives is
\begin{equation}
    \nabla_A=\partial_{Y^A}-Y_A (Y\cdot \partial_Y)\,.
\end{equation}
When acting with a derivative on a tensor (transverse to $Y_A$), the indices must be re-projected onto dS through the action of the metric $G_{AB}=\eta_{AB}-Y_AY_B$.

We will need to integrate points over the late time surface 
and over the whole of de Sitter. In the language of embedding space, we introduce the following shorthand notations
\begin{align}
	\int_P(\ldots)&\equiv\frac{2}{\text{Vol GL}(1,\mathbb{R})^+}\int d^{5}P\ \delta(P^2)\theta(P^0)(\ldots)\,.
    \label{eq:intP}\\
    \int_Y(\ldots)&\equiv\int d^{5}Y\ \delta(Y^2-1)(\ldots)\,.
    \label{eq:intY}
\end{align}
Finally, let us introduce the scalar invariant that can be built from two bulk points $\sigma\equiv Y_1\cdot Y_2$. Its values are related to the relative positions in spacetime of the two points as follows
\begin{equation}
	\text{Timelike:}\ \sigma>1\,,\qquad \text{Spacelike:}\ \sigma<1\,,\qquad \text{Lightlike:}\ \sigma=1\,.
\end{equation}
In conformally flat coordinates, it takes the form
\begin{equation}
	\sigma=\frac{\eta_1^2+\eta_2^2-|\mathbf{y}_1-\mathbf{y}_2|^2}{2\eta_1\eta_2}\,.
    \label{eq:sigmadef}
\end{equation}
Throughout this paper we will use the antisymmetrization symbol. The convention is
\begin{equation}
    T_{[\mu_1\cdots \mu_n]}\equiv\sum_{i}\text{sign}(\pi_i)T_{\pi_i\{\mu_1\ldots\mu_n\}}\,,
\end{equation}
where $\pi_i\{\mu_1\ldots\mu_n\}$ is the $i$-th permutation of the set $\{\mu_1\ldots\mu_n\}$ and sign$(\pi_i)$ is its signature.
Notice that with this definition 
$B_{[\mu\nu]} =2 B_{\mu\nu}$ when $B$ is antisymmetric.

\subsection{Fields}
\label{subsec:fields}
Tensor fields in embedding space $T_{A_1\cdots A_J}(Y)$ can be seen as uplifts of tensor fields in de Sitter space if they have support only at $Y^2=1$ and they satisfy the tangential condition
\begin{equation}
    Y^{A_i}T_{A_1\cdots A_i\cdots A_J}(Y)=0\,, \qquad \forall i\,,
\end{equation}
which ensures that they point along the de Sitter manifold.
The tensor fields in local coordinates $x^\mu$ can be reconstructed from the projection
\begin{equation}
    T_{\mu_1\cdots\mu_J}(x)=\frac{\partial Y^{A_1}}{\partial x^{\mu_1}}\cdots \frac{\partial Y^{A_J}}{\partial x^{\mu_J}}T_{A_1\cdots A_J}(Y)\,.
\end{equation}

\subsection{States}
\label{subsec:states} Given the isomorphism with the Euclidean conformal algebra, the Hilbert space associated to each UIR (reviewed in appendix \ref{sec:UIRs}) can be constructed precisely as in CFT \cite{Sun:2021thf}.
One can introduce a set of states $|\Delta,P\rangle_{A_1\ldots A_s}$, where $P$ runs over points in the late-time conformal boundary of de Sitter\footnote{This does not mean the states ``live" on the late time surface. We are simply using the fact that the conformal group acts naturally on the late time surface, making it convenient to use $P$ to label the states in the Hilbert space of the UIR labeled by $(\Delta,s)$.}, which form a basis of the UIR labeled by $(\Delta,s)$. 
They satisfy the following properties
\begin{itemize}
    \item Spin $s$ condition: $|\Delta,P\rangle_{A_1\cdots A_s}$ is a symmetric traceless tensor of $SO(1,4)$.
    \item Homogeneity: $|\Delta,\lambda P\rangle_{A_1\cdots A_s}=\lambda^{-\Delta}|\Delta,P\rangle_{A_1\cdots A_s}$ with $\lambda>0$. Based on this property we sometimes call $\Delta$ the \textbf{scaling dimension}.
    \item Tangential condition: $P^{A_i}|\Delta,P\rangle_{A_1\ldots A_i\ldots A_s}=0,\quad \forall i=1,\ldots s$.
\end{itemize}
The states in local coordinates $|\Delta,\mathbf{y}\rangle_{i_1\cdots i_s}$ are retrieved by pulling back $|\Delta,P\rangle_{A_1\cdots A_s}$:
\begin{equation}
    |\Delta,\mathbf{y}\rangle_{i_1\cdots i_s}=\frac{\partial P^{A_1}}{\partial \mathbf y^{i_1}}\cdots \frac{\partial P^{A_s}}{\partial \mathbf y^{i_s}}|\Delta,P\rangle_{A_1\cdots A_s}\,.
\end{equation}
In this paper we will focus on parity-preserving theories. In odd number of spatial dimensions, we can define parity as the simultaneous reflection of all spatial coordinates.
We denote its action by the operator $\Theta$. In embedding space, a parity transformation is realized as
\begin{equation}
    \begin{aligned}
        (Y^0,Y^i,Y^4)&\stackrel{\Theta}{\to} (Y^0,-Y^i,Y^4)\,,\\
        (P^0,P^i,P^4)&\stackrel{\Theta}{\to} (P^0,-P^i,P^4)\,,\\
        (\eta,\mathbf{y}^i)&\stackrel{\Theta}{\to} (\eta,-\mathbf{y}^i)\,.
        \label{eq:embedparity}
    \end{aligned}
\end{equation}
Then, states carry a parity quantum number
\begin{equation}
    \Theta|\Delta,P,\pm\rangle_{A_1\ldots A_s}=\pm(-1)^s|\Delta,\Theta P,\pm\rangle_{A_1\ldots A_s}\,.
\end{equation}
\subsection{EM fields in de Sitter}
\label{subsec:freemaxwell}
The main focus of this paper is the UIR of the photon, which in four dimensions is the only spin 1 representation in the exceptional type II series. The associated Hilbert space is realized by the single-particle states of the electromagnetic field in free Maxwell theory in de Sitter \cite{Cotaescu:2008hv,Allen:1985wd,Youssef:2010dw,Domazet:2014bqa}
\begin{equation}
	S=-\frac{1}{4}\int d^4x\sqrt{g}F^{\mu\nu}F_{\mu\nu}\,, \qquad F_{\mu\nu}=\partial_{[\mu}A_{\nu]}\,.
\end{equation}
In four dimensions, Maxwell theory is conformal, and the antisymmetric field strength is a conformal primary with $\mathbf{\Delta}=2$\footnote{We use $\mathbf{\Delta}$ to indicate the scaling dimension of a bulk CFT primary. Bulk unitarity implies $\mathbf{\Delta}\geq2$. This must be distinguished from $\Delta$, which is a complex label for the quadratic Casimir of UIRs of de Sitter.}. Its two-point function is thus Weyl covariant and the conformal factor appears as $\Omega^{\mathbf{\Delta}-J}(x)=1$ when all indices are down. This means the two-point function of the field strength in conformally flat coordinates in de Sitter is identical to the one in flat space \cite{Glavan:2022dwb}
\begin{equation}
	\langle0| F_{\mu\nu}(y_1)F_{\rho\sigma}(y_2)|0\rangle=\frac{1}{\pi^2}\frac{1}{(y^2_{12})^2}\left[\eta_{\mu[\rho}\eta_{\sigma]\nu}+2\frac{y_{12,[\mu}\eta_{\nu][\rho}y_{12,\sigma]}}{y^2_{12}}\right]\,,
\label{eq:2ptff}
\end{equation}
where here $y_{12,\mu}=(\eta_{12},\mathbf{y}_{12})$, $y^2_{12}=-\eta_{12}^2+\mathbf{y}_{12}^2$ and $\eta_{12}\equiv\eta_1-\eta_2$ (analogously for $\mathbf{y}_{12}$) and $\eta_{\mu\nu}$ is the Minkowski metric. Throughout this paper we work with Wightman functions. This choice can be systematically implemented by taking each instance of the time variables and replacing them as follows: $\eta_1\to \eta_1 e^{i\epsilon}$ and $\eta_2\to\eta_2 e^{-i\epsilon}$ where $\epsilon$ is a small positive number. This ensures the first coordinate lies in the past tubular domain $\mathcal{T}_-$ and the second in the future tubular domain $\mathcal{T}_+$. For a more detailed discussion with the same notation and in a similar context, see \cite{Loparco:2023rug}. 

The equations of motion are Maxwell's equations in the vacuum
\begin{equation}
	\mathrm{d}*F=0\,, \qquad \mathrm{d}F=0\,,
\end{equation}
where to make contact with later sections we are using the language of differential forms, so $*$ is the Hodge dual and $\mathrm{d}$ is the exterior derivative\footnote{We use the following conventions
\begin{equation}
    (*A)_{\mu_1\cdots\mu_{4-p}}\equiv\frac{1}{p!}\epsilon^{\nu_1\cdots\nu_p}_{\qquad \ \mu_1\cdots\mu_{4-p}}A_{\nu_1\cdots\nu_p}\,, \qquad (dA)_{\mu_1\ldots\mu_p\mu_{p+1}}\equiv\partial_{[\mu_1}A_{\mu_2\ldots\mu_{p+1}]}
\end{equation} and the following identity will be useful both in flat space and in de Sitter embedding space
\begin{equation}
    \epsilon_{\alpha_1\ldots \alpha_m\mu_1\ldots \mu_n}\epsilon^{\alpha_1\ldots \alpha_m\nu_1\ldots \nu_n}=-m!\delta^{[\nu_1}_{\mu_1}\cdots \delta^{\nu_n]}_{\mu_n}\,.
\end{equation}}.
\paragraph{The physical EM fields}
We will be interested in the electromagnetic fields measured by a free falling observer in de Sitter. To such an observer we can associate the locally inertial frame defined by the set of tetrads $e^\mu_a$ with $a=0,1,2,3$ satisfying
\begin{equation}
    g_{\mu\nu}e^\mu_ae^\nu_b=\eta_{ab}
\end{equation}
If we start from conformally flat coordinates (\ref{eq:coordinates}), they are given by $e^\mu_a=\delta^\mu_a\eta$. Then, the physical electric and magnetic fields the observer will measure in their locally inertial frame are \cite{Subramanian:2015lua}\footnote{Notice that now $\mathbf{E}_a=\mathbf{E}^a$ and $\mathbf{B}_a=\mathbf{B}^a$   in terms of factors of conformal time. This is not the case for $F_{\mu\nu}$ and $F^{\mu\nu}$.}
\begin{equation}
    \mathbf{E}_a\equiv F_{\mu\nu}e_a^\mu U^\nu\,, \qquad \mathbf{B}^a\equiv \frac{1}{2}\epsilon^{\mu\nu\rho\lambda}F_{\mu\nu}U_\rho e_{\lambda}^{a}\,,
\end{equation}
For two observers at rest, with fixed comoving relative spatial distance $\mathbf{y}$ and with four-velocities $U^\mu=(\eta,0,0,0)$, we obtain the two-point functions at equal time
\begin{equation}
\begin{aligned}
    \langle0|\mathbf{E}^a (\eta,\mathbf{y})\mathbf{E}^b(\eta,0)|0\rangle&=\frac{1}{\pi^2}\frac{\eta^4}{\mathbf{y}^4}\left[2\frac{\mathbf{y}^a\mathbf{y}^b}{\mathbf{y}^2}-\delta^{ab}\right]\,,\\
    \langle0|\mathbf{B}^a(\eta,\mathbf{y})\mathbf{B}^b(\eta,0)|0\rangle&=\frac{1}{\pi^2}\frac{\eta^4}{\mathbf{y}^4}\left[2\frac{\mathbf{y}^a\mathbf{y}^b}{\mathbf{y}^2}-\delta^{ab}\right]\,.
    \label{eq:EBlate}
\end{aligned}
\end{equation}
while the mixed two-point functions $\langle \mathbf{E}^a\mathbf{B}^b\rangle$ vanish at equal times. 

\paragraph{Embedding space}
It will also be useful to think of the field strength in embedding space. To discuss the equations of motion there, we will need the generalization of the Hodge dual to embedding space
\begin{equation}
	(\star T)_{A_1\ldots A_{4-p}}\equiv\frac{1}{p!}\epsilon_{A_1\ldots A_{4-p}}^{\qquad\quad\ B_1\ldots B_p C}Y_{C}T_{B_1\ldots B_p}\,.
    \label{eq:definitionhodge}
\end{equation}
Notice that, even if the Levi Civita tensor has one more index in embedding space, the Hodge dual still sends $p$ forms into $4-p$ forms. Pulling back to local coordinates, this reduces to the usual notion of Hodge dual. We also generalize the exterior derivative
\begin{equation}
	(\mathbb{d}T)_{A_1\ldots A_pB}\equiv G^C_B 
     G^{C_1}_{A_1}\dots
        G^{C_p}_{A_p}
      \partial_{[C} T_{
      C_1\ldots C_p]}\,.
      \label{eq:definitionext}
\end{equation}
Then, the equations of motion of the field strength in embedding space are\footnote{We use $F$ to indicate the embedding space realization of the field strength as well as the pullback in local coordinates. Context will be enough to distinguish the two.}
\begin{equation}
	\mathbb{d}\star F=0\,, \qquad \mathbb{d}F=0\,.
    \label{eq:embedeoms}
\end{equation}
In embedding space, the two-point function of the field strength can be written as
\begin{equation}
	\langle 0|F^{AB}(Y_1)F^{CD}(Y_2)|0\rangle=\Pi^{ABCD}G_2(Y_1\cdot Y_2)\,,
    \label{eq:FFfromPi2}
\end{equation}
where $G_2(\sigma)$ is the  Wightman function of the scalar irrep with $\Delta=2$ in the Bunch-Davies vacuum (see eq. (\ref{eq:green2})) and $\Pi^{ABCD}$ is a differential operator enforcing (\ref{eq:embedeoms}), explicitly given in eq.  (\ref{eq:dsproj}) (see also eq. (\ref{eq:piplus2piminus2})).

\section{The K\"allén-Lehmann representation for antisymmetric tensors}
\label{sec:kallanlehmann}
In this section we derive the K\"allén-Lehmann representation for antisymmetric tensors with two indices in de Sitter. The K\"allén-Lehmann decomposition in de Sitter was first derived for scalars in \cite{Bros:1990cu,Bros:1995js,Bros:1998ik}, then revisited from the Hilbert space perspective in \cite{Hogervorst:2021uvp,DiPietro:2021sjt} and generalized to symmetric traceless tensors in \cite{Loparco:2023rug} and to spinors in \cite{Schaub:2023scu}. Our motivation to generalize it to antisymmetric bosonic operators with two indices is to study operators with the same symmetries as the electromagnetic field strength. Since there exist group theoretic arguments (which we review in Appendix \ref{subsec:grouptheory}) which state that the photon UIR appears in the decomposition of tensor products of massive irreps \cite{MartindS,Penedones:2023uqc}, we expect to find contributions from photon states in the K\"allén-Lehmann decomposition of composite operators built with massive fields, even if the theory being considered contains no $U(1)$ gauge field\footnote{Notice that the appearance of photon states in the tensor product of two massive states does not imply every composite operator made of two massive fields will create photons. For example, $:\phi_1\phi_2:$ does not. The expectation (which is then met by the examples we study in Section \ref{sec:photonsfrommass0}) is that there should be \emph{some} composite operator that creates photon states. We find that $\partial_{[\mu}\phi_1\partial_{\nu]}\phi_2$ does that.}. 

We start by reviewing the decomposition for such operators in flat space \ref{subsec:flatspacekl} and then present our derivation in de Sitter in \ref{subsec:dSkl}.
\subsection{Flat space}
\label{subsec:flatspacekl}
The K\"allén-Lehmann representation is a decomposition of two-point functions into a basis in which each element is associated to a specific UIR of the isometry group. In flat space, the isometries form the Poincar\'e group, and traceless symmetric UIRs are labeled by the mass $m^2$ and the $SO(3)$ spin $s$. We are also going to keep track of the parity quantum number, which we indicate as $+$ or $-$. States further carry a $SO(3)$ index, while tensor fields carry spacetime indices transforming under the Lorentz group $SO(3,1)$. The possible elements appearing in the K\"allén-Lehmann representation of a certain kind of tensor field is thus dependent on its reduction to $SO(3)$. 

The $SO(3)$ content of a single spacetime index can be expressed in terms of $SO(3)$ Young tableaux as $\mathbb{Y}_0\oplus\mathbb{Y}_1$\footnote{We emphasize that the Young tableaux used in this section are of the group $SO(3)$, while in Appendix \ref{sec:UIRs} and in Section \ref{subsec:dSkl} we use $SO(4)$ Young tableaux.}. Using the tensor product decomposition of Young tableaux
\begin{equation}
    \mathbb{Y}_n\otimes\mathbb{Y}_l=\bigoplus^{\text{min}(n,l)}_{a=0}\bigoplus^{\text{min}(n,l)-a}_{b=0}\mathbb{Y}_{n+l-2a-b,b}
\end{equation}
where $\mathbb{Y}_{n,m}$ denotes the two-row Young tableau with $n$ boxes in the first row and $m$ boxes in the second. We can thus write the $SO(3)$ content of a two-index tensor as
\begin{equation}
    (\bullet\oplus{\tiny\yng(1)})\otimes(\bullet\oplus{\tiny\yng(1)})=\bullet\oplus{\tiny\yng(1)}\oplus{\tiny\yng(1)}\oplus\raisebox{-3.5pt}{{\tiny\yng(1,1)}}\oplus{\tiny\yng(2)}\,.
\end{equation}
We are specifically interested in antisymmetric tensor fields. The only representations that are relevant are thus $\raisebox{-3.5pt}{\tiny\yng(1,1)}$ and the antisymmetric combination of the two ${\tiny\yng(1)}$'s. 
\begin{equation}
    (\bullet\oplus{\tiny\yng(1)})\otimes(\bullet\oplus{\tiny\yng(1)})\Big|_{\text{antisymm.}}={\tiny\yng(1)}\oplus\raisebox{-3.5pt}{\tiny\yng(1,1)}\,.
    \label{eq:tworeps}
\end{equation}
Since for $SO(3)$ the two representations ${\tiny\yng(1)}$ and $\raisebox{-3.5pt}{\tiny\yng(1,1)}$ are equivalent, we expect only spin 1 states to contribute.  A projector on the subspace of spin 1 states thus suffices to cover all possible contributions to a two-point function of antisymmetric fields
\begin{equation}
    \mathbb{1}_{s=1}=\sum_{\pm}\int_0^\infty dm^2\int_{\mathbf{p}}|m^2,\mathbf{p},\pm\rangle_a^{\ \ a}\langle m^2,\mathbf{p},\pm|\,,
    \label{eq:flatidentity}
\end{equation}
where $a$ is a spatial index and we introduced the shorthand notation for the integral over the Lorentz invariant phase space (LIPS)
\begin{equation}
    \int_{\mathbf{p}}\equiv\int\frac{d^3\mathbf{p}}{(2\pi)^3}\frac{1}{2E_{m,\mathbf{p}}}\,, \qquad E_{m,\mathbf{p}}\equiv\sqrt{m^2+p^2}\,, \qquad p\equiv|\mathbf{p}|\,.
\end{equation}
Moreover, we kept a parity label because this is a good quantum number in parity invariant theories.

Now consider inserting this resolution of the identity in the middle of the vacuum two-point function of an antisymmetric tensor field $B^{\mu\nu}(x)$
\begin{equation}
    \langle0|B^{\mu\nu}(x_1)B^{\rho\sigma}(x_2)|0\rangle=\sum_{\pm}\int_0^\infty \!dm^2\!\int_{\mathbf{p}}\langle0|B^{\mu\nu}(x_1)|m^2,\mathbf{p},\pm\rangle_{a}^{\ \ a}\langle m^2,\mathbf{p},\pm|B^{\rho\sigma}(x_2)|0\rangle\,.
\end{equation}
By Poincar\'e invariance, these matrix elements are fixed to take the form
\begin{equation}
    \langle0|B^{\mu\nu}(x)|m^2,\mathbf{p},\pm\rangle_{a}=c^{\pm}(m^2)e^{-ip\cdot x}\Psi^{\mu\nu}_{\pm,a}\,,
\end{equation}
where $c^{\pm}(m^2)$ are theory-dependent functions of the mass, $p\cdot x=-E_{m,\mathbf{p}}t+\mathbf{p}\cdot\mathbf{x}$, $a$ is a spatial index and $\Psi^{\mu\nu}_{\pm,a}$ are two tensor structures which are antisymmetric in $\mu,\nu$ and are eigenstates of parity with eigenvalue $\pm$. 

In practice, it is more convenient to treat $a$ as a spacetime index orthogonal to the 4-momentum $p^\mu$ so that it becomes a spatial index in the center-of-mass frame.
Explicitly, we can take the two tensor structures to be
\begin{equation}
    \Psi^{\mu\nu}_{+,a}=p^{[\mu}\delta^{\nu]}_a\,, \qquad \Psi^{\mu\nu}_{-,a}=\epsilon^{\mu\nu\alpha\beta}p_{\alpha}\eta_{\beta a}\,.
\end{equation}
Notice that $\Psi^{\mu\nu}_{\pm,\alpha} p^\alpha=0$ as it should.
These tensor structures can be used to build projectors onto parity odd and even spin 1 states
\begin{equation}
    \eta^{\alpha\beta} \Psi^{\mu\nu}_{\pm,\alpha}\Psi^{\rho\sigma*}_{\pm,\beta}=\widetilde\Pi^{\mu\nu\rho\sigma}_{\pm}\,, \qquad 
\eta^{\alpha\beta}\Psi^{\mu\nu}_{\pm,\alpha}\Psi^{\rho\sigma*}_{\mp,\beta}=0\,.
\end{equation}
Explicitly, 
\begin{equation}
    \widetilde\Pi^{\mu\nu\rho\sigma}_+=p^{[\mu}\eta^{\nu][\rho}p^{\sigma]}\,, \qquad \widetilde\Pi^{\mu\nu\rho\sigma}_-=\epsilon^{\mu\nu\alpha\beta}p_{\alpha}p_\gamma \epsilon^{\rho\sigma\gamma}_{\quad\beta}\,.
\end{equation}
All this implies that the K\"allén-Lehmann representation for antisymmetric tensors in flat space must take the form \cite{Hinterbichler:2022agn}
\begin{equation}
    \langle0|B^{\mu\nu}(x_1)B^{\rho\sigma}(x_2)|0\rangle=\sum_\pm\int_0^\infty dm^2\int_{\mathbf{p}}e^{-ip\cdot x_{12}}\varrho^{\pm}(m^2)\widetilde\Pi_{\pm}^{\mu\nu\rho\sigma}\,,
\end{equation}
where we introduced the notation $x_{12}^\mu\equiv x_1^\mu-x_2^\mu\,,$ and we defined the \textbf{spectral densities} $\varrho^{\pm}(m^2)\equiv|c^{\pm}(m^2)|^2$ which are positive if $B^{\mu\nu}$ is a hermitian operator. Its hermitianity infact would imply $\left(\langle0|B^{\mu\nu}(x)|m^2,\mathbf{p},\pm\rangle_a\right)^*=\ _a\langle m^2,\mathbf{p},\pm|B^{\mu\nu}(x)|0\rangle$.

Now we can carry out the integral over the LIPS by making use of the building block
\begin{equation}
    \int_{\mathbf{p}}e^{-ip\cdot x_{12}}=\frac{1}{(2\pi)^2}\frac{m}{|x_{12}|}K_1(m|x_{12}|)=G_{m^2}(x_{12})\,,
\end{equation}
where $K_\nu(x)$ is a Bessel function of the second kind and we recognize the scalar Wightman function $G_{m^2}(x_{12})$. We trade factors of momentum for derivatives with respect to $x_1$ and $x_2$ and finally we land on 
\begin{mdframed}[backgroundcolor=shadecolor,linewidth=0pt]
\begin{equation}
    \langle0|B^{\mu\nu}(x_1)B^{\rho\sigma}(x_2)|0\rangle=\sum_\pm\int_0^\infty dm^2\varrho^{\pm}(m^2)\Pi^{\mu\nu\rho\sigma}_\pm G_{m^2}(x_{12})\,.
    \label{eq:KLflat}
\end{equation}
\end{mdframed}
\noindent where the two terms labeled by $\pm$ correspond to parity even and parity odd states, and explicitly
\begin{equation}
    \Pi^{\mu\nu\rho\sigma}_+=\partial_1^{[\mu}\eta^{\nu][\sigma}\partial_2^{\rho]}\,, \qquad \Pi^{\mu\nu\rho\sigma}_-=\epsilon^{\mu\nu\alpha\beta}\partial_{1,\alpha}\partial_{2,\gamma} \epsilon^{\rho\sigma\gamma}_{\quad\ \beta}\,.    
    \label{eq:finalprojflat}
\end{equation}
The second projector can be further simplified
\begin{equation}
    \Pi_-^{\mu\nu\rho\sigma}=m^2\eta^{\mu[\rho}\eta^{\sigma]\nu}+\partial_1^{[\rho}\eta^{\sigma][\nu}\partial_2^{\mu]}
\end{equation}
Notice that these two projectors are related by Hodge duality with respect to the two points $x_1$ and $x_2$
\begin{equation}
    \Pi_-=*_1*_2\Pi_+\,.
\end{equation}
Finally, let us remark that the normalization is chosen such that, for the photon field strength,
\begin{equation}
    \varrho_F^\pm(m^2)=\frac{1}{2}\delta(m^2)\,,
\end{equation}
where $\delta(\cdot)$ is a Dirac delta.

Let us now move on to treat the de Sitter case.
\subsection{de Sitter}
\label{subsec:dSkl}
In de Sitter spacetime, tensor fields carry indices transforming in $SO(1,4)$. To understand which representations can appear when decomposing the two-point function of a certain tensor, we use the classification of UIRs in terms of $SO(4)$ content as reviewed in Appendix \ref{sec:UIRs}. The decomposition for an antisymmetric tensor with two indices that we found in (\ref{eq:tworeps}) is still valid for this case, but we must now interpret it in terms of $SO(4)$ YTs instead of $SO(3)$.
\begin{equation}
    (\bullet\oplus{\tiny\yng(1)})\otimes(\bullet\oplus{\tiny\yng(1)})\Big|_{\text{antisymm.}}={\tiny\yng(1)}\oplus\raisebox{-3.5pt}{\tiny\yng(1,1)}
\end{equation}
Looking among the various UIRs in \ref{sec:UIRs}, we see that possible contributions can come from the principal series $\mathcal{P}_{\Delta,0}$ and $\mathcal{P}_{\Delta,1}$, the complementary series $\mathcal{C}_{\Delta,0}$ and $\mathcal{C}_{\Delta,1}$, the massless scalar irrep $\mathcal{V}_{1,0}$ and the photon UIR $\mathcal{U}_{1,0}$.

Let us first consider the scalar representations. When inserting the resolution of the identity in the two-point function of an antisymmetric tensor $B^{AB}(Y)$, we will find matrix elements of the kind
\begin{equation}
    \langle0|B^{AB}(Y)|\Delta,P,\pm\rangle\,,
\end{equation}
where $|0\rangle$ is the Bunch-Davies vacuum.

This quantity must be antisymmetric in $A,B$, it should only depend on $P$, $Y$, the metric and the Levi Civita symbol, and it should be tangential with respect to both $P$ and $Y$. Moreover, the state $|\Delta,P,\pm\rangle$ must satisfy the homogeneity property reviewed in \ref{subsec:states}. It is impossible to construct such a quantity, and so we conclude that no scalar UIRs can contribute to the K\"allén-Lehmann decomposition of an antisymmetric tensor. 

Let us thus focus on the spin 1 UIRs $\mathcal{P}_{\Delta,1}$, $\mathcal{C}_{\Delta,1}$ and $\mathcal{U}_{1,0}$. We sum over all of them in the resolution of the identity. Schematically\footnote{In reality, the Hilbert spaces associated to principal, complementary or exceptional representations all have different inner products, so one should sum each representation with its own specific weight. As we showed in Appendix B of \cite{Loparco:2023rug}, the complementary series part turns out to be precisely the analytic continuation of the principal series part. Eq. (\ref{eq:identity}) should thus be interpreted as a prescription to construct complementary and photon state contributions as analytic continuations of the principal series part. Numerical checks in Section \ref{sec:photonsfrommass0} are compatible with the validity of this prescription.}, 
\begin{equation}
    \mathbb{1}_{s=1}=\sum_{\pm}\left(\int_{\frac{3}{2}+i\mathbb{R}}[d\Delta]+\int_1 ^2d\Delta+\delta_{\Delta,2}\right)\int_P
    |\Delta,P,\pm\rangle^A_{\ \ A}\langle\Delta,P,\pm|
    \label{eq:identity}
\end{equation}
where the first two contributions in parenthesis are from the principal and complementary series and the last one is the photon UIR, and we are summing over the parity quantum number $(\pm1)$. In reality, the sum over parity is redundant for the photon UIR, because in general exceptional type II UIRs are irreducible with respect to $O(1,4)$ \cite{Sun:2021thf}, but for now we keep it in order to give a unified treatment to all representations. $\delta_{a,b}$ is a Kronecker delta and the integral over $P$ is defined in (\ref{eq:intP}). We do not include the Bunch-Davies vacuum contribution $|0\rangle\langle0|$ because the vev $\langle0|B^{AB}(Y)|0\rangle$ has to vanish by symmetry.
Finally, we are using the notation $[d\Delta]\equiv -id\Delta$.

All the matrix elements we will find when introducing (\ref{eq:identity}) in a two-point function of an antisymmetric field $B^{AB}(Y)$ will then have the form 
\begin{equation}
    \mathcal{B}^{AB,C}_{\Delta,\pm}(Y,P)\equiv\langle0|B^{AB}(Y)|\Delta,P,\pm\rangle^C\,.
\end{equation}
The object $\mathcal{B}$ must be antisymmetric in $A,B$ and the tangentiality and homogeneity conditions imply that
\begin{equation}
\begin{aligned}
    &\mathcal{B}^{AB,C}_{\Delta,\pm}(Y,\lambda P)=\lambda^{-\Delta}\mathcal{B}_{\Delta,\pm}^{AB,C}(Y,P)\,,\qquad Y_A\mathcal{B}_{\Delta,\pm}^{AB,C}(Y,P)=0\,, \\
    &P_C\mathcal{B}_{\Delta,\pm}^{AB,C}(Y,P)=0\,, \qquad \qquad\qquad\qquad Y_B\mathcal{B}_{\Delta,\pm}^{AB,C}(Y,P)=0\,.
\end{aligned}
\end{equation}
Moreover, it has to have eigenvalue $\pm1$ under parity, and an $i\epsilon$ prescription is necessary to make sense of it when $Y$ is timelike separated from $P$. In terms of local coordinates (\ref{eq:coordinates}), we make the choice $\eta\to e^{i\epsilon}\eta$, with $\epsilon>0$. We refer to our previous work \cite{Loparco:2023rug} to see how this specific choice of prescription for the matrix elements will lead down the line to working with Wightman functions in the Bunch-Davies vacuum.  Taking into account all these facts, and keeping our $i\epsilon$ prescription implicit to avoid cluttering, the matrix elements can be chosen as
\begin{equation}
\begin{aligned}
    \mathcal{B}^{AB,C}_{\Delta,+}(Y,P)&= \kappa^{+}(\Delta)G^{C[B}\nabla^{A]}\frac{1}{(-2Y\cdot P)^{\Delta}}\,, \\
    {\mathcal{B}}_{\Delta,-}^{AB,C}(Y,P)&={\kappa}^{-}(\Delta)\epsilon^{ABCDE}Y_D\nabla_E\frac{1}{(-2Y\cdot P)^{\Delta}}\,,
\end{aligned}
\end{equation}
where
$G^{AB}=\eta^{AB}-Y^AY^B$ is the induced metric and $\kappa^{\pm}(\Delta)$ are some constants that depend on the theory, the operator and the UIR. 
Notice that, up to the normalization constant, the two tensor structures are related by embedding space hodge duality (\ref{eq:definitionhodge}).

To progress, notice that $\frac{1}{(-2Y\cdot P)^\Delta}$ is precisely the scalar bulk-to-boundary propagator, or the plane waves of de Sitter. Integrating over $P$ the products of matrix elements thus returns the Wightman two-point function thanks to the split representation \cite{Sleight:2019hfp}. See also \cite{Bros:1995js} for a general in-depth discussion of the analytic properties of plane waves and $i\epsilon$ prescriptions in de Sitter. Altogether, we obtain
\begin{align}
    \int_P \mathcal{B}^{AB,E}_{\Delta,+}(Y_1,P)\left(\mathcal{B}^{CD,}_{\Delta,+E}(Y_2,P)\right)^*&=\varrho^{+}(\Delta)G_{1,E}^{\ \ \ \ [A}\nabla_1^{B]}G_{2}^{\ E[C}\nabla_2^{D]}G_\Delta(Y_1\cdot Y_2)\,, \label{eq:dSproj}\\
    \int_P {\mathcal{B}}_{\Delta,-}^{AB,E}(Y_1,P)\left({\mathcal{B}}_{\Delta,-\ CD,E}(Y_2,P)\right)^*&=\varrho^{-}(\Delta)\delta^{[A}_C\delta^{B}_D\delta^{G}_H\delta^{F]}_IY_{1,F}\nabla_{1,G}Y_{2}^H\nabla_{2}^IG_{\Delta}(Y_1\cdot Y_2)\,,\nonumber
\end{align}
\noindent where $G_\Delta$ is the scalar Wightman two-point function in de Sitter
\begin{equation}
    G_\Delta(\sigma)\equiv\frac{\Gamma(\Delta)\Gamma(3-\Delta)}{(4\pi)^2}\ _2F_1\left(\Delta,3-\Delta,2,\frac{1+\sigma}{2}\right)\,, \qquad \sigma\equiv Y_1\cdot Y_2\,,
    \label{eq:freeprop}
\end{equation}
and, if $B^{AB}$ is hermitian, similarly to the flat space case, $\varrho^\pm(\Delta)$ are nonnegative quantities obtained by absorbing some ulterior constants into $|\kappa^\pm(\Delta)|^2$. We call them spectral densities.

Given (\ref{eq:dSproj}) and in analogy with the flat space case, we thus define the projectors
\begin{equation}
    \Pi^{ABCD}_+\equiv G_{1,E}^{\ \ \ \ [A}\nabla_1^{B]}G_{2}^{\ E[C}\nabla_2^{D]}\,, \qquad \Pi^{AB}_{-\ \  CD}\equiv\delta^{[A}_C\delta^{B}_D\delta^{G}_H\delta^{F]}_IY_{1,F}\nabla_{1,G}Y_{2}^H\nabla_{2}^I\,.
    \label{eq:dsproj}
\end{equation}
It can be checked that, as in flat space, 
\begin{equation}
	\Pi_-=\star_1\star_2\Pi_+\,.
\end{equation}
The K\"allén-Lehmann representation for antisymmetric tensors in dS$_4$ can then be obtained by inserting (\ref{eq:identity}) into a two-point function and carrying out the integrals over $P$ using (\ref{eq:dSproj}). The resulting expression is
\begin{mdframed}[backgroundcolor=shadecolor,linewidth=0pt]
\begin{equation}
\begin{aligned}
    \langle0| B^{AB}(Y_1)B^{CD}(Y_2)|0\rangle=&\sum_\pm\int_{\frac{3}{2}+i\mathbb{R}}[d\Delta]\ \varrho^{\mathcal{P},\pm}(\Delta)\Pi^{ABCD}_\pm G_\Delta(\sigma)\\
    &+\sum_\pm\int_{1}^2d\Delta\ \varrho^{\mathcal{C},\pm}(\Delta)\Pi^{ABCD}_\pm G_\Delta(\sigma)\\
    &+\varrho^{\gamma}\Pi^{ABCD}G_2(\sigma)\,,
    \label{eq:KLdS}
\end{aligned}
\end{equation}
\end{mdframed}
\noindent
where $G_2(\sigma)$ is the Wightman function of a scalar with $\Delta=2$ (equivalently we could have written $\Delta=1$), explicitly
\begin{equation}
    G_2(\sigma)=G_1(\sigma)=\frac{1}{8\pi^2}\frac{1}{1-\sigma}\,.\label{eq:green2}
\end{equation}
As can be noticed from its simple form, this is the Wightman function of a conformally coupled scalar. The action of the projector turns it into the Wightman function of the field strength in free Maxwell theory. There is only one projector in the last line of (\ref{eq:KLdS}) because
\begin{equation}
   \Pi^{ABCD}G_2(\sigma) \equiv  \Pi^{ABCD}_+G_2(\sigma)=\Pi^{ABCD}_-G_2(\sigma)\,,
   \label{eq:piplus2piminus2}
\end{equation}
so we have grouped the two terms into a single contribution from the photon UIR and we called the positive coefficient in front $\varrho^\gamma$, where $\gamma$ stands to represent the photon. The deeper reason behind (\ref{eq:piplus2piminus2}) is that exceptional type II representations are irreducible with respect to O$(1,4)$ \cite{Sun:2021thf}, in other words the free photon is self-dual and parity maps the two helicities into each other.

The normalization of the projectors is chosen such that, if $B^{AB}=F^{AB}$ is the field strength of the photon in free Maxwell theory, then $\varrho_F^\gamma=1$ (and all other densities of course vanish).

\subsection{Inversion formulae}
We have constructed the K\"allén-Lehmann representation for antisymmetric tensors with two indices in flat space (\ref{eq:KLflat}) and in de Sitter (\ref{eq:KLdS}). Here we focus on deriving inversion formulae to extract the spectral densities from a given two-point function.
\paragraph{Flat space} 
 The projectors (\ref{eq:finalprojflat}) satisfy the following useful properties\footnote{It can be checked that, for example,
 \begin{equation}
     \left((*\mathrm{d}*)_1(*\mathrm{d}*)_2\Pi_+\right)^{\mu\rho}=4\partial_{1,\nu}\partial_{2,\sigma}\Pi_+^{\mu\nu\rho\sigma}
 \end{equation}}
\begin{equation}
\begin{aligned}
    (*\mathrm{d}*)_1(*\mathrm{d}*)_2\Pi_+&=4m^4\Pi_1\,, \qquad &&(*\mathrm{d}*)_1(*\mathrm{d}*)_2\Pi_-=0\,,\\
    (*\mathrm{d})_1(*\mathrm{d})_2\Pi_+&=0\,,\qquad &&(*\mathrm{d})_1(*\mathrm{d})_2\Pi_-=4m^4\Pi_1\,,
    \label{eq:projids}
\end{aligned}
\end{equation}
where $\Pi_1$  is a projector on spin 1 states 
\begin{equation}
    \Pi_1^{\mu\nu}=\eta^{\mu\nu}+\frac{1}{m^2}\partial_1^\mu \partial_2^\nu\,.
\end{equation}
For a given antisymmetric operator $B^{\mu\nu}$, let us define the two conserved currents
\begin{equation}
    j\equiv*\mathrm{d}*B\,, \qquad \tilde j\equiv*\mathrm{d}B\,.
\end{equation}
Then, using (\ref{eq:projids}), we can write
\begin{equation}
\begin{aligned}
    \langle0|j^\mu(x_1)j^\nu(x_2)|0\rangle&=4\int_0^\infty dm^2\ m^4\varrho^+(m^2)\Pi^{\mu\nu}_1 G_{m^2}(x_{12})\,,\\
    \langle0|\tilde j^\mu(x_1)\tilde j^\nu(x_2)|0\rangle&=4\int_0^\infty dm^2\ m^4\varrho^-(m^2)\Pi^{\mu\nu}_1 G_{m^2}(x_{12})\,.
\end{aligned}
\end{equation}
obtained by acting with $(*\mathrm{d}*)_1(*\mathrm{d}*)_2$ and $(*\mathrm{d})_1(*\mathrm{d})_2$ on (\ref{eq:KLflat}). We have recasted the  decomposition of $B^{\mu\nu}$ as the decomposition of two spin 1 currents. The final step to extract the densities is to integrate against the appropriate plane wave
\begin{mdframed}[backgroundcolor=shadecolor,linewidth=0pt]
\begin{equation}
    \begin{aligned}
        \varrho^+(m^2)&=\frac{1}{4 m^4}\tilde\Pi_{1}^{\mu\nu}(p)\int d^4x_{12}\ e^{-i p\cdot x_{12}}\langle0|j_\mu(x_1)j_\nu(x_2)|0\rangle\\
        \varrho^-(m^2)&=\frac{1}{4m^4}\tilde \Pi_{1}^{\mu\nu}(p)\int d^4x_{12}\ e^{-i p\cdot x_{12}}\langle0|\tilde j_\mu(x_1)\tilde j_\nu(x_2)|0\rangle
    \end{aligned}
\end{equation}
\end{mdframed}
\noindent where $\tilde \Pi_{1}^{\mu\nu}(p)=\eta^{\mu\nu}+\frac{p^\mu p^\nu}{m^2}$ 
and $p^2=-m^2$.
\paragraph{de Sitter}
The situation is totally analogous in de Sitter, where 
\begin{align}
    &(\star\mathbb{d}\star)_1(\star\mathbb{d}\star)_2\Pi_+=4(\Delta-1)^2(\Delta-2)^2\Pi_1\,, \quad &&(\star\mathbb{d}\star)_1(\star\mathbb{d}\star)_2\Pi_-=0\,,\\
    &(\star \mathbb{d})_1(\star \mathbb{d})_2\Pi_+=0\,,\quad &&(\star \mathbb{d})_1(\star \mathbb{d})_2\Pi_-=4(\Delta-1)^2(\Delta-2)^2\Pi_1\,,\nonumber
\end{align}
where
\begin{equation}
    \Pi_1^{AB}=\frac{1}{\Delta\bar\Delta}\left(\Delta\bar\Delta\eta^{AB}+\bar\Delta Y_1^B\partial_{Y_1}^A+\Delta Y_2^A\partial_{Y_2}^{B}+\sigma \partial_{Y_1}^A\partial_{Y_2}^B\right)\,,
\end{equation}
is the projector on spin 1 states, or in other words $\Pi_1^{AB}G_\Delta(\sigma)$ is the propagator of a free Proca field \cite{Loparco:2023rug}.
These differential operators have the further property of killing the photon contribution
\begin{equation}
    (\star\mathbb{d}\star)_1(\star\mathbb{d}\star)_2\Pi^{ABCD}G_2(\sigma)=0\,, \qquad (\star \mathbb{d})_1(\star \mathbb{d})_2\Pi^{ABCD}G_2(\sigma)=0\,.
\end{equation}
This follows directly from the Maxwell equations, keeping in mind eq. (\ref{eq:FFfromPi2}). Analogously to the flat space case, for any operator $B^{AB}$ we thus define 
\begin{equation}
    j\equiv \star \mathbb{d}\star B\,, \qquad \tilde j\equiv \star\mathbb{d}B\,,
\end{equation}
such that we can write
\begin{equation}
\begin{aligned}
    \langle0|j^A(Y_1)j^B(Y_2)|0\rangle&=4\int_{\frac{3}{2}+i\mathbb{R}}[d\Delta](\Delta-1)^2(\Delta-2)^2\varrho^{\mathcal{P},+}(\Delta)\Pi^{AB}_1G_{\Delta}(\sigma)\\
    \langle0|\tilde j^A(Y_1)\tilde j^B(Y_2)|0\rangle&=4\int_{\frac{3}{2}+i\mathbb{R}}[d\Delta](\Delta-1)^2(\Delta-2)^2\varrho^{\mathcal{P},-}(\Delta)\Pi^{AB}_1G_{\Delta}(\sigma)
\end{aligned}
\end{equation}
where we are assuming for simplicity that at most we have contributions from the principal series and the photon UIR. We discuss how to know which UIRs appear in a two-point function in Appendix \ref{sec:criterion}.

This has precisely the form of the Källén-Lehmann decomposition of a spin 1 operator in de Sitter, which was studied in \cite{Loparco:2023rug}. In that paper we showed how to invert such decompositions. In particular, we make use of the continuation to Euclidean Anti-de Sitter space \cite{Sleight:2019hfp,Sleight:2021plv,Sleight:2020obc} to obtain the formulae 
\begin{mdframed}[backgroundcolor=shadecolor,linewidth=0pt]
\begin{equation}
\begin{aligned}
    \varrho^\mathcal{P,+}(\Delta)&=n_\Delta\int _{X_2}\langle0|j_B(X_2)j_A(X_1)|0\rangle \Pi_1^{AB}G_{\Delta}(X_1\cdot X_2)\\
    \varrho^\mathcal{P,-}(\Delta)&=n_\Delta\int_{X_2}\langle0|\tilde j_B(X_2)\tilde j_A(X_1)|0\rangle \Pi_1^{AB}G_{\Delta}(X_1\cdot X_2)\,.
    \label{eq:inversionsdS}
\end{aligned}
\end{equation}
\end{mdframed}
\noindent where $\int_X\equiv\int d^5X\delta(X^2+1)\theta(X^0)$ and 
\begin{equation}
    n_\Delta\equiv\frac{4 (3-2 \Delta ) \sin (2 \pi  \Delta )}{3 (\Delta -3)(\Delta-2)^2(\Delta-1)^2 \Delta }
\end{equation}
In general, these inversion formulae are generalizations of the scalar ones first derived in \cite{Bros:2009bz}. We elaborate more on how to perform these inversions practically in Appendix \ref{sec:criterion}.
\subsection{Two-point functions at late times}
\label{subsec:latetimes}
We would like to understand how to extract the late time behavior of the two-point function of an antisymmetric operator knowing its K\"allén-Lehmann representation (\ref{eq:KLdS}). Using the conformally flat coordinates (\ref{eq:coordinates}),  late times means taking $\eta_1,\eta_2\to0^-$. 

In a free theory, the two-point function of the elementary field has a late time behavior that is governed by two powers of conformal time, $\Delta$ and $3-\Delta$, where $\Delta$ is the scaling dimension of the free field being considered. More generically, and nonperturbatively, the powers of conformal time that govern the late-time behavior of a two-point function are fixed by the poles of the principal series spectral densities or by the complementary series contributions appearing in the K\"allén-Lehmann decomposition \cite{Hogervorst:2021uvp,Loparco:2023rug}\footnote{In those works ``boundary operators" are proposed to be in one-to-one correspondance with the powers in the late-time expansion of a two-point function, in analogy to QFT in AdS. However, these powers do not correspond with the values $\Delta$ takes in the UIRs of $SO(1,4)$, and so their Hilbert space interpretation is unclear. Perhaps the better way to construct such boundary operators is as proposed in \cite{SalehiVaziri:2024joi}.}. Here we generalize this fact to the case of antisymmetric tensors. For simplicity, in this section we assume that all complementary series contributions appear as a discrete sum.\footnote{To the best of our knowledge, there is no example of a QFT two-point function in which the complementary series appears as a continuum.}

The first step in our derivation is to use the following identity to split the de Sitter Wightman functions into functions that have a definite power law behavior at late times \cite{Sleight:2019hfp}
\begin{equation}
    G_{\Delta}(\sigma)=\frac{(\Delta-\frac{3}{2})\Gamma(\Delta-\frac{3}{2})\Gamma(\frac{3}{2}-\Delta)}{2\pi}\left(G^{\text{AdS}}_\Delta(\sigma)-G^{\text{AdS}}_{3-\Delta}(\sigma)\right)
    \label{eq:dStoAdS1}
\end{equation}
where
\begin{equation}
    G^{\text{AdS}}_\Delta(\sigma)=\frac{2^{-\Delta-1}\Gamma(\Delta)}{\pi^\frac{3}{2}\Gamma(\Delta-\frac{1}{2})}\frac{1}{(-1-\sigma)^\Delta}\ _2F_1\left(\Delta,\Delta-1,2\Delta-2,\frac{2}{1+\sigma}\right)
\end{equation}
is the propagator of a free scalar in EAdS with mass $m^2=\Delta(\Delta-3)$, and the domain of validity of this identity is $\Delta\in\mathbb{C}$ and $\sigma\in\mathbb{C}\backslash[-1,\infty)$.

The upshot is that at late times, corresponding to $\sigma\to-\infty$, we have $G_\Delta^{\text{AdS}}(\sigma)\sim|\sigma|^{-\Delta}$.
Plugging (\ref{eq:dStoAdS1}) into the K\"allén-Lehmann decomposition (\ref{eq:KLdS}) we thus get
\begin{equation}
\begin{aligned}
    \langle 0|B^{AB}(Y_1)B^{CD}(Y_2)|0\rangle=&2\sum_\pm\int_{\frac{d}{2}+i\mathbb{R}}[d\Delta]N_\Delta\varrho^{\mathcal{P},\pm}(\Delta)\Pi_\pm^{ABCD}G^{\text{AdS}}_\Delta(\sigma)\\
    &+\sum_{\pm}\sum_{\Delta\in\{\Delta_C^\pm\}}N_{\Delta}\varrho^{\mathcal{C},\pm}_{\Delta}\Pi_\pm^{ABCD}\left(G_{\Delta}^{\text{AdS}}(\sigma)-G^{\text{AdS}}_{3-\Delta}(\sigma)\right)\\
    &+N_2\varrho^\gamma\left(\Pi^{ABCD}_++\Pi_-^{ABCD}\right)G^{\text{AdS}}_2(\sigma)
    \label{eq:klads}
\end{aligned}
\end{equation}
where $\{\Delta_C^\pm\}$ is the set of scaling dimensions of the complementary series contributions with parity even or odd\footnote{Here we take the elements of this set to be $\Delta_C^\pm\in(1,\frac{3}{2})$ without loss of generality}. Moreover, we used the symmetry of the integral over the principal series to simplify the expression and we used the following fact 
\begin{equation}
   \Pi_+^{ABCD}G_1^{\text{AdS}}(\sigma)=-\Pi_-^{ABCD}G_2^{\text{AdS}}(\sigma)\,,
\end{equation}
and defined the factor
\begin{equation}
    N_\Delta\equiv\frac{(\Delta-\frac{3}{2})\Gamma(\Delta-\frac{3}{2})\Gamma(\frac{3}{2}-\Delta)}{2\pi}\,.
    \label{eq:defndelta}
\end{equation}
Let us emphasize that eq. (\ref{eq:klads}) is only valid for large spacelike separation, specifically for $\sigma<-1$. 

\begin{figure}
\centering
\includegraphics[scale=1.45]{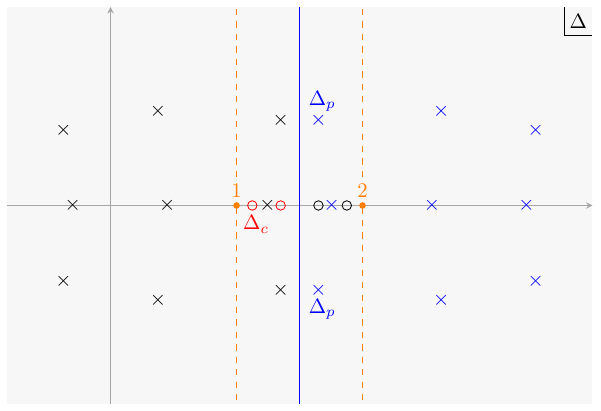}
\caption{A representation of the salient features of the K\"allén-Lehmann decomposition. The blue vertical line is the contour of integration over the principal series, the crosses are the poles of the principal series spectral density, the empty circles are complementary series contributions, which come as a discrete sum in all known examples, and the full orange circles represent the photon UIR. In eq. (\ref{eq:latetimeB}) we indicate the set of scaling dimensions corresponding to the red circles as $\{\Delta_C\}$ and the blue crosses as $\{\Delta_P\}$. $\Delta_c$ and $\Delta_p$ are the scaling dimensions with lowest real part of the sets $\{\Delta_C\}$ and $\{\Delta_P\}$ respectively. In the particular configuration shown in this picture, the dominant late time power law is set by $\Delta_c$. This figure could apply to both the parity even $(+)$ and odd $(-)$ contributions.}
\label{fig:deltapdeltac}
\end{figure}

The advantage of writing the K\"allén-Lehmann representation in this form is that the functions $G^{\text{AdS}}_\Delta$ decay exponentially for large Re$\Delta$, so the contour of integration over the principal series can be closed to the right, picking up the residues on the poles in the spectral densities\footnote{We discuss this fact further at the end of this section.}. Doing that, gives
\begin{equation}
\begin{aligned}
    \langle 0|B^{AB}(Y_1)B^{CD}(Y_2)|0\rangle=&-4\pi\sum_{\pm}\sum_{\Delta\in\{\Delta_P^\pm\}}N_{\Delta}\underset{\Delta'=\Delta}{\text{Res}}\left[\varrho^{\mathcal{P},\pm}(\Delta')\right]\Pi_\pm^{ABCD}G_{\Delta}^{\text{AdS}}(\sigma)\\
    &+\sum_{\pm}\sum_{\Delta\in\{\Delta_C^\pm\}}N_{\Delta}\varrho^{\mathcal{C},\pm}_{\Delta}\Pi_\pm^{ABCD}\left(G_{\Delta}^{\text{AdS}}(\sigma)-G^{\text{AdS}}_{3-\Delta}(\sigma)\right)\\
    &+N_2\varrho^\gamma\left(\Pi^{ABCD}_++\Pi_-^{ABCD}\right)G^{\text{AdS}}_2(\sigma)\,.
    \label{eq:latetimeB}
\end{aligned}
\end{equation}
where $\{\Delta_P^\pm\}$ is the set of poles with Re$\Delta>\frac{3}{2}$ of the principal series density $\varrho^{\mathcal{P},\pm}$. For the sake of clarity, we represent an example of the sets $\{\Delta_P^\pm\}$ and $\{\Delta_C^\pm\}$ in Figure \ref{fig:deltapdeltac}.

To proceed and really go to late times we need to specify the components of the two-point function and pull back to local coordinates. To simplify the notation and relate to physical observables, let us define the electric and magnetic components of the generic operator $B^{\mu\nu}$ as\footnote{We use $\mathbf{E}^{(B)}$ and $\mathbf{B}^{(B)}$ to indicate the electric and magnetic components associated to the operator $B_{\mu\nu}$, instead $\mathbf{E}$ and $\mathbf{B}$ indicate the electromagnetic fields of free Maxwell theory.}
\begin{equation}
   \mathbf{E}^{(B)}_a\equiv B_{\mu\nu}e^\mu_a U^\nu \,, \qquad \mathbf{B}^{(B)}_a\equiv \frac{1}{2}\epsilon^{\mu\nu\rho\lambda}B_{\mu\nu}U_\rho e^a_\lambda 
   \label{eq:EBdefinitions}
\end{equation}
Moreover, we will consider conformally flat coordinates (\ref{eq:coordinates}), for which $e^\mu_a=\delta^\mu_a\eta$ and we will focus on an observer at rest $U^\mu=(\eta,0,0,0)$. Keeping the details of how to pull back the tensor structures in Appendix \ref{subsec:bases2}, we obtain
\begin{mdframed}[backgroundcolor=shadecolor,linewidth=0pt]
\begin{equation}
\begin{aligned}
    \langle 0|\mathbf{E}^{(B)}_a(\eta,\mathbf{y})\mathbf{E}^{(B)}_b(\eta,0)|0\rangle\stackrel{\eta\to0}{\longrightarrow}&-4\pi\sum_\pm c_{\Delta_p^\pm}^\pm\underset{\Delta=\Delta_p^\pm}{\text{Res}}\left[\varrho^{\mathcal{P},\pm}(\Delta)\right]\left(\frac{\eta^2}{\mathbf{y}^2}\right)^{\Delta_p ^\pm+\frac{1}{2}\mp\frac{1}{2}}I_{ab}^\pm \\
    &+\sum_{\pm}c^\pm_{\Delta_c^\pm}\varrho_{\Delta_c^\pm}^{\mathcal{C},\pm}\left(\frac{\eta^2}{\mathbf{y}^2}\right)^{\Delta_c ^\pm+\frac{1}{2}\mp\frac{1}{2}}I_{ab}^\pm+c_2\varrho^\gamma\frac{\eta^4}{\mathbf{y}^4}I_{ab}^+
    \label{eq:lateB0i}
\end{aligned}
\end{equation}
\end{mdframed}
\noindent where $\Delta_p^\pm$ and $\Delta_c^\pm$ are the scaling dimensions with lowest real part in the sets $\{\Delta_P^\pm\}$ and $\{\Delta_C^\pm\}$ respectively, and
\begin{equation}
\begin{aligned}
&I_{ab}^+\equiv\delta_{ab}-\frac{\mathbf{y}_{a}\mathbf{y}_{b}}{\mathbf{y}^2}\,, \qquad &&c^+_\Delta\equiv\frac{(\Delta-1)\Gamma(\frac{3}{2}-\Delta)\Gamma(\Delta+1)}{4\pi^\frac{5}{2}}\,,\\
&I_{ab}^-\equiv \delta_{ab}\Delta-(\Delta+1)\frac{\mathbf{y}_{a}\mathbf{y}_{b}}{\mathbf{y}^2}\,,&&c^-_\Delta\equiv-\frac{2\Gamma(\frac{3}{2}-\Delta)\Gamma(\Delta+1)}{\pi^\frac{5}{2}}\,.
\label{eq:iij}
\end{aligned}
\end{equation}
For the ``magnetic" components instead we obtain
\begin{mdframed}[backgroundcolor=shadecolor,linewidth=0pt]
\begin{equation}
\begin{aligned}
    \langle0|\mathbf{B}^{(B)}_a(\eta,\mathbf{y})\mathbf{B}^{(B)}_b(\eta,0)|0\rangle\stackrel{\eta\to0}{\longrightarrow}&-4\pi\sum_\pm c_{\Delta_p^\pm}^\mp\underset{\Delta=\Delta_p^\pm}{\text{Res}}\left[\varrho^{\mathcal{P},\pm}(\Delta)\right]\left(\frac{\eta^2}{\mathbf{y}^2}\right)^{\Delta_p ^\pm+\frac{1}{2}\pm\frac{1}{2}}I_{ab}^\mp \\
    &+\sum_{\pm}c^\mp_{\Delta_c^\pm}\varrho_{\Delta_c^\pm}^{\mathcal{C},\pm}\left(\frac{\eta^2}{\mathbf{y}^2}\right)^{\Delta_c ^\pm+\frac{1}{2}\pm\frac{1}{2}}I_{ab}^\mp+c_2\varrho^\gamma\frac{\eta^4}{\mathbf{y}^4}I_{ab}^+
     \label{eq:lateBij}
\end{aligned}
\end{equation}
\end{mdframed}
\noindent
Let summarize the notable properties that these equations manifest, and let us point the reader to the examples in Section \ref{sec:photonsfrommass0} for further clarity.
\begin{itemize}
    \item If there is a complementary series contribution with scaling dimension $\Delta_c^+$ or a pole in the principal series density with $\frac{3}{2}<\text{Re}\Delta_p^+<2$, the two-point function of $\mathbf{E}^{(B)}$ dominates at late times and large distances over the two-point function of the electric field in free Maxwell theory. That is because the factors of $(\eta/|\mathbf{y}|)$ in (\ref{eq:lateB0i}) appear with power $2\Delta<4$ in these cases. The simplest example in which this happens is the case of the field strength of a free massive vector field, explored in Section \ref{subsec:proca}.
    \item For similar reasons, if there is a complementary series contribution with scaling dimension $\Delta_c^-$ or a pole in the principal series density with $\frac{3}{2}<\text{Re}\Delta_p^-<2$, the two-point function of $\mathbf{B}^{(B)}$ dominates at late times over the two-point function of the magnetic field in free Maxwell theory. An example in which this happens is the field strength of a massive pseudo-vector, also explored in Section \ref{subsec:proca}.
    \item There can be cancelations between some of the terms appearing in equations (\ref{eq:lateB0i}) and (\ref{eq:lateBij}). For example, in many of our examples, the photon UIR contributions get canceled by poles in the principal series. That happens if
    \begin{equation}
        4\pi\underset{\Delta=2}{\text{Res}}\left[\varrho^{\mathcal{P},+}(\Delta)\right]=\varrho^\gamma\,,\qquad 4\pi\underset{\Delta=2}{\text{Res}}\left[\varrho^{\mathcal{P},-}(\Delta)\right]=\varrho^\gamma
    \end{equation}
   
    \item If these cancelations do not happen and if there are no complementary series and principal series contributions with $\frac{3}{2}<\text{Re}\Delta_p^\pm<2$, the two-point function of $\mathbf{B}^{(B)}$ and/or of $\mathbf{E}^{(B)}$ can behave at late times exactly like the two-point function of the electric and magnetic fields in free Maxwell theory (up to the constant prefactor $\varrho^\gamma$). For example, the two-point function of $:\partial_{[i}\phi_1\partial_{j]}\phi_2:$ with $\phi_1$ and $\phi_2$ being two massless free scalars, behaves at late times exactly like $F_{ij}$ (see Section \ref{subsec:masslessscalars}).
\end{itemize}
These properties are very surprising from the point of view of a particle physicist. We are used to think of photons as the longest range excitation, with massive states instead showing exponential decay with the distance. This is certainly true on subhorizon scales in de Sitter. What we are finding is that, on superhorizon scales, some massive two-point functions dominate over massless ones. A possible intuition for this phenomenon comes from thinking of energy densities: in an expanding FRW universe with scale factor $a$, radiation dilutes as $a^{-4}$ while matter dilutes as $a^{-3}$, leaving massive states dominating over photons at late times.

\paragraph{An important caveat} Let us emphasize a subtle but important assumption that is needed to carry out this derivation. When we close the contour over the principal series to go from  (\ref{eq:klads}) to (\ref{eq:latetimeB}), we are assuming that the spectral densities do not grow too fast as Re$\Delta\to\infty$, otherwise they would overcome the exponential decay of the propagators and prevent us from dropping the arc at infinity.

It has been recently shown \cite{SalehiVaziri:2024joi} that the spectral density of a scalar operator is polynomially bounded for large $|\Delta|$ in the whole complex $\Delta$ plane, if the associated two-point function has a CFT-like singularity in the UV\footnote{In \cite{Chakraborty:2023eoq}, vertex operators were studied. Two-point functions of these operators in $d+1>2$ have more severe singularities than CFT two-point functions. Consequently, the spectral densities grow exponentially instead of polynomially, but with a behavior that is anyways overtaken by the propagators. Our arguments thus apply to these kinds of operators too.}. We assume a similar behavior for the antisymmetric tensor two-point functions we are considering, and we checked a posteriori that all spectral densities we derive in Section \ref{sec:photonsfrommass0} satisfy this property.

\subsection{A bound on magnetic fields at late times}
\label{subsec:pmf}
Magnetic fields are observed at every scale in our Universe \cite{Giovannini:2003yn,Subramanian:2015lua}. There are tantalizing hints that they might even be present in the cosmic voids \cite{Neronov_2010,Takahashi:2013lba,Hosking:2022umv}. Such magnetic fields are sometimes hypothesized to have a primordial origin, possibly tracing back to fluctuations in the magnetic component of the $U(1)_Y$ gauge field during inflation, later imprinted in the $U(1)$ magnetic field after electroweak symmetry breaking \cite{Joyce:1997uy}. 

As explained in Section \ref{subsec:freemaxwell}, the physical magnetic field measured by free falling observers in their locally inertial frame in free Maxwell theory in de Sitter has the following two-point function
\begin{equation}
    \langle 0| \mathbf{B}^a(\eta,\mathbf{y})\mathbf{B}^b(\eta,0)|0\rangle=-\frac{1}{\pi^2} \frac{H^4\eta^4}{\mathbf{y}^4}\left(\delta^{ab}-2\frac{\mathbf{y}^a\mathbf{y}^b}{\mathbf{y}^2}\right)\,.
\end{equation}
where we reinstated the appropriate factor of Hubble, and $a,b$ are tetrad indices, raised and lowered with the Minkowski metric.

If we approximate inflation as a period of de Sitter expansion that lasted $\sim60$ e-folds,  the resulting fluctuations are extremely suppressed and cannot seed the magnetic fields we observe in the Universe \cite{Subramanian:2015lua}. Nevertheless, there have been numerous works attempting to modify free Maxwell theory in order to alleviate the late time decay (see \cite{Demozzi:2009fu,Turner:1987bw,Durrer:2013pga,Subramanian:2015lua,Kandus:2010nw,Maleknejad:2012fw,Kobayashi:2014zza} for some examples). 

Based on what we proved on the late time scaling of two-point functions of antisymmetric operators (\ref{eq:lateBij}), we can say that, nonperturbatively, in a unitary and parity preserving QFT on rigid de Sitter, the late-time behavior of the two-point function of the magnetic component of any antisymmetric operator $B_{\mu\nu}$ has to be
\begin{mdframed}[backgroundcolor=shadecolor,linewidth=0pt]
\begin{equation}
    \langle 0| \mathbf{B}^{(B)}_a(\eta,\mathbf{y})\mathbf{B}^{(B)}_b(\eta,0)|0\rangle\stackrel{\eta\to0}{\longrightarrow} c\left(\frac{\eta^2}{\mathbf{y}^2}\right)^\Delta\left(\delta_{ab}-2\frac{\mathbf{y}_a\mathbf{y}_b}{\mathbf{y}^2}\right)\,, \qquad \Delta>1
    \label{eq:Bbound}
\end{equation}
\end{mdframed}
\noindent where $c$ is some dimensionful constant to match with the dimensions of the operator $B_{\mu\nu}$.

Of course, here we are assuming a rigid de Sitter spacetime. To understand what would happen with a slight breaking of the de Sitter isometries, as in inflation, one would need to carry out a more careful analysis, which we leave for future studies.

Let us point out that the bound (\ref{eq:Bbound}) is saturated by the magnetic field of a dual photon with an infinitesimal but nonzero mass (we explore this in more detail in Section \ref{subsec:proca}), which has two-point function at late times
\begin{equation}
    \langle 0| \mathbf{B}_a(\eta,\mathbf{y})\mathbf{B}_b(\eta,0)|0\rangle_{\text{dual Proca}}\stackrel{\eta\to0}{\longrightarrow}m^2\frac{H^2\eta^{2}}{\mathbf{y}^2}\left(\delta_{ab}-2\frac{\mathbf{y}_a\mathbf{y}_b}{\mathbf{y}^2}\right)\,.
\end{equation}
This case is especially interesting because the electric field of a dual massive photon instead still decays at late times like that of a photon
\begin{equation}
    \langle 0| \mathbf{E}_a(\eta,\mathbf{y})\mathbf{E}_b(\eta,0)|0\rangle_{\text{dual Proca}}\stackrel{\eta\to0}{\longrightarrow} c\frac{\eta^{4}}{\mathbf{y}^4}\left(\delta_{ab}-2\frac{\mathbf{y}_a\mathbf{y}_b}{\mathbf{y}^2}\right)
\end{equation}
This is interesting from a phenomenological perspective because usually mechanisms that generate magnetic fields at late times are accompanied by the generation of even larger electric fields \cite{Kobayashi:2014zza}. A totally analogous bound can be derived from (\ref{eq:lateB0i}) for the components of the electric field
\begin{equation}
    \langle 0| \mathbf{E}^{(B)}_a(\eta,\mathbf{y})\mathbf{E}^{(B)}_b(\eta,0)|0\rangle\stackrel{\eta\to0}{\longrightarrow}c\left(\frac{\eta^2}{\mathbf{y}^2}\right)^\Delta\left(\delta_{ab}-2\frac{\mathbf{y}_a\mathbf{y}_b}{\mathbf{y}^2}\right)\,, \qquad \Delta>1\,.
\end{equation}
This bound is saturated by the electric component of the field strength of a regular Proca field with an infinitesimal but nonzero mass, which behaves as
\begin{equation}
\begin{aligned}
    \langle 0| \mathbf{E}_a(\eta,\mathbf{y})\mathbf{E}_b(\eta,0)|0\rangle_{\text{Proca}}&\stackrel{\eta\to0}{\longrightarrow} m^2\frac{H^2\eta^{2}}{\mathbf{y}^2}\left(\delta_{ab}-2\frac{\mathbf{y}_a\mathbf{y}_b}{\mathbf{y}^2}\right)\\
 \langle 0| \mathbf{B}_a(\eta,\mathbf{y})\mathbf{B}_b(\eta,0)|0\rangle_{\text{Proca}}&\stackrel{\eta\to0}{\longrightarrow}c\frac{\eta^{4}}{\mathbf{y}^4}\left(\delta_{ab}-2\frac{\mathbf{y}_a\mathbf{y}_b}{\mathbf{y}^2}\right)
\end{aligned}
\end{equation}
It would be very interesting to understand whether these facts have any  phenomenological relevance, but that is beyond the scope of this paper.
\section{Photons without photons: free theories}
\label{sec:photonsfrommass0}
Armed with all the technical results of the previous sections, here we discuss some interesting facts regarding massive free theories in de Sitter. It has been known for some time \cite{MartindS,Penedones:2023uqc} that photon states appear in the tensor product of single particle states in the exceptional type I or principal and complementary series (we review the arguments in \ref{subsec:grouptheory}). What we present is the concrete realization of this statement in QFT: a variety of composite operators formed with free massive fields have a non-zero photon contribution $\varrho^\gamma$ in their K\"allén-Lehmann representation, implying that they interpolate between the vacuum and states in the  photon UIR. Moreover, we also find that any two-index antisymmetric conformal primary in a conformal field theory in de Sitter creates photon states. We extract the spectral densities for each case with the methods outlined in Appendix \ref{sec:criterion}.

In this section we also study the late time behavior of the two-point functions of these operators. We find that in several cases the late time and large distance behaviors of these two-point functions dominate over those of the electric and magnetic fields in free Maxwell theory.  

For a summary of the results of this section, see Table \ref{tab:summary}.
\subsection{Free scalars}
\label{subsec:masslessscalars}
As a first example, let us consider the case of two different free scalar fields $\phi_1$ and $\phi_2$ with massess $m_i^2=\Delta_i(3-\Delta_i)$ with $i=1,2$. This is an interesting case because, reflecting the group theory treatement of Appendix \ref{subsec:grouptheory}, we will see that there is no photon contribution until we tune both masses to be 0. Moreover, the late time behavior will dominate over that of the Maxwell field strength for some range of the masses. 

The composite operator of interest is 
\begin{equation}
    B^{AB}(Y)=\nabla^{[A}\phi_1\nabla^{B]}\phi_2(Y)\,,
\end{equation}
where normal ordering is implicit.

The two-point function of this operator can be computed by carrying out the appropriate Wick contractions. In Appendix \ref{sec:criterion} we present a criterion (\ref{eq:criterion}) to discern which representations have non-vanishing spectral densities in the K\"allén-Lehmann decomposition of a two-point function. Applying it to this case, we see that there is no complementary series contribution when 
\begin{equation}
    \frac{1}{2}<\text{Re}[\Delta_1]+\text{Re}[\Delta_2]<\frac{11}{2}\,,\qquad \text{only principal series UIRs}\,.
    \label{eq:regimenphinphi}
\end{equation}
We thus start from this regime, which includes all cases where at least one of the two scalars is in the principal series, and a portion of the cases in which both scalars are in the complementary series. We apply our inversion formulae (\ref{eq:inversionsdS}) and obtain that the only nonvanishing spectral density in this regime is\footnote{The notation $\Gamma(a\pm i\mu)$ in the denominator stands for the product $\Gamma(a+ i\mu)\Gamma(a- i\mu)$. The numerator contains a product over 8 $\Gamma-
$functions.}
\begin{equation}
\begin{aligned}
    \varrho_{\partial\phi_1\partial\phi_2}^{\mathcal{P},+}(\frac{3}{2}+i\mu)&=\frac{\mu\sinh\pi\mu}{12\pi^5\Gamma(1+\frac{3}{2}\pm i\mu)}\prod_{\pm,\pm,\pm}\Gamma\left(\frac{1+\frac{3}{2}\pm i\mu\pm i\mu_1\pm i\mu_2}{2}\right)\,.
\end{aligned}
\end{equation}
where we are using the notation $\Delta=\frac{3}{2}+i\mu$.
The parity odd spectral density vanishes, consistently with the fact that one cannot build parity odd states with two scalars.

The form of this spectral density is of course similar to that of $\varrho_{\phi_1\phi_2}$, which was first derived in \cite{Bros:2009bz}. This kind of structure, with a product of eight gamma functions, will keep returning in the following examples.

As discussed in Section \ref{subsec:latetimes}, the late time behavior of a two-point function is dominated by the pole of the principal series density with Re$\Delta>\frac{3}{2}$ that is closest to the Re$\Delta=\frac{3}{2}$ axis. We call it $\Delta_p^+$, where the plus indicates it is a pole of $\varrho_{\partial\phi_1\partial\phi_2}^{\mathcal{P},+}$.
As we analytically continue $\Delta_1$ and $\Delta_2$ on the complementary series (take $\Delta_i\in(0,\frac{3}{2})$ without loss of generality) and towards the extremes of the regime (\ref{eq:regimenphinphi}), we have
\begin{equation}
    \Delta_p^+=\Delta_1+\Delta_2+1
\end{equation}
Notice that, when we reach the regime $\frac{1}{2}<\Delta_1+\Delta_2<1$, we have
\begin{equation}
    \frac{3}{2}<\Delta_p^+<2
\end{equation}
Following the discussion in \ref{subsec:latetimes}, in this regime the two-point function of the electric components of $\partial_{[\mu}\phi_1\partial_{\nu]}\phi_2$, defined in (\ref{eq:EBdefinitions}), goes at late times as 
\begin{mdframed}[backgroundcolor=shadecolor,linewidth=0pt]
\begin{equation}
    \langle0|\mathbf{E}_a^{(\partial\phi_1\partial\phi_2)}(\eta,\mathbf{y})\mathbf{E}_b^{(\partial\phi_1\partial\phi_2)}(\eta,0)|0\rangle\stackrel{\eta\to0}{\longrightarrow} c_{\Delta_1,\Delta_2}\left(\frac{\eta^2}{\mathbf{y}^2}\right)^{\Delta_p^+}\left(\delta_{ab}-2\frac{\mathbf{y}_{a}\mathbf{y}_{b}}{\mathbf{y}^2}\right)\,,
    \label{eq:latemassivescalars}
\end{equation}
\end{mdframed}
\noindent
which shows a slower large distance decay than the two-point function of the electric field in free Maxwell theory (\ref{eq:EBlate}). The constant $c_{\Delta_1\Delta_2}$ is given by (using eq. (\ref{eq:lateB0i}))
\begin{equation}
    c_{\Delta_1\Delta_2}\equiv\frac{(\Delta_1+\Delta_2)\Gamma(\Delta_1+1)\Gamma(\Delta_2+1)\Gamma(\frac{3}{2}-\Delta_1)\Gamma(\frac{3}{2}-\Delta_2)}{8\pi^5}
\end{equation}
The purely spatial components $\partial_{[i}\phi_1\partial_{j]}\phi_2$ are instead still subleading with respect to $F_{ij}$ (see (\ref{eq:lateBij})).
We can then continue beyond (\ref{eq:regimenphinphi}) towards lighter scalar fields. As we surpass the threshold value $\Delta_1+\Delta_2=\frac{1}{2}$, poles at
\begin{equation}
    \Delta=\Delta_1+\Delta_2+1\,,\qquad \Delta=2-\Delta_1-\Delta_2
\end{equation}
cross the integration contour, and the residues on their positions need to be added to the K\"allén-Lehmann decomposition, just like in the scalar case \cite{Bros:2009bz}. Since they are related by shadow symmetry, their contribution is the same. They correspond to a complementary series UIR, and so now there is a new non-vanishing spectral density
\begin{equation}
\begin{aligned}
    \varrho^{\mathcal{C},+}_{\partial\phi\partial\phi,\Delta_1+\Delta_2+1}=-4\pi\underset{\Delta=\Delta_1+\Delta_2+1}{\text{Res}}\left[\varrho^{\mathcal{P},+}_{\partial\phi_1\partial\phi_2}(\Delta)\right]
\end{aligned}
\end{equation}
Now the late time behavior is governed by the complementary series contribution with lowest scaling dimension, $\Delta_c^+=\Delta_1+\Delta_2+1$, and so the expression at late times is still the analytic continuation of (\ref{eq:latemassivescalars}).

We can continue further, all the way to $m_1^2=m_2^2=0$. In that limit, we have $\Delta_1=\Delta_2=0$ and the complementary series contribution with $\Delta=\Delta_1+\Delta_2+1=1$ matching the scaling dimension associated to the photon UIR. We thus obtain the following K\"allén-Lehmann representation for massless scalars
\begin{equation}
\begin{aligned}    \langle0|\nabla_1^{[A}\phi_1\nabla_1^{B]}\phi_2(Y_1)\nabla_2^{[C}\phi_1\nabla_2^{D]}\phi_2(Y_2)|0\rangle=&\int_{\frac{3}{2}+i\mathbb{R}}[d\Delta]\varrho^{\mathcal{P},+}_{\partial\phi_1\partial\phi_2}(\Delta)\Pi^{ABCD}_+G_\Delta(\sigma)\\
    &+\varrho^\gamma_{\partial\phi_1\partial\phi_2}\Pi^{ABCD}G_2(\sigma)\,,
    \label{eq:masslesskl}
\end{aligned}
\end{equation}
where now
\begin{mdframed}[backgroundcolor=shadecolor,linewidth=0pt]
\begin{equation}
    \varrho^{\mathcal{P},+}_{\partial\phi_1\partial\phi_2}(\Delta)=\frac{(5-\Delta) (\Delta +2) (2 \Delta -3) \cot (\pi  \Delta )}{384 \pi ^2}\,, \qquad \varrho^\gamma_{\partial\phi_1\partial\phi_2}=\frac{1}{8\pi^2}\,.
    \label{eq:rhomassless}
\end{equation}
\end{mdframed}
and the photon spectral density is simply $\varrho_{\partial\phi_1\partial\phi_2}^\gamma=\varrho^{\mathcal{C},+}_{\partial\phi_1\partial\phi_2,1}$. We thus see that this operator creates states in the photon UIR when acting on the vacuum. 

At late times, formula (\ref{eq:latemassivescalars}) cannot be trusted anymore due to the fact that the coefficient $c_{\Delta_1,\Delta_2}$ vanishes as $\Delta_1\to\Delta_2\to0$. At the same time, the photon UIR contribution to the late-time expansion of $\mathbf{E}_a^{(\partial\phi_1\partial\phi_2)}$ is canceled by the pole in the principal series density
\begin{equation}
    4\pi\underset{\Delta=2}{\text{Res}}\left[\varrho^{\mathcal{P},+}_{\partial\phi_1\partial\phi_2}(\Delta)\right]=\varrho^\gamma
\end{equation}
so the two-point function of $\mathbf{E}_a^{(\partial\phi_1\partial\phi_2)}$ becomes subleading to the two-point function of a pure electric field $\mathbf{E}_a$ at late times. Instead, in the same limit, the magnetic components go as
\begin{mdframed}[backgroundcolor=shadecolor,linewidth=0pt]
\begin{equation}
     \langle 0|\mathbf{B}^{(\partial\phi_1\partial\phi_2)}_a(\eta,\mathbf{y})\mathbf{B}^{(\partial\phi_1\partial\phi_2)}_b(\eta,0)|0\rangle\stackrel{\eta\to0}{\longrightarrow} \varrho^\gamma_{\partial\phi_1\partial\phi_2}\langle0| \mathbf{B}_a(\eta,\mathbf{y})\mathbf{B}_b(\eta,0)|0\rangle_{\rm free \ EM}
     \label{eq:latephiiphij}
\end{equation}
\end{mdframed}
\paragraph{Flat space limit} The flat space limit is achieved by setting $\Delta\to i mR$, restoring all factors of the de Sitter radius $R$ necessary to match units and finally taking $R\to \infty$.  Under this procedure the spectral weight on the photon UIR simply goes to zero
\begin{equation}
    \varrho^\gamma_{\partial\phi_1\partial\phi_2}=\frac{1}{8\pi^2R^4}\to0
\end{equation}
and the principal series spectral densities account for all the flat space UIRs
\begin{equation}
    \varrho_{\partial\phi_1\partial\phi_2}^+(m^2)=\frac{m^2}{192\pi^2}\,, \qquad \varrho_{\partial\phi_1\partial\phi_2}^-(m^2)=0\,.
\end{equation}
Notice the absence of a photon contribution in flat space, which would be signaled by a Dirac delta supported at $m^2=0$.
\subsection{Free vector and free scalar}
\label{subsec:vectorscalar}
As a second example, we consider a massive vector $V^A$ and a massive scalar $\phi$ in a free theory with masses $m_V^2=(\Delta_V-1)(2-\Delta_V)$ and $m_\phi^2=\Delta_\phi(3-\Delta_\phi)$ taken to be on the principal series. We study the normal ordered operator 
\begin{equation}
    B^{AB}(Y)=V^{[A}\nabla^{B]}\phi(Y)\,.
\end{equation}
Applying our criterions (\ref{eq:criterion}) we find that the two-point function of this operator never contains complementary series contributions as long as the masses are on the principal series. Using our inversion formulae we obtain the two principal series spectral densities
\begin{equation}
\begin{aligned}
    \varrho_{V\partial\phi}^{\mathcal{P},+}(\frac{3}{2}+i\mu)&=\frac{\mu\sinh\pi\mu f(\mu,\mu_V,\mu_\phi)}{768\pi^5(1+4\mu^2)(1+4\mu_V^2)\Gamma(1+\frac{3}{2}\pm i\mu)}\prod_{\pm,\pm,\pm}\Gamma\left(\frac{\frac{3}{2}\pm i \mu\pm i \mu_\phi\pm i\mu_V}{2}\right)\,,\\
    \varrho_{V\partial\phi}^{\mathcal{P},-}(\frac{3}{2}+i\mu)&=\frac{2\mu\sinh(\pi\mu)}{3\pi^5(1+4\mu^2)\Gamma(1+\frac{3}{2}\pm i\mu)}\prod_{\pm,\pm,\pm}\Gamma\left(\frac{1+\frac{3}{2}\pm i\mu\pm i\mu_V\pm i \mu_\phi}{2}\right)\,,
\end{aligned}
\end{equation}
where 
\begin{equation}
\begin{aligned}
	f(\mu,\mu_V,\mu_\phi)=&64\mu^6+16\mu^4(33-8\mu_\phi^2)+3(4\mu_\phi^2+9)^2\\
&+4\mu^2(16\mu_\phi^2+48\mu_\phi^2(2\mu_V^2+3)-48\mu_V^2(\mu_V^2+4)+243)\\
&+16(15-16\mu_\phi^2)\mu_V^4+32\mu_\phi^2\mu_V^2(4\mu_\phi^2+9)+128\mu_V^6\,.
\end{aligned}
\end{equation}
and a nonvanishing photon contribution
\begin{mdframed}[backgroundcolor=shadecolor,linewidth=0pt]
\begin{equation}
    \varrho^\gamma_{V\partial\phi}=\frac{(\mu_V^2-\mu_\phi^2)(1+(\mu_V-\mu_\phi)^2)(1+(\mu_V+\mu_\phi)^2)}{96\sinh(\pi(\mu_V-\mu_\phi))\sinh(\pi(\mu_V+\mu_\phi))}\,.
\end{equation}
\end{mdframed}
It is interesting to observe that this operator creates photon states even if it is composed of massive elementary fields.

At late times, the two-point function of this operator has a subleading behavior with respect to the Maxwell field strength, because the leading poles have Re$\Delta_p^\pm=3>2$, and the photon UIR contribution is canceled by poles in the principal series densities
\begin{equation}
   4\pi\ \underset{\Delta=2}{\text{Res}}\left[\varrho_{V\partial\phi}^{\mathcal{P},+}(\Delta)\right]=4\pi\ \underset{\Delta=2}{\text{Res}}\left[\varrho_{V\partial\phi}^{\mathcal{P},-}(\Delta)\right]= \varrho^\gamma_{V\partial\phi}
\end{equation}
Nevertheless, if we continue to the complementary series, we observe that there is a range of the parameters, namely $1<\Delta_V+\Delta_\phi<2$, for which the pole with lowest real part in the parity even principal series density, $\Delta_p^+=\Delta_V+\Delta_\phi$, satisfies $\frac{3}{2}<\Delta_p^+<2$, meaning that at late times the electric components (defined in (\ref{eq:EBdefinitions})) go like
\begin{mdframed}[backgroundcolor=shadecolor,linewidth=0pt]
\begin{equation}
    \langle0|\mathbf{E}^{(V\partial\phi)}_a(\eta,\mathbf{y})\mathbf{E}^{(V\partial\phi)}_b(\eta,0)|0\rangle\stackrel{\eta\to0}{\longrightarrow} c_{\Delta_V,\Delta_\phi}\left(\frac{\eta^2}{\mathbf{y}^2}\right)^{\Delta_p^+}\left(\delta_{ab}-2\frac{\mathbf{y}_{a}\mathbf{y}_{b}}{\mathbf{y}^2}\right)\,.
\end{equation}
\end{mdframed}
\noindent This two-point function presents a slower large distance decay than the late times two-point function of the electric field in free Maxwell theory, just like in the case (\ref{eq:latemassivescalars}).
\paragraph{Flat space limit}
As for the previous example, we take the flat space limit by setting $\mu\to mR$ as well as $\mu_\phi\to m_\phi R$ and $\mu_V\to m_V R$, introducing the correct factors of $R$ to restore units and taking $R\to\infty$. The photon contribution vanishes exponentially as
\begin{equation}
    \varrho^\gamma_{V\partial\phi}\stackrel{R\to\infty}{\longrightarrow}\frac{R^4}{24}|m_V^2-m_\phi^2|^3e^{-2\pi\ \text{max}(m_\phi,m_V)\ R}
\end{equation}
The principal series spectral densities instead account for all massive states in flat space
\begin{align}
        \varrho_{V\partial\phi}^{+}(m^2)&=\theta(m^2-(m_\phi+m_V)^2)\frac{\left(m^2-(m_\phi+m_V)^2\right)^\frac{1}{2}\left(m^2-(m_\phi-m_V)^2\right)^\frac{1}{2}}{192\pi^2m_V^2m^6}f(m,m_\phi,m_V)\,,\nonumber\\
        \varrho_{V\partial\phi}^{-}(m^2)&=\theta(m^2-(m_\phi+m_V)^2)\frac{\left(m^2-(m_\phi+m_V)^2\right)^\frac{3}{2}\left(m^2-(m_\phi-m_V)^2\right)^\frac{3}{2}}{96\pi^2m^6}\,.
\end{align}
where we used the Heaviside $\theta-$function   and 
\begin{equation}
    f(m,m_\phi,m_V)\equiv (m_\phi^2-m_V^2+m^2)^2(2m_V^2+m^2)+4m^2m_\phi^2(m_V^2-m^2) 
\end{equation}
Again we see that there is no photon contribution in flat space.
\subsection{Free vectors}
\label{subsec:massivevectors}
An ulterior example is the case of two massive vectors $V_1^A$ and $V_2^A$ with masses $m_i^2=(\Delta_i-1)(2-\Delta_i)$ initially taken to be on the principal series. We study the normal ordered operator
\begin{equation}
    B^{AB}(Y)=V_1^{[A}V_2^{B]}\,.
\end{equation}
Again this operator does not create states in complementary series irreps if $\Delta_i$ are on the principal series. With the inversion formulas, we obtain
\begin{equation}
\begin{aligned}
    \varrho_{V_1V_2}^{\mathcal{P},+}(\frac{3}{2}+i\mu)&=\frac{4\mu\sinh\pi\mu[53+4(\mu^2+2(\mu_1^2+\mu_2^2))]}{3\pi^5\Gamma(1+\frac{3}{2}\pm i\mu)(1+4\mu^2)(1+4\mu_1^2)(1+4\mu_2^2)}\prod_{\pm,\pm,\pm}\Gamma\left(\frac{1+\frac{3}{2}\pm i\mu\pm i\mu_1\pm i\mu_2}{2}\right)\\
    \varrho_{V_1V_2}^{\mathcal{P},-}(\frac{3}{2}+i\mu)&=\frac{\mu\sinh\pi\mu f(\mu,\mu_1,\mu_2)}{12\pi^5\Gamma(1+\frac{3}{2}\pm i\mu)(1+4\mu^2)(1+4\mu_1^2)(1+4\mu_2^2)}\prod_{\pm,\pm,\pm}\Gamma\left(\frac{\frac{3}{2}\pm i\mu\pm i\mu_1\pm i\mu_2}{2}\right)\,,
\end{aligned}
\end{equation}
with
\begin{align}
    f(\mu,\mu_1,\mu_2)=&16 \Big(\mu ^4 \left(2 \mu_1^2+2 \mu_2^2+1\right)+\mu ^2 \left(-4 \mu_1^4+\mu_1^2 \left(16 \mu_2^2+19\right)-4 \mu_2^4+19 \mu_2^2+9\right)\nonumber\\
    &+2 \left(\mu_1^6-\mu_1^4 \left(\mu_2^2-8\right)-\mu_1^2 \mu_2^2 \left(\mu_2^2+7\right)+\mu_2^4 \left(\mu_2^2+8\right)\right)\Big)\\
    &+9 \left(58 \mu_1^2+58 \mu_2^2+27\right)\nonumber
\end{align}
and a nonzero photon contribution
\begin{mdframed}[backgroundcolor=shadecolor,linewidth=0pt]
\begin{equation}
    \varrho_{V_1V_2}^\gamma
    =\frac{(\mu_1^2-\mu_2^2)(1+(\mu_1-\mu_2)^2)(1+(\mu_1+\mu_2)^2)(13+2(\mu_1^2+\mu_2^2))}{12(1+4\mu_{1}^2)(1+4\mu_2^2)\sinh(\pi(\mu_1+\mu_2))\sinh(\pi(\mu_1+\mu_2))}\,.
\end{equation}
\end{mdframed}
Once again, this operator creates states in the photon UIR  despite being composed of massive fields. At late times, the photon contribution is canceled by poles in the principal series spectral densities
\begin{equation}
4\pi\ \underset{\Delta=2}{\text{Res}}\left[\varrho^{\mathcal{P},+}_{V_1V_2}(\Delta)\right]=4\pi\ \underset{\Delta=2}{\text{Res}}\left[\varrho^{\mathcal{P},-}_{V_1V_2}(\Delta)\right]=\varrho^{\gamma}_{V_1V_2}\,.
\end{equation}
Moreover, in this case the principal series poles with lowest real parts Re$\Delta>\frac{3}{2}$ are at Re$\Delta_p^+=4$ and Re$\Delta_p^-=3$. Analytically continuing the two vectors to the complementary series, we cannot push these poles below $\Delta_p^+=3$ and $\Delta_p^-=2$ without violating the unitarity bounds $1<\Delta_{1,2}<2$, meaning that there is no regime of masses for which this two-point function behaves at late times and large distances with a slower decay than the two-point functions of $F_{\mu\nu}$.
\paragraph{Flat space limit}
In the flat space limit, the photon density once again decays exponentially
\begin{equation}
    \varrho^{\gamma}_{V_1V_2}\stackrel{R\to\infty}{\longrightarrow}\frac{R^4}{24}|m_1^2-m_2^2|^3\frac{m_1^2+m_2^2}{m_1^2m_2^2}e^{-2\pi\ \text{max}(m_1,m_2)\ R}
\end{equation}
The principal series densities, instead, become the continuum of massive states
\begin{equation}
\begin{aligned}
    \varrho_{V_1V_2}^{+}(m^2)&=\theta(m^2-(m_1+m_2)^2)\frac{(m^2-(m_1+m_2)^2)^\frac{3}{2}(m^2-(m_1-m_2)^2)^\frac{3}{2}(m^2+2(m_1^2+m_2^2))}{192\pi^2m^6m_1^2m_2^2}\,,\\
\varrho_{V_1V_2}^{-}(m^2)&=\theta(m^2-(m_1+m_2)^2)\frac{(m^2-(m_1+m_2)^2)^\frac{1}{2}(m^2-(m_1-m_2)^2)^\frac{1}{2}f(m,m_1,m_2)}{96\pi^2m^6m_1^2m_2^2}\,,
\end{aligned}
\end{equation}
where
\begin{equation}
    f(m,m_1,m_2)\equiv m^4(m_1^2+m_2^2)+(m_1^2-m_2^2)^2(m_1^2+m_2^2)-2m^2(m_1^4-4m_1^2m_2^2+m_2^4)
\end{equation}
and no photon contribution is present in flat space.
\subsection{CFT}
\label{subsec:cft}
This phenomenon also happens for antisymmetric conformal primaries in CFTs in de Sitter. 
The conformal group of dS$_4$ is $SO(2,4)$, so it is useful to consider the super-embedding of dS$_4$ into $\mathbb{R}^{4,2}$ with metric $\eta=\text{diag}(-1,1,\ldots,1,-1)$ and to restrict ourselves to its light-cone 
\begin{equation}
    \mathcal{Y}^2=Y^2-(\mathcal{Y}^5)^2=0\,,
\end{equation}
where $Y^A\in\mathbb{R}^{4,1}$ and $\mathcal{Y}^{\mathcal{A}}\in\mathbb{R}^{4,2}$. Then, the two-point function of some antisymmetric operator $\mathcal{O}^{\mathcal{A}\mathcal{B}}(\mathcal{Y})$ with conformal dimension $\mathbf{\Delta}$\footnote{We use $\mathbf{\Delta}$ for the label of the quadratic Casimir of $SO(2,4)$ to distinguish it from $\Delta$, the label of the quadratic Casimir of $SO(1,4)$.} must satisfy
\begin{itemize}
    \item Homogeneity: $\mathcal{O}^{\mathcal{A}\mathcal{B}}(\lambda\mathcal{Y})=\lambda^{-\mathbf{\Delta}}\mathcal{O}^{\mathcal{A}\mathcal{B}}(\mathcal{Y})$
    \item Tangentiality: $\mathcal{Y}_{\mathcal{A}}\mathcal{O}^{\mathcal{A}\mathcal{B}}(\mathcal{Y})=0\,.$
\end{itemize}
There are three tensor structures that satisfy homogeneity and have the right index symmetry
\begin{equation}
    T_1=\frac{\eta^{\mathcal{C}[\mathcal{A}}\eta^{\mathcal{B}]\mathcal{D}}}{(-2\mathcal{Y}_1\cdot\mathcal{Y}_2)^{\mathbf{\Delta}}}\,, \qquad T_2=\frac{\mathcal{Y}_1^{[\mathcal{A}}\eta^{\mathcal{B}][\mathcal{C}}\mathcal{Y}_2^{\mathcal{D}]}}{(-2\mathcal{Y}_1\cdot\mathcal{Y}_2)^{\mathbf{\Delta+1}}}\,, \qquad T_3=\frac{\mathcal{Y}_1^{[\mathcal{A}}\mathcal{Y}_2^{\mathcal{B}]}\mathcal{Y}_1^{[\mathcal{C}}\mathcal{Y}_2^{\mathcal{D}]}}{(-2\mathcal{Y}_1\cdot\mathcal{Y}_2)^{\mathbf{\Delta}+2}}\,.
\end{equation}
Imposing tangentiality, we land on a unique linear combination of these tensor structures for the two-point function of an antisymmetric CFT primary
\begin{equation}
    \langle0|\mathcal{O}^{\mathcal{A}\mathcal{B}}(\mathcal{Y}_1)\mathcal{O}^{\mathcal{C}\mathcal{D}}(\mathcal{Y}_2)|0\rangle=\frac{c_{\mathcal{O}}}{2}\frac{\mathcal{W}_{12}^{\mathcal{A}][\mathcal{C}}\mathcal{W}_{21}^{\mathcal{D}][\mathcal{B}}}{(-2\mathcal{Y}_1\cdot\mathcal{Y}_2)^{\mathbf{\Delta}}}\,, \qquad \mathcal{W}_{ij}^{\mathcal{A}\mathcal{B}}=\eta^{\mathcal{A}\mathcal{B}}-\frac{\mathcal{Y}_i^{\mathcal{A}}\mathcal{Y}_j^{\mathcal{B}}}{\mathcal{Y}_i\cdot\mathcal{Y}_j}\,,
\end{equation}
where $c_{\mathcal{O}}$ is a normalization factor.
Specifying to the de Sitter section of the lightcone, $(\mathcal{Y}^5=1)$, we get
\begin{equation}
    \langle0|\mathcal{O}^{AB}(Y_1)\mathcal{O}^{CD}(Y_2)|0\rangle=\frac{c_{\mathcal{O}}}{2}\frac{W^{AC}W^{BD}}{(2-2Y_1\cdot Y_2)^{\mathbf{\Delta}}}\,, 
    \label{eq:CFT2p}
\end{equation}
where $[AB]$ and $[CD]$ are antisymmetrized and 
\begin{equation}
    W^{AB}\equiv \eta^{AC}-\frac{Y_1^AY_1^C+Y_2^AY_2^C-Y_2^AY_1^C-(Y_1\cdot Y_2)Y_1^AY_2^C}{1-Y_1\cdot Y_2}\,.
\end{equation}
For completeness we also report the expression in conformally flat coordinates
\begin{equation}
    \langle0|\mathcal{O}_{\mu\nu}(y_1)\mathcal{O}_{\rho\sigma}(y_2)|0\rangle=\frac{c_{\mathcal{O}}}{(y_{12}^2)^{\mathbf{\Delta}}}(\eta_1\eta_2)^{\mathbf{\Delta}-2}\left(\eta_{\mu[\rho}\eta_{\sigma]\nu}+2\frac{y_{12,[\mu}\eta_{\nu][\rho}y_{12,\sigma]}}{y_{12}^2}\right)\,.
\end{equation}
Comparing with eq. (\ref{eq:2ptff}) we recognize that the field strength is a conformal primary with $\mathbf{\Delta}=2$ and $c_F=\pi^{-2}$.

Applying the inversion formulae (\ref{eq:inversionsdS}) we obtain the following principal series spectral densities
\begin{equation}
\begin{aligned}
    \varrho_{\mathcal{O}}^{\mathcal{P},+}(\Delta)&=c_{\mathcal{O}}\frac{\pi(2\Delta-3)\cos(\pi\Delta)}{2^{2\mathbf{\Delta}-3}(\Delta-1)(\Delta-2)}\frac{\Gamma(\mathbf{\Delta}-\Delta)\Gamma(\mathbf{\Delta}-\bar\Delta)}{\Gamma(\mathbf{\Delta}-2)\Gamma(\mathbf{\Delta}+1)}\\
    \varrho_{\mathcal{O}}^{\mathcal{P},-}(\Delta)&=\varrho^{\mathcal{P},+}_{\mathcal{O}}(\Delta)\,,
\end{aligned}
\end{equation}
and the photon contribution
\begin{mdframed}[backgroundcolor=shadecolor,linewidth=0pt]
\begin{equation}
    \varrho^\gamma_{\mathcal{O}}=\frac{2^{5-2\mathbf{\Delta}}\pi^2c_{\mathcal{O}}}{\mathbf{\Delta}(\mathbf{\Delta}-1)}\,.
    \label{eq:cftphoton}
\end{equation}
\end{mdframed}
The positivity of these densities implies the unitarity bound for the scaling dimension of an antisymmetric CFT primary $\mathbf{\Delta}>2$. The equality $\varrho_{\mathcal{O}}^{\mathcal{P},+}(\Delta)=\varrho^{\mathcal{P},-}_{\mathcal{O}}(\Delta)$ is due to the fact that the two-point function of an antisymmetric conformal primary is self dual. In other words, if we define $\tilde{\mathcal{O}}\equiv*\mathcal{O}$, we have
\begin{equation}
    \langle0|\tilde{\mathcal{O}}^{\mu\nu}(y_1)\tilde{\mathcal{O}}^{\rho\sigma}(y_2)|0\rangle=\langle0|\mathcal{O}^{\mu\nu}(y_1)\mathcal{O}^{\rho\sigma}(y_2)|0\rangle\,.
\end{equation}
Notice that the presence of the photon UIR (\ref{eq:cftphoton}) is independent of $\mathbf{\Delta}$: any two-index antisymmetric conformal primary creates states in the photon UIR when acting on the vacuum in de Sitter. Moreover, if we plug in $\mathbf{\Delta}=2$ and $c_F=\pi^{-2}$ we retrieve $\varrho_F^\gamma=1$, consistent with our choice of normalization of this spectral density.

At late times, the photon contribution is canceled by poles in the principal series spectral densities
\begin{equation}
    4\pi \underset{\Delta=2}{\text{Res}}\left[\varrho_{\mathcal{O}}^{\mathcal{P},+}(\Delta)\right]=4\pi \underset{\Delta=2}{\text{Res}}\left[\varrho_{\mathcal{O}}^{\mathcal{P},-}(\Delta)\right]=\varrho^\gamma_{\mathcal{O}}\,,
\end{equation}
and the leading poles in the spectral densities, at $\Delta_p^\pm=\mathbf{\Delta}>2$, lead to a late time behavior that is subleading to that of the two-point function of the photon field strength.
\paragraph{Flat space limit} The photon UIR density in the flat space limit goes like
\begin{equation}
   \varrho^\gamma_{\mathcal{O}}\stackrel{R\to\infty}{\longrightarrow}c_{\mathcal{O}}R^{4-2\mathbf{\Delta}}\frac{4^{3-\mathbf{\Delta}}\pi^2}{\mathbf{\Delta}(\mathbf{\Delta}-1)}\,,
   \label{eq:cftphotonflat}
\end{equation}
where we see that it survives only if $\mathbf{\Delta}=2$, meaning if the operator is the photon field strength. The principal series densities instead reduce to
\begin{equation}
\begin{aligned}
    \varrho_{\mathcal{O}}^{+}(m^2)&=c_{\mathcal{O}}\frac{2^{5-2\mathbf{\Delta}}\pi^2m^{2\mathbf{\Delta}-6}}{\Gamma(\mathbf{\Delta}-2)\Gamma(\mathbf{\Delta}+1)}\\
    \varrho_{\mathcal{O}}^{-}(m^2)&=\varrho_{\mathcal{O}}^{+}(m^2)\,.
\end{aligned}
\end{equation}
While for $2<\mathbf{\Delta}<3$ there is support for these spectral densities at $m^2=0$, this does not indicate the creation of photon states in flat space: they are measure zero in the K\"allén-Lehmann integrals and their presence would be indicated by a delta function, which is what (\ref{eq:cftphotonflat}) reduces to if $\mathbf{\Delta}=2$.
\subsection{Field strengths of massive vector and pseudo-vector}
\label{subsec:proca}
As a final example we would like to discuss the theories of a free massive vector \cite{Gazeau:1999xn} and of a free pseudo-vector with both masses parametrized as $m_V^2=(\Delta_V-1)(2-\Delta_V)\,.$ In particular, consider the field strength operators $H^{AB}_\pm\equiv\nabla^{[A}V^{B]}_\pm$, where the $\pm$ here indicates the intrinsic parity of the vector field. These operators do not create states in the photon UIR unless $m_V^2=0$, but their two-point functions have very interesting late time behaviors. 

Given that the field strengths are operators built of one elementary field, it is not surprising that their two-point functions are
\begin{equation}
    \langle0|H^{AB}_\pm(Y_1)H^{CD}_\pm(Y_2)|0\rangle=\Pi_\pm^{ABCD}G_\Delta(Y_1\cdot Y_2)\,,
\end{equation}
so that their spectral densities are
\begin{equation}
    \varrho_{H_\pm}^{\mathcal{P},\pm}(\Delta)=\frac{1}{2}\left(\delta(\Delta-\Delta_V)+\delta(\Delta-\bar\Delta_V)\right)\,, \quad \varrho^{\mathcal{P},\mp}_{H_\pm}(\Delta)=0\,, \quad \varrho^\gamma_{H_\pm}=0\,,
\end{equation}
where we assumed $V^A_\pm$ to be in the principal series\footnote{Analogously, if $V_\pm$ where to be on the complementary series the principal series density would vanish and the complementary series $(\pm)$ density would be a Dirac delta.}.  

Analogously to what we did in free Maxwell theory, let us define the ``electric and magnetic fields" associated to these massive vectors in locally inertial coordinates
\begin{equation}
    \mathbf{E}^{\pm}_a\equiv H_{\pm,\mu\nu}e^\mu_a U^\nu\,, \qquad \mathbf{B}^{\pm,a}\equiv\frac{1}{2}\epsilon^{\mu\nu\rho\lambda}H_{\pm,\mu\nu}U_{\rho}e_\lambda^a
\end{equation}
Then, to extract the late time behavior we use the functional identity (\ref{eq:dStoAdS1}) and pull back to conformally flat coordinates (\ref{eq:coordinates}). Using (\ref{eq:lateads}) we get, for an observer with four-velocity $U^\mu=(\eta,0,0,0)$,
\begin{equation}
\begin{aligned}
    \langle0|\mathbf{E}^\pm_{a}(\eta,\mathbf{y})\mathbf{E}^\pm_{b}(\eta,0)|0\rangle&\stackrel{\eta\to0^-}{\longrightarrow}\left(c_1^\pm\left(\frac{\eta^2}{\mathbf{y}^2}\right)^{\Delta_V+\frac{1}{2}\mp\frac{1}{2}}+c_2^\pm\left(\frac{\eta^2}{\mathbf{y}^2}\right)^{\frac{7}{2}-\Delta_V\mp\frac{1}{2}}\right)I^\pm_{ab}+\ldots\\
    \langle0|\mathbf{B}^\pm_a(\eta,\mathbf{y})\mathbf{B}^\pm_b(\eta,0)|0\rangle&\stackrel{\eta\to0^-}{\longrightarrow}\left(c_3^\pm\left(\frac{\eta^2}{\mathbf{y}^2}\right)^{\Delta_V+\frac{1}{2}\pm\frac{1}{2}}+c_4^\pm\left(\frac{\eta^2}{\mathbf{y}^2}\right)^{\frac{7}{2}-\Delta_V\pm\frac{1}{2}}\right)I_{ab}^\mp+\ldots
\end{aligned}
\end{equation}
where $I^\pm_{ab}$ were defined in (\ref{eq:iij}),
$c_i^\pm$ are some unimportant constants and the dots stand for subleading contributions at late times.

Let us point out the notable aspects of these two-point functions. If we compare with the electric and magnetic field two-point functions of free Maxwell theory (\ref{eq:EBlate}), which we report here for convenience
\begin{equation}
\begin{aligned}
\langle0|\mathbf{E}^a(\eta,\mathbf{y})\mathbf{E}^b(\eta,0)|0\rangle&=\frac{1}{\pi^2}\frac{H^4\eta^4}{\mathbf{y}^4}\left(\delta^{ab}-2\frac{\mathbf{y}^a\mathbf{y}^b}{\mathbf{y}^2}\right)\,,\\
\langle0|\mathbf{B}^a(\eta,\mathbf{y})\mathbf{B}^b(\eta,0)|0\rangle&=\frac{1}{\pi^2}\frac{H^4\eta^4}{\mathbf{y}^4}\left(\delta^{ab}-2\frac{\mathbf{y}^a\mathbf{y}^b}{\mathbf{y}^2}\right)\,,
\label{eq:maxwelleb}
\end{aligned}
\end{equation}
we can state that, for any value of the mass in the principal or complementary series, the electric field of a massive vector $(+)$ and the magnetic field of a massive pseudo-vector $(-)$ dominate over those of a photon at late times and large distances. In particular, the most dominant behavior is achieved with infinitesimal values of the masses, for which the two-point functions go as
\begin{mdframed}[backgroundcolor=shadecolor,linewidth=0pt]
\begin{equation}
\begin{aligned}
\langle0|\mathbf{E}^a_+(\eta,\mathbf{y})\mathbf{E}^b_+(\eta,0)|0\rangle&\stackrel{\eta\to0^-}{\longrightarrow} m^2_V\frac{H^2\eta^2}{\mathbf{y}^2}\left(\delta^{ab}-2\frac{\mathbf{y}^a\mathbf{y}^b}{\mathbf{y}^2}\right)\\
\langle0|\mathbf{B}^a_-(\eta,\mathbf{y})\mathbf{B}^b_-(\eta,0)|0\rangle&\stackrel{\eta\to0^-}{\longrightarrow} m^2_V\frac{H^2\eta^2}{\mathbf{y}^2}\left(\delta^{ab}-2\frac{\mathbf{y}^a\mathbf{y}^b}{\mathbf{y}^2}\right)
\label{eq:massiveeb}
\end{aligned}
\end{equation}
\end{mdframed}
\noindent Only in the case in which the masses are exactly zero, we retrieve the two-point functions (\ref{eq:maxwelleb}), from terms that are subleading in the late time expansion. 

Notice that the difference between (\ref{eq:massiveeb}) and (\ref{eq:maxwelleb}) is sizable. For $60$ e-folds of quasi-de Sitter inflation, fluctuations in these massive electromagnetic fields would dominate over their massless counterparts by a factor of $e^{120}(m_V/H)^2$ 
\section{Photons without photons: one loop}
\label{sec:photonsfrommass1}
We have shown interesting properties of some composite operators in free theories in de Sitter, namely that they create states with the quantum numbers of a photon and that they have two-point functions which at late times and large distances dominate over those of the Maxwell field strength. In this section we want to check whether these properties persist once we consider some classes of interactions, namely those of the kind $g B^2$ where $B$ is any of the operators we studied and $gFB$ where $F$ is the field strength of the photon.

In order to compute two-point functions at one-loop we found it useful to continue to Euclidean de Sitter, or the sphere. The in-in formalism in the Bunch-Davies vacuum in de Sitter or perturbation theory on the sphere are completely equivalent, as proven at all orders in \cite{Higuchi:2008tn}. In \cite{Loparco:2023akg} it was also proven non-perturbatively that two-point functions can be continued safely from the sphere to de Sitter, under reasonable assumptions that can be checked to hold case by case. 

Since the continuation from the sphere is not the most common way of approaching perturbation theory in de Sitter, we first review the computation (and interpretation of the results) of two-point functions at one-loop in a theory of scalars with a cubic interaction. We believe that we are phrasing the results of this computation in a novel way that has not been discussed before.
We relegate some technical details to Appendix \ref{sec:sphere}. For more recent results on loops in de Sitter, see \cite{Cacciatori:2024zrv}.
\subsection{Warm-up: $\phi\chi^2$ at one loop}
\label{sec:cubicscalar}
Consider a theory of two weakly interacting real scalars $\phi$ and $\chi$, with action
\begin{equation}
    S=-\frac{1}{2}\int d^{d+1}x\sqrt{g}\left(\partial_\mu\phi\partial^\mu\phi+m_\phi^2\phi^2+\partial_\mu\chi\partial^\mu\chi+m_\chi^2\chi^2+g\phi\chi^2\right)
\end{equation}
It will be useful to adopt the parametrizations $\Delta_\phi\bar\Delta_\phi=m^2_\phi$ and $\Delta_\phi=\frac{d}{2}+i\mu_\phi$. We will follow what was done in \cite{Marolf:2010zp} in a language that is closer to more recent treatments such as \cite{DiPietro:2021sjt, Chakraborty:2023qbp,Chakraborty:2025myb}.

We are interested in the leading correction to the bulk two-point function of $\phi$.
This will require renormalization, which we will perform in dimensional regularization, and the introduction of the following counterterm Lagrangian, at leading order in the coupling
\begin{equation}
    \mathcal{L}_{\text{ct}}=-\frac{1}{2}g^2\left(\delta_\phi\partial_\mu\phi\partial^\mu\phi+\delta_m\phi^2\right)\,,
\end{equation}
where for later convenience we extracted explicitly the dependence on the coupling at the order of perturbation theory which we are interested in.

It will be convenient to split the counterterms as $\delta=Z+c$ where $Z$'s are infinite terms to cancel the divergences in the diagrams and $c$'s are arbitrary finite renormalization constants, free parameters which have to be fixed with a measurement. Since there is no unambiguous definition of ``physical mass" in de Sitter, we leave $c$ unfixed and discuss the interpretation at the end. We have not introduced counterterms for the coupling or for $\chi$ because we will focus on the two-point function of $\phi$.

To perform our computations, we will analytically continue to the Euclidean sphere. Scalar two-point functions on the Euclidean sphere can be decomposed into a complete basis as
\begin{equation}
    \langle\mathcal{O}(Y_1)\mathcal{O}(Y_2)\rangle=\sum_{J=0}^{\infty}[\mathcal{O}]_Jf_J(Y_1\cdot Y_2)\,,
    \label{eq:harmonics}
\end{equation}
 where $[\mathcal{O}]_J$ are some 
 coefficients analogous to the Fourier transform of 
 the two-point function of $\mathcal{O}$ in flat space, and $f_J(\sigma)$ is proportional to a Gegenbauer polynomial, given explicitly in (\ref{eq:gegenbauer}). 
 
 This representation presents a polynomial ambiguity, meaning that if we perform the shift
 \begin{equation}
     [\mathcal{O}]_J\to [\mathcal{O}]_J+\sum_{n=0}^m a_n[J(J+d)]^n
     \label{eq:polyambi}
 \end{equation}
 the sum (\ref{eq:harmonics}) converges to the same two-point function, for any $m$ and any $a_n$\footnote{This is true when the two-point function is evaluated at finite separation. This ambiguity essentially corresponds to the addition of contact terms, which have the form of delta functions and their derivatives in position space.}. This ambiguity is crucial, because the constants $a_n$ are directly related to the counterterms $\delta$.
 
 Gegenbauer polynomials satisfy an orthogonality relation, which in term of these $f_J$ functions reads
 \begin{equation}
 \int_{Y_2}f_J(Y_1\cdot Y_2)f_{J'}(Y_2\cdot Y_3)=\delta_{JJ'}f_J(Y_1\cdot Y_3)\,.
 \label{eq:orthoffs}
 \end{equation}
 This relation is very useful to perform diagrammatic computations on the sphere.
 
 At order $g^0$, the ``momentum space" two-point function of $\phi$ is (up to the polynomial ambiguity) \cite{Marolf:2010zp,Chakraborty:2023eoq}
\begin{equation}
    [\phi]^{(0)}_J=\frac{1}{J(J+d)+m^2_\phi}\,.
\end{equation}
Notice that it presents poles at $J=-\Delta_\phi$ and $J=-\bar\Delta_\phi$, and that it resembles very closely the Feynman propagator of a free scalar in flat momentum space.

\begin{figure}
\centering
\includegraphics[scale=1.6]{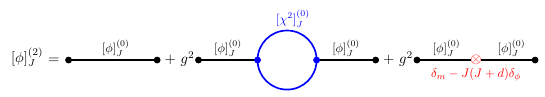}
\caption{A diagrammatic representation of eq. (\ref{eq:phi2}). Because of the orthogonality of Gegenbauer polynomials on the sphere, diagrams on the sphere factorize.}
\label{fig:diagrams1}
\end{figure}

Because of the orthogonality relation (\ref{eq:harmonics}), computing the two-point function of $\phi$ at order $g^2$ reduces to computing a simple product of ``momentum space" propagators, represented by the diagrams in Figure \ref{fig:diagrams1}
\begin{equation}
    [\phi]^{(2)}_J=[\phi]_J^{(0)}\left[1+g^2\left([\chi^2]_J^{(0)}+\delta_m-J(J+d)\delta_\phi\right)[\phi]^{(0)}_J\right]+O(g^4)\,.
    \label{eq:phi2}
\end{equation}
The notorious difficulty of computing loops in de Sitter is then reduced to computing $[\chi^2]^{(0)}_J$, the ``momentum space" two-point function of the composite operator $\chi^2$ in the free theory. This was first done explicitly in \cite{Marolf:2010zp} and we report their result in (\ref{eq:gsqrd}) in terms of a sum of$\ _7F_6$ generalized hypergeometric functions. Let us point out though that a simpler way to express this quantity is through a dispersive integral of the K\"allén-Lehmann density of the operator $\chi^2$ \cite{Chakraborty:2025myb}\footnote{We thank Shota Komatsu for suggesting to us this dispersive representation. It appeared afterwards in \cite{Chakraborty:2025myb}.}
\begin{equation}
    [\chi^2]_J^{(0)}=\int_{\frac{d}{2}+i\mathbb{R}}[d\Delta]\frac{\varrho^{\mathcal{P}(0)}_{\chi^2}(\Delta)}{J(d+J)+\Delta(d-\Delta)}
\end{equation}
where the equation is to be trusted for Re$[J]>-\frac{d}{2}$ and the extension to the rest of the complex $J$ plane can be obtained by analytic continuation. The upshot of this representation is that the spectral density $\varrho^{\mathcal{P}(0)}_{\chi^2}(\Delta)$ has a much simpler expression in terms of a product of Gamma functions (\ref{eq:rhochi2}). For more details, see Appendix \ref{subsec:bases2}. 

Similarly to flat space, when probing the behavior of the two-point function of an elementary field close to its poles, all chains of bubble diagrams start contributing equally. Their contributions form a geometric series that can be resummed, leading to the following improved result
\begin{equation}
    [\phi]_J^{(2)}=\frac{[\phi]^{(0)}_J}{1-g^2[\phi]^{(0)}_J\left([\chi^2]_J^{(0)}+\delta_m-J(J+d)\delta_\phi\right)}\,.
\end{equation}
In analogy with flat space, $[\chi^2]^{(0)}$ has the role of the \textit{self-energy}. Near $d=3$ it has a divergence of the form
\begin{equation}
    [\chi^2]_J^{(0)}=-\frac{1}
    {8\pi^2}\frac{1}{d-3}+[\tilde \chi^2]_J\,,
    \label{eq:chi2div}
\end{equation}
where $[\tilde \chi^2]_J$ is finite. The fact that the divergence has the same numerical prefactor as in flat space is a good consistency check, and is due to the fact that the UV properties of a theory are insensitive to the spacetime curvature. 

Equation (\ref{eq:chi2div}) fixes the divergent parts of the counterterms to be
\begin{equation}
    Z_m=\frac{1}
    {8\pi^2}\frac{1}{d-3}\,, \qquad Z_\phi=0\,,
\end{equation}
and the renormalized two-point function in $d+1=4$ is thus
\begin{equation}
\begin{aligned}
    [\phi]_J^{(2)}=\frac{1}{J(J+3)+m_\phi^2-g^2\left([\tilde \chi^2]_J+c_m-c_\phi J(J+3)\right)}\,.
    \label{eq:phi(2)}
\end{aligned}
\end{equation}
where $c_m$ and $c_\phi$ are renormalization constants.

The decomposition in spherical harmonics (\ref{eq:harmonics}) only converges when the two points are on the sphere. To analytically continue back to any two-point configuration in de Sitter, we can perform a Watson-Sommerfeld transform \cite{Marolf:2010zp,Marolf:2011sh,Hogervorst:2021uvp,DiPietro:2021sjt,Chakraborty:2023qbp}, obtaining
\begin{equation}
    \langle\phi(Y_1)\phi(Y_2)\rangle^{(2)}=\int_J[\phi]^{(2)}_JC_J(-\sigma)
    \label{eq:interactJ}
\end{equation}
where $\int_J$ is defined in (\ref{eq:shortJ}) and $C_J$ are Gegenbauer polynomials, explicitly given in (\ref{eq:gegenbauer}). This integral representation now converges for all $\sigma\in\mathbb{C}\backslash[1,\infty)$, where the excluded interval is the branch cut of the two-point function at time-like separation.

As we discuss in appendix \ref{sec:sphere}, this kind of integral representation makes manifest the leading late-time behavior of a two-point function. In fact, the powers of the late-time fall-offs are encoded in the nontrivial poles of the integrand, as expressed by eq. (\ref{eq:finallatescalar}). 

The late-time fall-offs of a two-point function are closely related to physical observables. It thus makes sense to impose renormalization conditions on them. For example, similarly to flat space, we can fix $c_\phi$ by requiring that the residues on the poles of the momentum-space two-point function (\ref{eq:phi(2)}) stay the same as in free theory
\begin{equation}
    \underset{J=J_*}{\text{Res}}\left[\left(J+\frac{3}{2}\right)[\phi]_J^{(2)}\right]\stackrel{!}{=}\underset{J=J_*}{\text{Res}}\left[\left(J+\frac{3}{2}\right)[\phi]_J^{(0)}\right]=\frac{1}{2}\,.
\end{equation}
This is satisfied by
\begin{equation}
    c_\phi=0\,.
\end{equation}
Then, the positions of the poles are
\begin{equation}
    J=-\Delta_\phi+g^2\frac{[\tilde \chi^2]_{-\Delta_\phi}+c_m}{3-2\Delta_\phi}+\mathcal{O}(g^4)\,,\qquad J=-\bar\Delta_\phi+g^2\frac{[\tilde \chi^2]_{-\bar\Delta_\phi}+c_m}{3-2\bar\Delta_\phi}+\mathcal{O}(g^4)\,.
    \label{eq:Jpoles}
\end{equation}
More explicitly, at late times the two-point function goes like
\begin{equation}
    \langle\phi(\eta,\mathbf{y})\phi(\eta,0)\rangle\stackrel{\eta\to0^-}{\longrightarrow}c_1\left(\frac{\eta^2}{\mathbf{y}^2}\right)^{\Delta_1}+c_2\left(\frac{\eta^2}{\mathbf{y}^2}\right)^{\Delta_2}+\ldots
    \label{eq:latetimecorrect}
\end{equation}
where $c_1$ and $c_2$ are some unimportant constants which can be reconstructed from (\ref{eq:finallatescalar}), and the two powers at late times are \cite{Marolf:2010zp,Chakraborty:2023qbp}
\begin{equation}
    \Delta_1\equiv\Delta_\phi-g^2\frac{[\tilde \chi^2]_{-\Delta_\phi}+c_m}{3-2\Delta_\phi}\,,\qquad \Delta_2\equiv\bar\Delta_\phi-g^2\frac{[\tilde \chi^2]_{-\bar\Delta_\phi}+c_m}{3-2\bar\Delta_\phi}\,,
    \label{eq:d1d2def}
\end{equation}
and the dots are subleading terms in the late-time expansion. 

If we think again about the polynomial ambiguity (\ref{eq:polyambi}) we see that we effectively chose $a_0=\delta_m$, $a_1=-\delta_\phi$ and we set all other $a_n$'s to zero. We can do that because the effect of the higher $a_n$'s on the position of the poles and on their residues is degenerate with the choice of $a_0$ and $a_1$. 

\paragraph{Physical interpretation} Let us discuss some physical consequences of this result. The first thing we notice is that if we sum the two powers in (\ref{eq:latetimecorrect}) we obtain
\begin{equation}
   \Delta_1+\Delta_2=3- \frac{4\pi g^2 \varrho^{\mathcal{P}\,(0)}_{\chi^2}(\Delta_\phi)}{(3- 2\Delta_\phi)^2}\,,
   \label{eq:d1d2sum}
\end{equation}
\noindent where we used the relation between coefficients of spherical harmonics and spectral densities reported in (\ref{eq:rhofromcharm}), and $\varrho^{\mathcal{P}\,(0)}_{\chi^2}(\Delta)$ is the spectral density of $\chi^2$ over the principal series in the free theory, given explicitly in eq. (\ref{eq:rhochi2}). 

The sum (\ref{eq:d1d2sum}) is strictly different than $3$, irrespective of whether $\phi$ and $\chi$ are in the principal or complementary series, and is independent of the choice of renormalization constant $c_m$. This is counterintuitive, since unitary irreps in de Sitter have asymptotic behaviors related by shadow symmetry ($\Delta\to 3-\Delta$). The point is that the effect of interactions on a two-point function in de Sitter \emph{can never be absorbed in a shift of the mass}. Elementary fields, renormalized by interactions, create a continuum of states and not one particular UIR of a specific mass, even if we are at weak couplings and if $\phi$ is lighter than $\chi$. This behavior is similar to that of a resonance in flat space, where the imaginary part of the self-energy cannot be reabsorbed in a shift of the mass and is interpreted as the width of the particle. 

\begin{figure}
\centering
\hspace{-1.5cm} 
\includegraphics[scale=0.72]{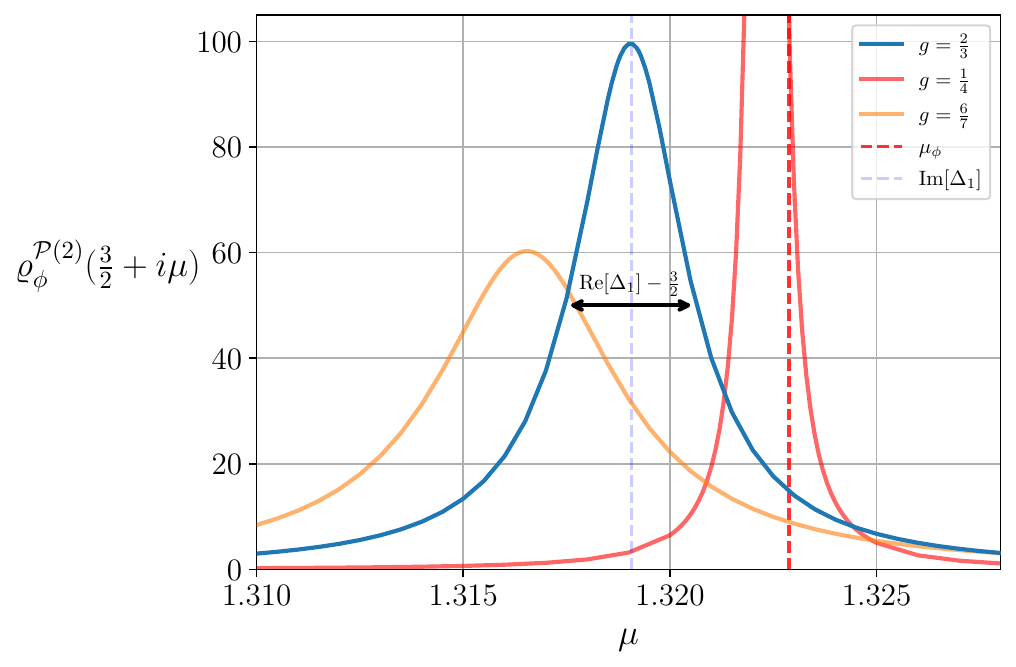}
\caption{The spectral density (\ref{eq:rhophi}) for $\phi$ starting in the principal series for various values of the coupling $g$. We use the parametrization $\Delta=\frac{3}{2}+i\mu$ and fix $\mu_\chi=\frac{1}{2}$, $m_\phi R=2$ and $c_m=0$. Taking $g\to0$ the spectral density approaches a Dirac delta function around $\mu=\mu_\phi$. For finite coupling $g$, the spectral density is instead peaked around the shifted value $\mu=\text{Im}[\Delta_1]$ (eq. (\ref{eq:imdelta})). The position of the peak depends on the renormalization constant $c_m$, while the width is set by Re$[\Delta_1]-\frac{3}{2}$ and is scheme-independent (see eq. (\ref{eq:redelta})).}
\label{fig:interactingrhos}
\end{figure}

Let us now specify to the cases in which $\phi$ starts in the principal or complementary series when $g=0$.
\begin{itemize}
    \item When $\phi$ starts in the \textbf{principal series} we have
    \begin{equation}
    \begin{aligned}
        \text{Re}[\Delta_1]=\text{Re}[\Delta_2]&=\frac{3}{2}+\frac{g^2}{2\mu_\phi}\text{Im}\left[[\tilde \chi^2]_{-\Delta_\phi}\right]\\
        &=\frac{3}{2}+\frac{g^2\pi}{2\mu_\phi^2}\varrho_{\chi^2}^{\mathcal{P}(0)}(\Delta_\phi)
        \label{eq:redelta}
    \end{aligned}
    \end{equation}
    where we used (\ref{eq:rhofromcharm}). The quantity on the rhs is strictly greater than $\frac{3}{2}$ and does not depend on the renormalization constant, meaning that the fact that the field $\phi$ decays at late times strictly faster than in free theory is  a (scheme-independent) physical prediction \cite{DiPietro:2021sjt,Arkani-Hamed:2015bza,Marolf:2010zp}. This quantity is related to the width of the peak of the spectral density, as shown in Figure \ref{fig:interactingrhos}.
    
    The imaginary parts of the two powers are instead
    \begin{equation}
        \text{Im}[\Delta_1]=-\text{Im}[\Delta_2]=\mu_\phi-\frac{g^2}{2\mu_\phi}\left(\text{Re}[[\tilde\chi]_{-\frac{d}{2}-i\mu_\phi}]+c_m\right)\,.
        \label{eq:imdelta}
    \end{equation}
    and so they depend on the renormalization condition which must be fixed with a measurement (for example, of the position of the peak of the spectral density as shown in Figure \ref{fig:interactingrhos}).
    \item When $\phi$ starts in the \textbf{complementary series}, instead, $\Delta_1,\Delta_2\in\mathbb{R}$ and it is impossible to say conclusively if the field decays faster or slower at late times wrt the free theory. The only conclusive statement that can be made is (\ref{eq:d1d2sum}) and the discussion that follows.
\end{itemize}
The reader might worry that these facts signal some tension with unitarity. To see that this is not the case, let us further continue the integral contour in (\ref{eq:interactJ}) to lie on the Re$J=-\frac{3}{2}$ line in order to retrieve the K\"allén-Lehmann representation. 
There are qualitative differences now depending on whether $\phi$ is in the principal or complementary series when the theory is free.
\paragraph{Principal series}

If $\phi$ is in the principal series in the free theory, the two poles (\ref{eq:Jpoles}) are not in the $-\frac{3}{2}<\text{Re}[J]<0$ interval. Moving the contour to Re$J=-\frac{3}{2}$, using the symmetries of the integrand and the relation (\ref{eq:rhofromcharm}), we can write the K\"allén-Lehmann decomposition as 
\begin{equation}
    \langle\phi(Y_1)\phi(Y_2)\rangle=\int_{\frac{d}{2}+i\mathbb{R}}[d\Delta]\varrho^{\mathcal{P}(2)}_\phi(\Delta)G_\Delta(\sigma)
    \label{eq:interactrho}
\end{equation}
where the spectral density is, at this order in perturbation theory,
\begin{equation}
    \varrho^{\mathcal{P}(2)}_\phi(\Delta)=\frac{g^2\varrho^{\mathcal{P}(0)}_{\chi^2}(\Delta)}{\left(\Delta\bar\Delta-\Delta_\phi\bar\Delta_\phi+g^2\left([\tilde \chi^2]_{-\Delta}+c_m\right)\right)\left(\Delta\bar\Delta-\Delta_\phi\bar\Delta_\phi+g^2\left([\tilde \chi^2]_{-\bar\Delta}+c_m\right)\right)}\,.
    \label{eq:rhophi}
\end{equation}
where the superscripts $(2)$ and $(0)$ indicate the order in perturbation theory.

We thus see that the field $\phi$ creates only states in the principal series, regardless of whether $\chi$ is in the principal or complementary series. In Figure \ref{fig:interactingrhos} we show a plot of the spectral density (\ref{eq:rhophi}). 
Using the parametrization $\Delta=\frac{3}{2}+i\mu$ we see that the spectral density is peaked around $\mu=\text{Im}[\Delta_1]$. As we take $g\to0$, it instead tends to a Dirac delta function around $\mu=\mu_\phi$. 

\begin{figure}
\centering
\includegraphics[scale=1.25]{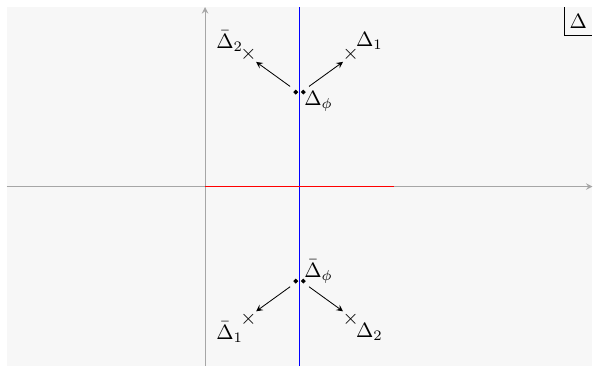}
\caption{The poles in the spectral density (\ref{eq:rhophi}) which have $O(1)$ residue. The dots represent the poles as $g\to0$, the free theory. In the strict $g=0$ case, they pinch the contour over the principal series and effectively become two delta functions. The crosses represent the position of the poles at order $g^2$, $\Delta_1$ and $\Delta_2$ (defined in eq. (\ref{eq:d1d2def})) and their shadows. The blue line is the principal series, while the red line is the complementary series. For this figure we chose $\chi$ and $\phi$ to start off in the principal series in the free theory.}
\label{fig:poles}
\end{figure}

In Figure \ref{fig:poles}, instead, we compare the position of its poles in the free theory and at leading order in the coupling. 

\paragraph{Complementary series} 
\begin{figure}
\centering
\includegraphics[scale=0.7]{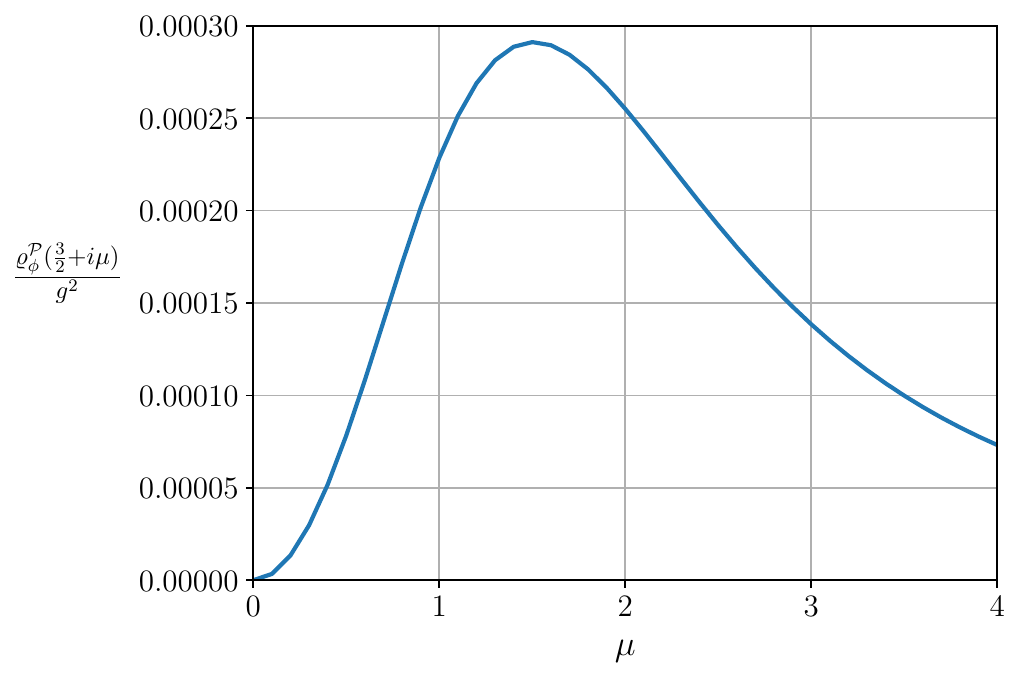}
\caption{The spectral density (\ref{eq:rhophi}) for $\phi$ starting in the complementary series. We use the parametrization $\Delta=\frac{3}{2}+i\mu$ and fix $\mu_\chi=\frac{1}{2}$, $m_\phi R=\frac{1}{2}$ and the renormalization constant $c_m=0$. Changing $g$ leads to indistinguishable differences in this plot, if we stay in the perturbative regime.}
\label{fig:interactingrhos2}
\end{figure}
If $\phi$ is in the complementary series in the free theory (without loss of generality, take $\Delta_\phi\in(\frac{3}{2},3)$), when continuing the integration contour in (\ref{eq:interactJ}) we pick up a pole, leading to
\begin{equation}
    \langle\phi(Y_1)\phi(Y_2)\rangle=\int_{\frac{3}{2}+i\mathbb{R}}[d\Delta]\varrho^{\mathcal{P}(2)}_\phi(\Delta)G_\Delta(\sigma)+\frac{\sin(\pi\Delta_*)}{\sin(\pi\Delta_\phi)}G_{\Delta_*}(\sigma)
    \label{eq:interactrhocomp2}
\end{equation}
where $\varrho^{\mathcal{P}(2)}_\phi(\Delta)$ is in (\ref{eq:rhophi}) and $\Delta_*\equiv\Delta_\phi-g^2\frac{[\tilde\chi^2]_{-\bar\Delta_\phi}+c_m}{3-2\Delta_\phi}$ 
corresponds to a small deformation of the free theory 
 complementary series UIR.

This analysis shows that if $\phi$ starts in the complementary series at $g=0$, as soon as interactions are turned on it starts creating states in the principal series and states in the complementary series irrep with $\Delta=\Delta_*$. Importantly, this irrep does not correspond directly to both late-time power laws that we found in (\ref{eq:latetimecorrect}). This is not an inconsistency: to determine the late-time behavior from (\ref{eq:interactrhocomp2}) we need to close the contour of integration over the principal series. This in turns leads to a cancellation between a pole in  principal series spectral density and one of the two asymptotic behaviors of $G_{\Delta_*}$, in total matching with (\ref{eq:latetimecorrect}).

\subsection{Photons from interacting massive theories}
\label{subsec:interacphot}
After our scalar warm-up, we move on to weakly interacting theories involving the operators discussed in Section \ref{sec:photonsfrommass0}. Explicitly, the class of actions we consider are of the form
\begin{equation}
    S=\int d^{d+1}x\sqrt{g}\left(K-\frac{g^2}{2}B^{\mu\nu}B_{\mu\nu}\right)
\end{equation}
where $B_{\mu\nu}$ is any of the operators studied in Section \ref{sec:photonsfrommass0} and $K$ is a sum of the appropriate kinetic and mass terms of the fundamental fields that make up $B_{\mu\nu}$. Some technical details are in Appendix \ref{subsec:bases2}. We are going to show that, in this class of theories, the properties we found in the absence of interactions actually persist at one loop.

To start, we need the decomposition in Gegenbauer polynomials of two-point functions of antisymmetric operators on the sphere. Analogously to the K\"allén-Lehmann decomposition, it is
\begin{equation}
    \langle B^{AB}(Y_1)B^{CD}(Y_2)\rangle=\sum_{J=0}^\infty\sum_{\pm}[B]_{\pm,J}f_{\pm,J}^{AB,CD}(Y_1,Y_2)
    \label{eq:Binharms}
\end{equation}
where $[B]_{\pm,J}$ are some 
coefficients which again have the interpretation of ``momentum space" two-point functions and $f^{AB,CD}_{\pm,J}$ is defined in (\ref{eq:fabcddef}). Once again, this representation presents polynomial ambiguities of the same form as (\ref{eq:polyambi}), and we will choose the constants appropriately to renormalize UV divergences.

For example, for the photon's field strength in the free theory we have, up to the polynomial ambiguities,
\begin{equation}
    [F]_{+,J}=[F]_{-,J}=\frac{1}{2(J+1)(J+2)}\,.
    \label{eq:freephotonmom}
\end{equation}
Moreover, we derived the following orthogonality properties, crucial to carry out integrals on the sphere in perturbation theory
\begin{equation}
\begin{aligned}
    \int_{Y_2}f_{\pm,J}^{AB,CD}(Y_1,Y_2)f_{\pm,J'\ CD}^{\qquad\  \  \ EF}(Y_2,Y_3)&=\delta_{JJ'}N_{J}f_{\pm,J}^{ABEF}(Y_1,Y_3)\\
    \int_{Y_2}f_{\pm,J}^{AB,CD}(Y_1,Y_2)f_{\mp,J'\ CD}^{\qquad\  \  \ EF}(Y_2,Y_3)&=0
    \label{eq:orthogonality}
\end{aligned}
\end{equation}
where the coefficient $N_J$ is proportional to the inverse of a photon propagator 
\begin{equation}
    N_J=2(J+1)(J+2)=\frac{1}{[F]_{\pm,J}}\,.
    \label{eq:NJ}
\end{equation}
We are interested in computing the two-point function of $B$ at one loop. It will be useful to focus directly on the ``momentum-space" two-point function $[B]_{\pm,J}$. In Appendix \ref{sec:sphere} we outline how to compute it in free theory starting from the spectral densities we presented in Section \ref{sec:photonsfrommass0}, and we give an explicit expression for the case of massless scalars in eq. (\ref{eq:masslessscalars6F5}). The important thing to know is that, when $B$ is a composite operator, the naive momentum-space coefficients $[B]_{\pm,J}$ already diverge in the free theory. This divergence can simply be cured fixing the ambiguous constants in (\ref{eq:polyambi}). For example, in the case in which $B$ is composed of massless scalars, like at the end of Section \ref{subsec:masslessscalars}, we find in Appendix \ref{subsec:interactmasslessscalars} that we need to fix
\begin{equation}
    a_0=-\frac{1}{d-3}\frac{5}{48\pi^2}+a'_0\,,\qquad a_1=\frac{1}{d-3}\frac{1}{96\pi^2}+a'_1
\end{equation}
where $a'_0$ and $a'_1$ are further ambiguous constants that do not diverge in $d=3$ and that are to be fixed with renormalization conditions. 

\begin{figure}
\centering
\includegraphics[scale=1.6]{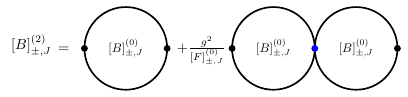}
\caption{The diagrams contributing to the two-point function of $B$ at leading order in the coupling. The inverse factors of the two-point function of $F$ appear due to the orthogonality relations (\ref{eq:orthogonality}). We represent the propagator of $B$ with two lines because in most of the examples we studied in Section \ref{sec:photonsfrommass0}, $B$ is composed of two elementary fields.}
\label{fig:diagrams4}
\end{figure}

Now let us call $[B]^{(0)}_{\pm,J}$ the free theory renormalized momentum space two-point functions (including the $a_0'$ and $a_1'$ ambiguous constants). To obtain the one-loop correction we need to sum diagrams as in Figure \ref{fig:diagrams4}. 
Just like for the scalar case explored in the previous Section, we can resum the chains of bubbles that all contribute equally near the free theory poles, and we obtain the following improved result
\begin{equation}
    [B]_{\pm,J}^{(2)}=\frac{[B]_{\pm,J}^{(0)}}{1-g^2\frac{[B]_{\pm,J}^{(0)}}{[F]_{\pm,J}}}
    \label{eq:B1}
\end{equation}
Then, the two-point function of $B$ in de Sitter is obtained by analytically continuing (\ref{eq:Binharms}) with a Watson-Sommerfeld transform
\begin{equation}
    \langle B^{AB}(Y_1)B^{CD}(Y_2)\rangle^{(2)}=\sum_\pm\int_J[B]^{(2)}_{J,\pm} \Pi_\pm^{ABCD}C_J(-\sigma)\,, 
    \label{eq:intBJ}
\end{equation}
where $\int_J$ is defined in (\ref{eq:shortJ}) and $C_J$ are Gegenbauer polynomials. 

As we explain more in detail in the Appendix \ref{subsec:bases2}, the analytic structure of the coefficients $[B]_{\pm,J}$ determines the late-time behavior of the two-point function. Moreover, it also determines whether an operator creates photon states, through the following relation to the photon spectral density:
\begin{equation}
    \varrho^\gamma=\sum_\pm\underset{J=-1}{\text{Res}}[B]_{\pm,J}\,.
    \label{eq:photonfromB}
\end{equation}
Just like in the scalar case, interactions will in general shift the positions of the poles of $[B]_{\pm,J}$. To understand how, let us write the behavior of $[B]_{\pm,J}$ in the free theory close to a pole at $J=J_*$ as
\begin{equation}
    [B]_{\pm,J}^{(0)}\sim\frac{r_*}{J-J_*}+a_0'
\end{equation}
where we emphasize the role of the renormalization constant $a_0'$ but we will from now on ignore $a_1'$ because its only further effect is to change the residue of the pole.

At leading order in the coupling, given (\ref{eq:B1}), the position of the pole shifts as 
\begin{equation}
\begin{aligned}
    J_*\to& J_*+g^2\frac{r_*}{a_0'+1}\frac{1}{[F]_{\pm,J_*}}+O(g^4)\,,\\
    &=J_*+g^2\frac{r_*}{a_0'+1}(J_*+1)(J_*+2)+O(g^4)
    \label{eq:Jpoles2}
\end{aligned}
\end{equation}
Notice that something special happens when $[B]_{\pm,J}^{(0)}$ has a pole at either of the positions associated with the free photon, $J=-1$ or $J=-2$. In that case, a cancellation happens in eq. (\ref{eq:Jpoles2}) which leads to the fact that the pole does not shift, independently of the choice of renormalization constant. Since such a pole is related to the photon spectral density through (\ref{eq:photonfromB}), we can state that \emph{the creation of states in the photon UIR persists at one loop}. 
\paragraph{K\"allén-Lehmann decomposition at one loop}
As we did in the previous section, let us derive the K\"allén-Lehmann representation at one loop for this two-point function. To do it, we continue the contour of integration in (\ref{eq:intBJ}) to lie on the Re$J=-\frac{3}{2}$ axis. 
We do this carefully in the Appendix \ref{subsec:bases2}, resulting in eq. (\ref{eq:rhofromharms2}). We obtain
\begin{equation}
\begin{aligned}
    \varrho_{B}^{\mathcal{P},\pm(2)}(\Delta)&=\frac{\varrho_{B}^{\mathcal{P},\pm(0)}(\Delta)}{\left(1+g^2(\Delta\bar\Delta-2)[B]_{\pm,-\Delta}^{(0)}\right)\left(1+g^2(\Delta\bar\Delta-2)[B]_{\pm,-\bar\Delta}^{(0)}\right)}\,,\\
    \varrho_{B,\Delta_c^{(2)}}^{\mathcal{C},\pm}&=\frac{\left(\frac{3}{2}-\Delta_c^{(2)}\right)}{\left(\frac{3}{2}-\Delta_c^{(0)}\right)}\varrho^{\mathcal{C},\pm}_{B,\Delta_c^{(0)}}\,,\qquad
    \varrho_{B}^{\gamma(2)}=\frac{\varrho_{B}^{\gamma(0)}}{1-g^2\varrho_{B}^{\gamma(0)}}
    \label{eq:densitiesB2}
\end{aligned}
\end{equation}
where $\Delta_c^{(0)}$ are the scaling dimensions of the complementary series contributions in the free theory, while $\Delta_c^{(2)}$ are the corresponding scaling dimensions in the interacting theory at order $g^2$. They are related via (\ref{eq:Jpoles2}), explicitly 
\begin{equation}
    \Delta_c^{(2)}=\Delta_c^{(0)}-\frac{2g^2}{c_0'+1}(1-\Delta_c^{(0)})(2-\Delta_c^{(0)})\left(\frac{3}{2}-\Delta_c^{(0)}\right)\varrho_{B,\Delta_c^{(0)}}^{\mathcal{C},\pm}\,,
\end{equation}
where, to avoid clutter, we are suppressing a $\pm$ subscript on each $\Delta_c$.

Notice that $\Delta_c^{(2)}$ depends on the renormalization constant $c_0'$, as does the position of all poles in $\varrho_{B}^{\mathcal{P},\pm(2)}(\Delta)$\footnote{Naively one could imagine that $\varrho_{B}^{\mathcal{P},\pm(0)}(\Delta)$ in the numerator reintroduces the poles of the free theory. In reality, the residue on those positions is zero because the $[B]^{(0)}$ coefficients in the denominator diverge too.}. Since the position of all poles depends on the same renormalization constant, the measurement of one pole fixes immediately the position of all other poles.

Notice that $\varrho^{\gamma(2)}_B>\varrho^{\gamma(0)}_B$, meaning that interactions actually enhance the creation of photons by the operator $B$.

\begin{figure}
\centering
\includegraphics[scale=0.7]{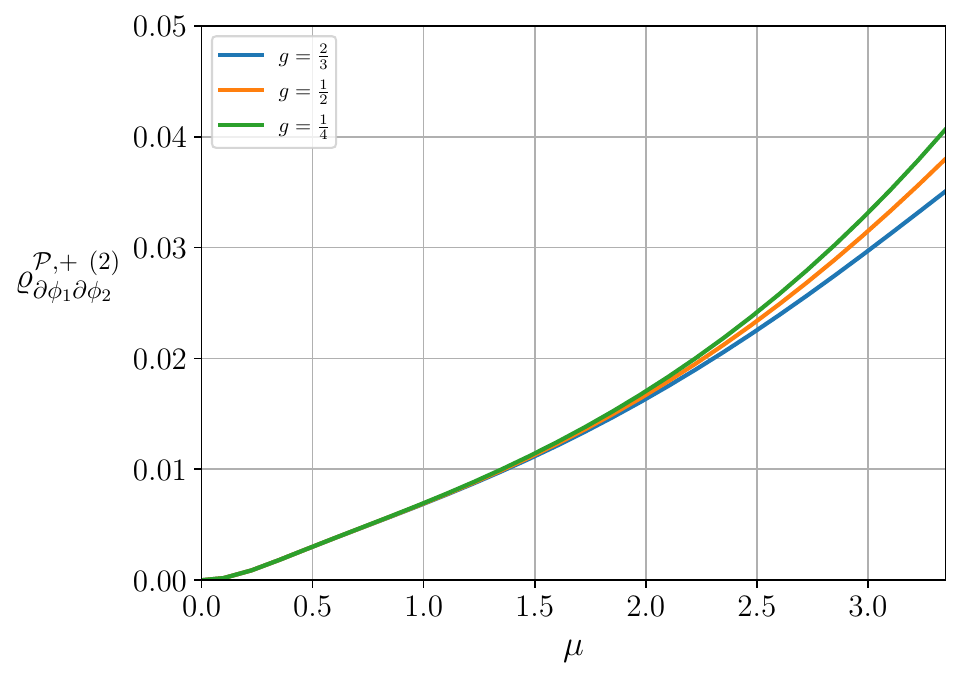}
\caption{The spectral density (\ref{eq:densitiesB2}) of the operator $\partial_{[\mu}\phi_1\partial_{\nu]}\phi_2$ at one loop for various values of the coupling. We use the parametrization $\Delta=\frac{3}{2}+i\mu$ and fix the renormalization constants $a_0'=a_1'=0$.}
\label{fig:rhodphidphi}
\end{figure}

As an illustrative example, we plot the spectral density $\varrho_{\partial\phi_1\partial\phi_2}^{\mathcal{P},+(2)}$ in Figure \ref{fig:rhodphidphi}.
\paragraph{Late time behavior at one loop}
The late time expansion of the two-point function of $B$, as discussed extensively in Section \ref{subsec:latetimes}, is governed by the positions of the poles in the principal series densities, by the complementary series contributions and by the photon term. The situation is then essentially the same as in free theory, up to the shift  (\ref{eq:Jpoles2}). In particular, if the photon contribution was canceled at late times in the free theory, as in all cases we explored with the exception of the massless scalars, this will continue to happen once interactions are turned on, because
\begin{equation}
    4\pi\underset{\Delta=2}{\text{Res}}\left[\varrho^{\mathcal{P},\pm(2)}_B(\Delta)\right]=\varrho_B^{\gamma(2)}\,.
\end{equation}
The rest of the discussion on the late time behavior is identical to the cases in the free theories explored in Section \ref{sec:photonsfrommass0} up to order $g^2$ shifts in the late time power laws. 
\subsection{Photons interacting with composite fields}
Here we discuss another class of theories, where a photon interacts with any of the operators we studied in Section \ref{sec:photonsfrommass0} through a nonminimal coupling
\begin{equation}
    S=\int d^{d+1}x\sqrt{g}\left(K-\frac{1}{4}F^{\mu\nu}F_{\mu\nu}-g F^{\mu\nu}B_{\mu\nu}\right)
    \label{eq:actionffbb}
\end{equation}
where again $K$ is the necessary sum of kinetic terms of the fundamental fields composing $B$. We will study the two point functions $\langle FF\rangle$, $\langle BB\rangle$ and the mixing $\langle FB\rangle$. Diagrammatically, we have
\begin{center}
\includegraphics[scale=1.5]{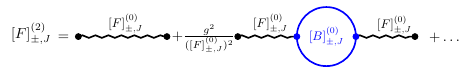}
\end{center}
\begin{center}
\includegraphics[scale=1.5]{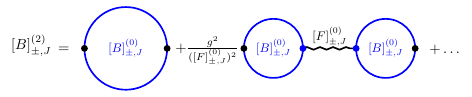}
\end{center}
\begin{center}
\includegraphics[scale=1.5]{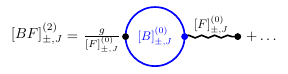}
\end{center}
where the inverse factors of the photon propagators come from (\ref{eq:orthogonality}). For each case we can resum chains of such diagrams to obtain the following improved results
\begin{equation}
    [F]_{\pm,J}^{(2)}=\frac{[F]_{\pm,J}^{(0)}}{1-g^2\frac{[B]^{(0)}_{\pm,J}}{[F]_{\pm,J}^{(0)}}}\qquad 
    [B]_{\pm,J}^{(2)}=\frac{[B]_{\pm,J}^{(0)}}{1-g^2\frac{[B]^{(0)}_{\pm,J}}{[F]_{\pm,J}^{(0)}}}\qquad
    [BF]_{\pm,J}^{(2)}=\frac{g[B]_{\pm,J}^{(0)}}{1-g^2\frac{[B]^{(0)}_{\pm,J}}{[F]_{\pm,J}^{(0)}}}
    \label{eq:FBBoneloop}
\end{equation}
where $[F]_{\pm,J}^{(0)}$ was given in (\ref{eq:freephotonmom}). Notice that $[B]_{\pm,J}^{(2)}$ in these theories turns out to have the same exact expression as in the previous ones (\ref{eq:B1}), because of the particular cancellation of the photon propagator factors.

The position of the poles at $J=J_*$ is precisely shifted at one loop as in (\ref{eq:Jpoles2})
\begin{equation}
    J_*\to J_*+g^2\frac{r_*}{a_0'+1}(J_*+1)(J_*+2)
\end{equation}
where $a_0'$ is a renormalization constant and $r_*$ is the residue on the pole in the free theory. 
In particular, the creation of photon states is protected at one loop also by these interactions, due to the relation (\ref{eq:photonfromB}).

\paragraph{K\"allén-Lehmann representation at one loop} For the $B$ operators, given (\ref{eq:FBBoneloop}) there will be no difference with the theories studied in the previous section, and the result is thus what we discussed there. For the photon field strength, instead, we find (again using (\ref{eq:rhofromharms2}))
\begin{equation} 
    \varrho_{F}^{\mathcal{P},\pm(2)}(\Delta)=g^2\varrho_{B}^{\mathcal{P},\pm(2)}(\Delta)\,, \qquad \varrho_{F,\Delta_c^{(2)}}^{\mathcal{C},\pm}=g^2\varrho_{B,\Delta_c^{(2)}}^{\mathcal{C},\pm}\,, \qquad \varrho_{F}^{\gamma(2)}=\frac{1}{1-g^2\varrho^{\gamma(0)}_B}\,,
    \label{eq:rhoFF}
\end{equation}
where $\varrho_{B}^{\mathcal{P},\pm(2)}$ and $\varrho_{B,\Delta_c^{(2)}}^{\mathcal{C},\pm}$ are the same as in (\ref{eq:densitiesB2}).

\begin{figure}
\centering
\hspace{-1.6cm} 
\includegraphics[scale=0.65]{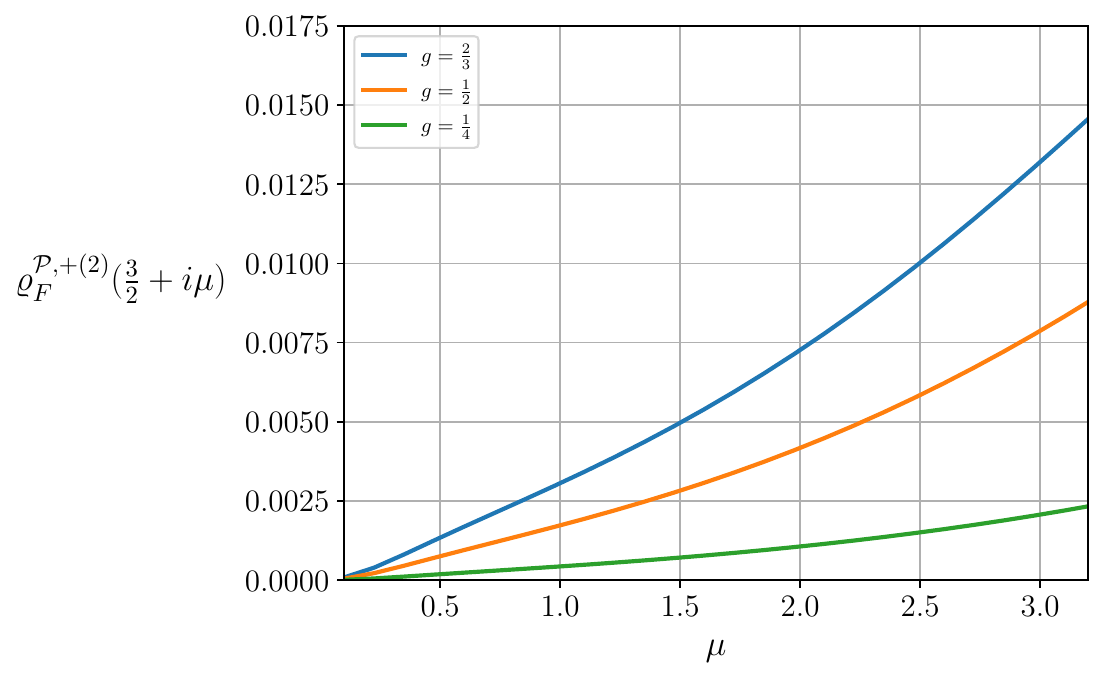}
\caption{The spectral density of the field strength of the photon over the principal series at one loop when it is interacting with massless scalars (\ref{eq:rhoFF}). It goes to zero as the free theory is approached.}
\label{fig:photonRho}
\end{figure}

We thus observe that the photon field strength starts creating states in the principal and complementary series. This is completely analogous to flat space, with the difference that the creation of electron-positron pairs 
in flat space QED by the photon field strength is gapped by their combined rest mass, while in de Sitter there is no gap. We plot its spectral density over the principal series in Figure (\ref{fig:photonRho}) for the case of massless scalars, in which $B_{\mu\nu}=\partial_{[\mu}\phi_1\partial_{\nu]}\phi_2$.

Finally, the mixed two-point function $\langle FB\rangle$ has the following spectral densities
\begin{equation}
    \varrho_{FB}^{\mathcal{P},\pm(2)}(\Delta)=g\varrho_{B}^{\mathcal{P},\pm(2)}(\Delta)\,,\qquad \varrho_{F,\Delta_c^{(2)}}^{\mathcal{C},\pm}=g\varrho_{B,\Delta_c^{(2)}}^{\mathcal{C},\pm}\,,\qquad \varrho_{FB}^{\gamma(2)}=g\varrho_{B}^{\gamma(2)}\,.
    \label{eq:rhoFB}
\end{equation}
\textbf{Late time behavior at one loop}
At late times, the photon field strength retains its original behavior, protected from interactions, plus order $g^2$ terms proportional to the corrected late time behavior of $B$
\begin{equation}
    \lim_{\eta\to0^-}\langle F_{\mu\nu}(y_1)F_{\rho\sigma}(y_2)\rangle^{(2)}=\lim_{\eta\to0^-}\langle F_{\mu\nu}(y_1)F_{\rho\sigma}(y_2)\rangle^{(0)}+g^2 \lim_{\eta\to0^-}\langle B_{\mu\nu}(y_1)B_{\rho\sigma}(y_2)\rangle^{(2)}\,.
\end{equation}
But let us ask the following question: is there a field redefinition that leaves the late time behavior of the photon field strength untouched? To answer that, consider Table \ref{tab:spectrals}, where we summarize the structure of spectral densities at one loop in this theory.
\begin{table}[h]
\centering
\begin{tabular}{c|c|c}
    $\varrho^{\mathcal{P},\pm}$  & $F$  & $B$ \\
      \hline
     $F$ &  $g^2x$ & $gx$\\
     \hline
     $B$ & $gx$ & $x$
\end{tabular}
\qquad\qquad 
\begin{tabular}{c|c|c}
    $\varrho^{\gamma}$  & $F$  & $B$ \\
      \hline
     $F$ &  $\frac{1}{1-g^2y}$ & $\frac{gy}{1-g^2y}$\\
     \hline
     $B$ & $\frac{gy}{1-g^2y}$ & $\frac{y}{1-g^2y}$
\end{tabular}
\caption{The matrix of spectral densities appearing in the K\"allén-Lehmann decomposition of the two-point functions $\langle BB\rangle$, $\langle FF\rangle$ and $\langle FB\rangle$ in the theory (\ref{eq:actionffbb}) at one loop. $x$ is $\varrho_B^{\mathcal{P},\pm(2)}$ (\ref{eq:densitiesB2}) and $y$ is $\varrho_B^{\gamma(0)}$.}
\label{tab:spectrals}
\end{table}

\noindent where the complementary series spectral density matrix looks exactly like the one for the principal series.  By diagonalizing these matrices we deduce that we can define $F'\equiv F-gB$ and $B'\equiv B+gF$ such that the matrices take the form of Table \ref{tab:spectrals2}.
\begin{table}[h]
\centering
\begin{tabular}{c|c|c}
    $\varrho^{\mathcal{P},\pm}$  & $F'$  & $B'$ \\
      \hline
     $F'$ &  $0$ & $0$\\
     \hline
     $B'$ & $0$ & $(1+2g^2)x$
\end{tabular}
\qquad\qquad 
\begin{tabular}{c|c|c}
    $\varrho^{\gamma}$  & $F'$  & $B'$ \\
      \hline
     $F'$ &  $1$ & $g$\\
     \hline
     $B'$ & $g$ & $y+g^2(1+2y)$
\end{tabular}
\caption{The matrix of spectral densities after the field redefinitions $F'\equiv F-gB$ and $B'\equiv B+gF$. Now $F'$ has a late-time behavior that is purely that of a photon, but it still mixes with the other field $B'$.}
\label{tab:spectrals2}
\end{table}

\noindent Now the two-point function of $F'$ behaves as a photon at late times, given that its only nonvanishing spectral density is $\varrho^\gamma_{F'}$, but it keeps mixing with $B'$, given that $\varrho^{\gamma}_{F'B'}\neq0$.

\paragraph{Summary} In this Section we asked whether the interesting properties of composite operators that we discovered in free theories actually persist once interactions are turned on. We considered interactions of the kind $B^2$ and $FB$ where $B$ is any of the composites studied in \ref{sec:photonsfrommass0}, and $F$ is a photon field strength. The crucial finding is that the poles in the momentum-space representation of $B$ at order $g^2$ are related to the poles in the free theory through
\begin{equation}
    J^{(2)}=J^{(0)}+g^2c(J^{(0)}+1)(J^{(0)}+2)+O(g^4)\,,
    \label{eq:Jshift}
\end{equation}
for some constant $c$. Moreover, the presence of a pole at $J=-1$ is related to the photon UIR spectral density through
\begin{equation}
    \varrho^\gamma_B=\sum_\pm\underset{J=-1}{\text{Res}}[B]_{\pm,J}\,.
\end{equation}
These two equations tell us that the creation of states in the photon UIR persists at one loop, because a pole at $J=-1$ does not shift.

We also checked that the late-time expansion is affected only infinitesimally by the interactions, meaning that the powers experience a shift of order $g^2$ and so, in particular, if they dominated over the photon in the free theory, they will keep doing so when interactions are turned on.

In the case of the $FB$-type interactions we find the same results. In that case, we also observed that a field redefinition $F'\equiv F-gB$ and $B'\equiv B+gF$ can be performed such that the operator $F'$ has the two-point function of a pure photon. However, even with this field redefinition, the mixed two-point function $\langle F'B'\rangle$ is nonzero.

\section{Conclusions and Future Directions}
In this work, we have demonstrated several surprising properties of photons in de Sitter spacetime that challenge our flat-space intuition. We have shown that photon states appear generically in the Hilbert space of any QFT in de Sitter, even without gauge symmetry. This phenomenon, which we characterized through the Källén-Lehmann representation for antisymmetric tensors, persists beyond free theory into weakly interacting regimes. Perhaps most remarkably, we found that certain operators exhibit two-point functions with slower late-time large-distance decay than the electromagnetic field itself, establishing that photons are not the dominant spin-1 excitations in the infrared regime of QFT in de Sitter.

These results open several promising avenues for future research:
\begin{itemize}
\item \textbf{What is a graviton in de Sitter spacetime?} It has been shown that states in the UIR of the graviton, labeled by $\Delta=0,3$ and $SO(3)$ spin 2, appear in the tensor product of states belonging to massive irreps \cite{MartindS}. It is therefore likely that many of the surprising properties we derived for photons are also shared by states in the graviton UIR. For instance, we expect that composite operators with the symmetries of the Weyl tensor, constructed from free or weakly coupled massive fields, should create states in the graviton UIR when acting on the Bunch-Davies vacuum. Studying the late-time behavior of their two-point functions would be very interesting, and may allow for a derivation of a bound on gravitational wave fluctuations analogous to our bound on magnetic fields \eqref{eq:Bbound}, at least in the $G \to 0$ limit.


\item \textbf{Spectral decompositions and inflation.} The decomposition of the Hilbert space that we rely on in this and our previous work assumes an exact de Sitter background. However, in slow-roll inflation de Sitter invariance is weakly broken. How do such perturbations affect the tools we have developed? Can our bounds and spectral decompositions be generalized systematically in this context?

\item \textbf{Parity-violating theories.} Throughout this paper we have worked in parity-invariant setups. It would be important to generalize the K\"allén-Lehmann decomposition to theories that violate parity. For instance, does the bound on magnetic fields weaken in the presence of parity violation?

\item \textbf{Photon ``decay"?} In flat space, the K\"allén-Lehmann spectral density is closely related to the decay rate of a resonance into daughter particles. A similar connection has been proposed in de Sitter spacetime \cite{Bros:2010rku}. A worthwhile direction is to clarify how decay rates should be defined in this context, perhaps by introducing Unruh-DeWitt-like detectors \cite{Conroy:2022rgp}, and to make the connection between decay rates and spectral densities more precise. Given the ubiquity of the photon UIR, we can ask at what rate do photons decay into other species — for example, into electron-positron pairs 
in QED \cite{Cotaescu:2010bn}. Moreover, since all particles can decay into one another in de Sitter, what governs the late-time balance between particle species? What are the most populated states when interactions are included? Preliminary results suggest, for instance, that complementary series scalars tend to decay preferentially into principal series states, and not vice versa, once again going against our flat space intuition.
\item \textbf{Gravity} The presence of dynamical gravity, even in the $G\to0$ limit, implies constraints on the Hilbert space of QFT in de Sitter \cite{Higuchi:1991tk,Higuchi:1991tm,Chakraborty:2023yed,Anninos:2024iwf}. It would be very interesting to understand how taking into account these constraints affects the results of our work.
\item \textbf{Higgs mechanism and tensor products in de Sitter}\  \ Can the photon states, that we found to be present in any QFT in dS,  get Higgsed? From a representation-theoretic perspective, states in the photon UIR can combine with scalar states to form long multiplets if the scalars are in the $\mathcal{V}_{1,0}$ UIR, as can be understood by looking at the $SO(4)$ content of these representations (see Appendix \ref{sec:UIRs}). One can then ask: \emph{are massless scalar states also populated in any QFT in dS?} From what is currently known, the answer seems to be no. Previous works \cite{Dobrev:1976vr,Dobrev:1977qv,Hirai:1965,Hirai:1966,MartindS,zhang2017tensorproductscomplementaryseries,Penedones:2023uqc} have studied the tensor products
$\mathcal{P}_{\Delta_1,s_1}\otimes\mathcal{P}_{\Delta_2,s_2}$, $\mathcal{C}_{\Delta_1,0}\otimes\mathcal{C}_{\Delta_2,0}$, $\mathcal{C}_{\Delta_1,0}\otimes\mathcal{P}_{\Delta_2,0}$ and $\mathcal{V}_{1,0}\otimes\mathcal{V}_{1,0}$, and they have shown that states in $\mathcal{V}_{1,0}$ do not appear in any of them. Moreover, these states are absent in the decomposition of traceless symmetric conformal multiplets into $SO(1,4)$ UIRs. It would be useful to complete the study of all tensor products of UIRs of $SO(1,4)$ to understand how many of the photon states in a generic QFT in dS$_4$ can get Higgsed.
\end{itemize}

\vspace{1em}
\noindent There are many other interesting questions in QFT and gravity in the context of cosmology. We argue that taking the Hilbert space structure seriously — particularly its decomposition into UIRs of the de Sitter isometry group — can reveal surprising physics and impose powerful, theory-independent constraints. We believe this perspective will continue to provide valuable insights into QFT in de Sitter spacetime.


\label{sec:conclusion}
\section*{Acknowledgements}
We are grateful to Prish Chakraborty, Victor Gorbenko, Shota Komatsu,  Ugo Moschella, Guilherme Pimentel, Jiaxin Qiao, Suvrat Raju, Riccardo Rattazzi, Veronica Sacchi, Cong Shen, Zimo Sun, Kamran Salehi Vaziri for useful discussions. In particular, we thank Prish Chakraborty and Veronica Sacchi for valuable detailed comments on an initial draft. We also acknowledge illuminating conversations regarding this project with the attendants of the SwissMAP 2024 general meeting, the workshop ``Cosmological Correlators in Taiwan" and the high energy group in ICTS Bengalore. 

JP and ML are supported by the Simons Foundation grant 488649 (Simons Collaboration on the Nonperturbative Bootstrap) and the Swiss National Science Foundation through the project
200020\_197160 and through the National Centre of Competence in Research SwissMAP.

\appendix
\section{Unitary Irreducible Representations of $SO(1,4)$}
\label{sec:UIRs}
de Sitter is a maximally symmetric spacetime. Its isometries form the group $SO(1,4)$, generated by $L_{AB}=-L_{BA}$ satisfying the Lorentz algebra
\begin{equation}
    [L_{AB},L_{CD}]=\eta_{BC}L_{AD}-\eta_{AC}L_{BD}+\eta_{AD}L_{BC}-\eta_{BD}L_{AC}\,.
\end{equation}
In a unitary representation, $L_{AB}$ are realized as anti-hermitian operators on a Hilbert space. The algebra $\mathfrak{so}(1,4)$ is isomorphic to the $3$-dimensional Euclidean conformal algebra, specifically
\begin{equation}
    L_{ij}=M_{ij}, \qquad L_{04}=D\,, \qquad L_{4i}=\frac{1}{2}(P_i+K_i)\,, \qquad L_{0i}=\frac{1}{2}(P_i-K_i)\,,
\end{equation}
where $D$ is the dilatation, $P_i$ are the translations, $K_i$ are the special conformal transformations and $M_{ij}$ are the rotations. We choose the quadratic Casimir of $SO(1,4)$ to be
\begin{equation}
    \mathcal{C}_2^{SO(1,4)}=\frac{1}{2}L_{AB}L^{AB}=D(3-D)+P_iK_i+\frac{1}{2}M_{ij}M^{ij}\,,
    \label{eq:quadraticC2}
\end{equation}
where $\frac{1}{2}M_{ij}M^{ij}$ is the quadratic Casimir of $SO(3)$.

The irreducible representations of $SO(1,4)$ can be classified in terms of a complex parameter $\Delta$ and the spin $s$ of $SO(3)$. 
Then, the eigenvalues of the quadratic Casimir (\ref{eq:quadraticC2}) on states in these representations are
\begin{equation}
    \mathcal{C}_2^{SO(1,4)}=\Delta(3-\Delta)-s(1+s)\,.
\end{equation}
For scalar representations, this coincides with the usual definition of mass in de Sitter. More generally, the convention is \cite{Sun:2021thf,Hinterbichler:2016fgl}
\begin{equation}
\begin{aligned}
    s=0:\qquad  m^2&\equiv\Delta(3-\Delta)\,,\\
    s\geq1:\qquad m^2&\equiv (\Delta+s-2)(1+s-\Delta)\,.
\end{aligned}
\end{equation}
In this convention, gauge fields are massless.

In four dimensions, there are four types of UIRs other than the trivial representation \cite{Dirac:1945cm,HarishChandra,Gelfand,Bargmann:1946me}. Here we report their classification in terms of $\Delta$ and $s$, their content when restricted to the maximal subgroup $SO(4)$ and the physical interpretation of their associated Hilbert spaces. We find it useful to use the language of two-row Young tableaux $\mathbb{Y}_{n,m}$ where $n$ is the number of boxes in the first row and $m$ is the number of boxes in the second row. When we write $\mathbb{Y}_{n,-m}$ we mean a representation that is related to $\mathbb{Y}_{n,m}$ by parity. In fact, the integers $n$ and $m$ correspond respectively to the eigenvalues of the two Cartans $L_{12}$ and $L_{34}$, and under parity (see eq. (\ref{eq:embedparity})) only $L_{34}$ switches sign \cite{Sun:2021thf}. We will follow the treatment of \cite{Sun:2021thf,Penedones:2023uqc}.
\begin{itemize}
    \item \textbf{Principal series} $\mathcal{P}_{\Delta,s}$: $\Delta\in\frac{3}{2}+i\mathbb{R}$ and $s\geq0$\,. The $SO(4)$ content is 
    \begin{equation}
        \mathcal{P}_{\Delta,s}\Big|_{SO(4)}=\bigoplus^\infty_{n=s}\bigoplus^s_{m=-s}\mathbb{Y}_{n,m}\,.
    \end{equation}
    This representation corresponds to the states of a single free ``heavy"  massive particle with $m^2>\frac{9}{4}H^2$ for $s=0$  and $m^2>\frac{(2s-1)^2}{4}H^2$ for $s\ge1$.
    \item \textbf{Complementary series} $\mathcal{C}_{\Delta,s}$: $0<\Delta<3$ for $s=0$ and $1<\Delta<2$ when $s\geq 1$. The $SO(4)$ content is the same as for the principal series:
    \begin{equation}
        \mathcal{C}_{\Delta,s}\Big|_{SO(4)}=\bigoplus^\infty_{n=s}\bigoplus^s_{m=-s}\mathbb{Y}_{n,m}\,.
    \end{equation}
  This representation corresponds to the states of a single free ``light"  massive particle with $0<m^2<\frac{9}{4}H^2$
    for $s=0$  and $0<m^2<\frac{(2s-1)^2}{4}H^2$ for $s\ge1$.
    \item \textbf{Type I exceptional series} $\mathcal{V}_{p,0}$: $\Delta=2+p$ and $s=0$ for $p\in\mathbb{Z}_{>0}$. The $SO(4)$ content is
    \begin{equation} 
        \mathcal{V}_{p,0}\Big|_{SO(4)}=\bigoplus^\infty_{n=p}\mathbb{Y}_{n}
        \label{eq:so4ofV}\,.
    \end{equation}
    This representation corresponds to the states of a single free scalar particle with a specific  shift-symmetry \cite{Bonifacio:2018zex}. This can be seen from the fact that the first $0<n<p$ modes in (\ref{eq:so4ofV}) are missing.
    \item \textbf{Type II exceptional series} $\mathcal{U}_{s,t}:$ $\Delta=2+t$ and $s\geq 1$ with $t=0,1,2\ldots s-1$. The $SO(4)$ content is
     \begin{equation} 
        \mathcal{U}_{s,t}^\pm\Big|_{SO(4)}=\bigoplus^\infty_{n=s}\bigoplus^s_{m=t+1}\mathbb{Y}_{n,\pm m}\,,
    \end{equation}
    where we highlighted the fact that for each $s$ and $t$ there are two irreducible representations labeled by $\pm$. These are related by parity, so that $\mathcal{U}_{s,t}=\mathcal{U}_{s,t}^+\oplus\mathcal{U}_{s,t}^-$ is irreducible with respect to $O(1,4)$.
    
    These representations are associated to the single-particle Hilbert spaces of  ``partially massless fields" when $t<s-1$ and massless gauge fields when $t=s-1$. Partially massless fields have a number of degrees of freedom which is in between that of a massive field and that of a gauge field. 
    
    The states of a free \textbf{photon} are associated to the UIR $\mathcal{U}_{1,0}$, which is the focus of this paper. In that case, $\mathcal{U}^+_{1,0}$ and $\mathcal{U}^-_{1,0}$ correspond to the two helicities of the photon.
    
\end{itemize}
There is an isomorphism between the representations labeled $\Delta$ and $3-\Delta$.
\section{Photons without photons: group theory explanation}
\label{subsec:grouptheory}
As first pointed out by Martin \cite{MartindS} and then further explored in \cite{Penedones:2023uqc}, in de Sitter, states in the photon UIR appear in various tensor products of states belonging to other UIRs. Tensor products of UIRs of $SO(1,4)$ were also studied in \cite{Dobrev:1976vr,Dobrev:1977qv,Hirai:1965,Hirai:1966,zhang2017tensorproductscomplementaryseries}. Here we will review the derivation of Martin in the case of two massless scalars. For pedagogical purposes, we start by considering the case of two massive scalars, where instead no photon state appears, and then take their masses to zero. We give a completely parallel treatment of this case in the language of spectral densities and QFT in section \ref{subsec:masslessscalars}.

We follow the notation in \cite{Penedones:2023uqc}. In this section we label states by the complex number $\Delta$ and the $SO(4)$ spin $n$, where $n$ is related to the quadratic Casimir of $SO(4)$ as 
\begin{equation}
    \mathcal{C}_2^{SO(4)}|\Delta,n\rangle_{a_1\ldots a_n}=n(2+n)|\Delta,n\rangle_{a_1\ldots a_n}\,,
\end{equation}
where $a_i=1,\ldots,4$ and states are traceless and symmetric in these indices. In terms of the quadratic Casimir of $SO(1,4)$, we have
\begin{equation}
    \mathcal{C}_2^{SO(1,4)}=-L_{0a}^2-\mathcal{C}_2^{SO(4)}\,.
\end{equation}
Remember that the reduction to $SO(4)$ of a scalar principal series and of a photon UIR are
\begin{equation}
    \mathcal{P}_{\Delta,0}\Big|_{SO(4)}=\bigoplus_{n=0}^\infty\mathbb{Y}_n\,, \qquad
    \mathcal{U}_{1,0}\Big|_{SO(4)}=\bigoplus_{n=1}^\infty\left(\mathbb{Y}_{n,-1}\oplus\mathbb{Y}_{n,1}\right)\,.
    \label{eq:oplusexc2}
\end{equation}
Moreover, consider the fact that the action of a generator $L_{0a}$ on a state in the principal series was computed in \cite{Penedones:2023uqc} to be
\begin{equation}
    L_{0a}|\Delta,n\rangle_{a_1\ldots a_n}=\alpha_n(\Delta)|\Delta,n+1\rangle_{aa_1\ldots a_n}+\beta_n(\Delta)(\delta_{a(a_1}|\Delta,n-1\rangle_{a_2\ldots a_n)}-\text{trace})\,,
    \label{eq:casimirondelta}
\end{equation}
with
\begin{equation}
    \alpha_n(\Delta)=\sqrt\frac{(n+1)(\Delta+n)(\bar\Delta+n)}{4+2n}\,,\qquad\beta_n(\Delta)=\sqrt\frac{n(\Delta+n-1)(\bar\Delta+n-1)}{2+2n}\,.
    \label{eq:alphabetan}
\end{equation}
Importantly, for the case of massless scalars the analogous coefficients are simply the analytic continuation of (\ref{eq:alphabetan}) to $\Delta=3$ and $\bar\Delta=0$.

At the same time, for a state in $\mathcal{U}_{1,0}$, which can be expressed as $|\gamma\rangle_{a,b}=-|\gamma\rangle_{b,a}$ \cite{Penedones:2023uqc}, the action of a generator should not lead outside of its Hilbert space. In particular, it must be that
\begin{equation}
    L_{0a}|\gamma\rangle_{a,b}=0
\end{equation}
because $L_{0a}|\gamma\rangle_{a,b}$ is a spin $1$ representation of $SO(4)$, and such representation does not appear in (\ref{eq:oplusexc2}). If a photon appears in the tensor product of two scalars, it must be that we can find some coefficients $c_n$ such that
\begin{equation}
    |\gamma\rangle_{a,b}\equiv\sum_{n=1}^\infty c_n\left(|\Delta_1,n\rangle_{aa_2\ldots a_n}|\Delta_2,n\rangle_{b}^{\ \ a_2\ldots a_n}-(a\leftrightarrow b)\right)\,, \qquad L_{0a}|\gamma\rangle_{a,b}=0\,.
    \label{eq:definitiongamma}
\end{equation}
Using (\ref{eq:casimirondelta}), this requirement can be recasted as a recursion relation for the $c_n$ coefficients for $n\geq1$ \cite{Penedones:2023uqc}
\begin{equation}
\begin{aligned}
    c_n\alpha_n(\Delta_1)+c_{n+1}\beta_{n+1}(\Delta_2)\frac{n+3}{n+1}&=0\,,\\
    c_n\alpha_n(\Delta_2)+c_{n+1}\beta_{n+1}(\Delta_1)\frac{n+3}{n+1}&=0\,.
\label{eq:recursion}
\end{aligned}
\end{equation}
Moreover, the condition $L_{0a}|\gamma\rangle_{a,b}=0$ also imposes that $c_1\beta_1=0$. This is the crucial difference between the case of massive scalars and that of massless scalars. For massive scalars, since $\beta_1\neq0$, this condition immediately implies that $c_1=0$ and further that all $c_n=0$, meaning there is no photon state in the tensor product of massive scalars. If instead we consider massless scalars, then one can see from (\ref{eq:alphabetan}) that $\beta_1=0$, meaning there is no restriction on $c_1$. The recursion relation (\ref{eq:recursion}) then can be solved as 
\begin{equation}
    c_n=\frac{C}{(n+1)(n+2)}\,,
    \label{eq:recursionsolution}
\end{equation}
for an arbitrary constant $C$. It can be then checked that the state $|\gamma\rangle_{a,b}$ found by plugging (\ref{eq:recursionsolution}) into (\ref{eq:definitiongamma}) is normalizable, and we can thus state that precisely one photon representation appears in the tensor product of two massless scalar representations.

We have reviewed the group theoretic derivation for the case of two massless scalars. Completely analogous proofs show that photon states also appear in the tensor product of a principal series scalar and a principal series vector, of two principal series vectors and in the decomposition of an antisymmetric two-index conformal primary. In Section \ref{sec:photonsfrommass0} we show how these statements are realized in free QFTs in de Sitter, and in Section \ref{sec:photonsfrommass1} we show that they persist at one loop for some classes of interactions. 
\section{Some details on inversion formulae}
\label{sec:criterion}
Here we derive the inversion formulae presented in (\ref{eq:inversionsdS}). Once the two-point functions of $j\equiv \star \mathbb{d}\star B$ and $\tilde j\equiv\star \mathbb{d}B$ are constructed, the techniques to be used are precisely those presented in \cite{Loparco:2023rug}, so this is merely a review of the derivation in that paper. 

Let us focus for example on the two-point function of $j$, which we have expressed as
\begin{equation}
    \langle0|j^A(Y_1)j^B(Y_2)|0\rangle=\int_{\frac{3}{2}+i\mathbb{R}}[d\Delta]c_\Delta\varrho^{\mathcal{P},+}(\Delta)\Pi_1^{AB}G_\Delta(\sigma)+\text{complementary series}
\end{equation}
where here $c_\Delta\equiv 4(\Delta-1)^2(\Delta-2)^2$.

The main idea is to analytically continue both sides of this equation to Euclidean Anti de Sitter space (EAdS). We will indicate vectors in the embedding space that lie in the EAdS hyperboloid with $X^A\in\mathbb{R}^{1,4}$. In contrast to points in de Sitter, these vectors satisfy $X^AX_A =-1$. Then, it is known that \cite{Sleight:2019hfp,Sleight:2021plv,Sleight:2020obc}
\begin{equation}
    \Pi_1^{AB}G_{\frac{3}{2}+i\mu}(X_1\cdot X_2)=\Gamma(i\mu)\Gamma(-i\mu)\Pi_1^{AB}\Omega_\mu(X_1\cdot X_2)\,,
\end{equation}
where $\Omega_\mu$ is a harmonic function in EAdS, meaning it is an egeinfunction of the Laplacian in EAdS with eigenvalue $\frac{9}{4}+\mu^2$. Such functions with $\mu\in\mathbb{R}$ form a complete and orthogonal basis of square integrable transverse two-point functions in EAdS \cite{Camporesi:1994ga,Costa:2014kfa}. This property of completeness conditioned on square integrability allows us to establish the following criterion for when terms from the complementary series appear in the K\"allén-Lehmann decomposition: 
\begin{equation}
    \int_{X_1}\langle0|j^A(X_1)j^B(X_2)|0\rangle\langle0|j_A(X_1)j_B(X_2)|0\rangle<\infty\longrightarrow\text{no complementary series}
    \label{eq:sqrtintgrb}
\end{equation}
We can phrase this criterion in a more useful form. If we contract the indices of the two-point function with some auxiliary null vectors $W^A$ tangent to EAdS ($W\cdot X=0$), we can write the two-point function as \cite{Costa:2014kfa}
\begin{equation}
    W_{1A}W_{2B}\langle0|j^A(X_1)j^B(X_2)|0\rangle=(W_1\cdot W_2)f_0(X_1\cdot X_2)+(W_1\cdot X_2)(W_2\cdot X_1)f_1(X_1\cdot X_2)\,.
\end{equation}
Let us further introduce the following notation for the fall-offs of $f_0$ and $f_1$ at spacelike infinity ($\sigma\to-\infty$):
\begin{equation}
    f_0(\sigma)\to|\sigma|^{-\omega_0}\,, \qquad f_1(\sigma)\to|\sigma|^{-\omega_1-1}\,.
\end{equation}
Then, we showed in \cite{Loparco:2023rug} that the square integrability condition (\ref{eq:sqrtintgrb}) for spin 1 two-point functions in dS$_4$ can be recasted as
\begin{mdframed}[backgroundcolor=shadecolor,linewidth=0pt]
\begin{equation}
    \underset{n}{\text{min}}[\text{Re}(\omega_n)]>\frac{5}{2}\quad\longrightarrow\quad \text{no complementary series}
    \label{eq:criterion}
\end{equation}
\end{mdframed}
\noindent For all the two-point functions we studied in Section \ref{sec:photonsfrommass0} we could find a regime of the parameters of the theory for which this criterion was satisfied. Then, by continuation in those parameters, we retrieved the complementary series contributions as poles that crossed the contour of integration over the principal series.

\paragraph{Finding the principal series density}
Let us thus assume we are in a regime where (\ref{eq:criterion}) is satisfied. We can express our two-point function continued to EAdS as
\begin{equation}
    \langle0|j^A(X_1)j^B(X_2)|0\rangle=\int_{\mathbb{R}}d\mu\ c_{\frac{3}{2}+i\mu}\Gamma(i\mu)\Gamma(-i\mu)\varrho^{\mathcal{P},+}(\frac{3}{2}+i\mu)\Pi_1^{AB}\Omega_{\mu}(X_1\cdot X_2)
\end{equation}
Now we will use the following orthogonality property of harmonic functions
\begin{equation}
    \int_{X_2}\Pi^{AB}_1\Omega_\mu(X_1\cdot X_2)\Pi_{1,B}^{\quad C}\Omega_{\mu'}(X_2\cdot X_3)=\frac{1}{2}\left(\delta(\mu+\mu')+\delta(\mu-\mu')\right)\Pi_1^{AC}\Omega_\mu(X_1\cdot X_3)
\end{equation}
together with the coincident point expression \cite{Costa:2014kfa}
\begin{equation}
    \Pi_{1A}^{A}\Omega_\mu(-1)=\frac{3 \mu  \left(4 \mu ^2+9\right) \tanh (\pi  \mu )}{64 \pi ^2}
\end{equation}
to derive the inversion formula \cite{Loparco:2023rug} (\ref{eq:inversionsdS})
\begin{equation}
    \varrho^\mathcal{P,+}(\Delta)=n_\Delta\int _{X_2}\Pi_1^{AB}G_{\Delta}(X_1\cdot X_2)\langle0|j_B(X_2)j_A(X_1)|0\rangle
    \label{eq:inv222}
\end{equation}
with $n_\Delta$ defined in (\ref{eq:defndelta}), and analogously for the parity odd spectral density. 
But how do we use such a formula in practice? First of all, one can fix some coordinates to reduce (\ref{eq:inv222}) to a scalar integral (see eq. (G.5) in \cite{Loparco:2023rug}). Then, the practical solvability of the integral is dependent on the form of the two-point function of $j$. We know how to carry out these integrals explicitly only for the two-point functions of some classes of operators:
\begin{itemize}
    \item CFT traceless symmetric and two-index antisymmetric primaries (\cite{Loparco:2023rug} and this paper)
    \item Composite operators made of two elementary fields in free theories \cite{Loparco:2023rug}
    \item Free fields in dimensional reduction (decomposing a two-point function in $d$ dimensional free theory on the Hilbert space of a $d'<d$ dimensional hypersurface) \cite{Loparco:2023akg}.
    \item Vertex operators constructed from compact massless scalars \cite{Chakraborty:2023eoq}.
\end{itemize}
In the first two cases, the computation of this integral can be carried out in a completely algorithmic way. We reported the necessary steps in great detail in Appendix H of \cite{Loparco:2023rug}, so we will refrain from repeating them here.
\paragraph{Finding the complementary series density}
As stated before, in all known cases the complementary series irreps appear as a discrete sum of terms, which can be retrieved by analytic continuation in some parameter. In practice, one starts from some range of the parameters of the theory in which the two-point function of interest only includes principal series (or in other words, it satisfies criterion (\ref{eq:criterion})), and then continues in those parameters. Poles in the principal series densities can cross the contour of integration under this procedure, and lead to complementary series contributions. We show a practical example of this in the case of the massive scalars in Section \ref{subsec:masslessscalars}. 

In general, we can thus write
\begin{equation}
    \varrho^{\mathcal{C},\pm}(\Delta)=\sum_{\{\Delta'\}}\delta(\Delta-\Delta')\varrho^{\mathcal{C},\pm}_{\Delta'}\,,\qquad \varrho^{\mathcal{C},\pm}_{\Delta'}=-4\pi\underset{\Delta=\Delta'}{\text{Res}}\left[\varrho^{\mathcal{P},\pm}(\Delta)\right]
\end{equation}
where $\{\Delta'\}$ is the set of poles in the principal series density that cross the contour of integration from right to left under analytic continuation.
\paragraph{Finding the photon UIR density}
The photon density can always be reconstructed numerically once all the other densities are known: one can subtract the K\"allén-Lehmann decomposition from the two-point function itself and extract the photon density from a fit, given that it is just a number and not a function.

It turns out that, in all the examples we studied, we could extract an analytic form for the photon density in terms of the other parameters of the theory. We first computed the principal series densities as described in the previous Sections. Then, we studied the late-time expansion of the electric and magnetic components of the two-point function of interest. Matching with (\ref{eq:lateB0i}) and (\ref{eq:lateBij}) we encountered two types of outcomes, which allowed us to derive an analytic form:
\begin{itemize}
    \item \textbf{Outcome 1}: There are no terms in the late time expansion of the two-point function that behave as an electric or magnetic field. At the same time, there are poles at $\Delta=2$ in the principal series spectral density. Then, the following cancellation must be taking place:
    \begin{equation}
        \varrho^\gamma=4\pi\underset{\Delta=2}{\text{Res}}\left[\varrho^{\mathcal{P},+}(\Delta)\right]=4\pi\underset{\Delta=2}{\text{Res}}\left[\varrho^{\mathcal{P},+}(\Delta)\right]\,.
    \end{equation}
    This is what happens in most of the cases studied in Section \ref{sec:photonsfrommass0}.
    \item \textbf{Outcome 2}: The magnetic component of the two-point function behaves as a magnetic field at late times, but the electric component does not behave as an electric field. Then, the following cancellation must happen
    \begin{equation}
        \varrho^\gamma=4\pi\underset{\Delta=2}{\text{Res}}\left[\varrho^{\mathcal{P},+}(\Delta)\right]\,.
    \end{equation}
    This is what happens in the case of the massless scalars \ref{subsec:masslessscalars}.
\end{itemize}
A third outcome which would allow for an analytic computation of $\varrho^\gamma$, but which we have not encountered in any example, is the following
\begin{itemize}
    \item \textbf{Outcome 3}: The electric component of the two-point function behaves as an electric field at late times, but the magnetic component does not behave as a magnetic field. Then, the following cancellation must happen
    \begin{equation}
        \varrho^\gamma=4\pi\underset{\Delta=2}{\text{Res}}\left[\varrho^{\mathcal{P},-}(\Delta)\right]\,.
    \end{equation}
\end{itemize}
\section{Perturbation theory on the sphere}
\label{sec:sphere}
In this Appendix we present all the technical details needed to perform the computations in Section \ref{sec:photonsfrommass1} of the main text. The part on scalar two-point functions follows \cite{Chakraborty:2023qbp}.
\subsection{Decompositions in complete bases: scalar two-point functions}
Any smooth two-point function on the sphere can be decomposed into a complete basis of Gegenbauer polynomials
\begin{equation}
    \langle\mathcal{O}(Y_1)\mathcal{O}(Y_2)\rangle=\frac{\Gamma(\frac{d}{2})}{2\pi^{\frac{d+2}{2}}}\sum_{J=0}^\infty\left(J+\frac{d}{2}\right)[\mathcal{O}]_JC_J(Y_1\cdot Y_2)
    \label{eq:harmonicdecompose}
\end{equation}
for some coefficients $[\mathcal{O}]_J$ and where $C_J$ are Gegenbauer polynomials on the $(d+1)$-dimensional sphere
\begin{equation}
    C_J(\sigma)=\frac{\Gamma(J+d)}{\Gamma(J+1)\Gamma(d)}\ _2F_1\left(-J,d+J,\frac{d+1}{2},\frac{1-\sigma}{2}\right)\,.
    \label{eq:gegenbauer}
\end{equation}
To make connection with the notation used in the main text, then,
\begin{equation}
    f_J(\sigma)\equiv\frac{\Gamma(\frac{d}{2})(J+\frac{d}{2})}{2\pi^{\frac{d+2}{2}}}C_J(\sigma)
\end{equation}
The coefficients $[\mathcal{O}]_J$ can be interpreted as the ``momentum space" two-point function of the operator $\mathcal{O}$.
For example, the free propagator of a scalar field $\phi$ of mass $m^2$ decomposes as (\ref{eq:harmonicdecompose}) with the coefficients
\begin{equation}
    [\phi]_J=\frac{1}{J(J+d)+m^2}\,,
\end{equation}
making manifest the similarity with momentum Feynman propagators in flat space. 
A useful property of this representation is that diagrams become simply products of the momentum space two-point functions $[\mathcal{O}]_J$ because of the orthogonality of Gegenbauer polynomials
\begin{equation}
    \int_{Y_2}C_{J}(Y_1\cdot Y_2)C_{J'}(Y_2\cdot Y_3)=\delta_{JJ'}\frac{2\pi^{\frac{d+2}{2}}}{\Gamma(\frac{d}{2})}\frac{1}{(J+\frac{d}{2})}C_{J}(Y_1\cdot Y_3)\,,
\end{equation}
where $\int_Y$ is defined in (\ref{eq:intY}). From this, the orthogonality property (\ref{eq:orthoffs}) of the main text follows.

\begin{figure}
\centering
\includegraphics[scale=1.4]{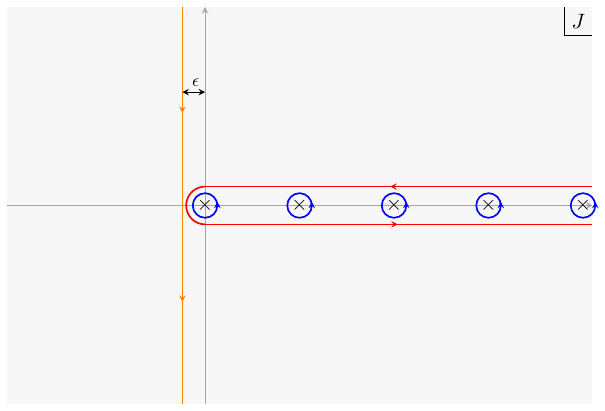}
\caption{A representation of the steps needed to go from the sum over Gegenbauer polynomials (\ref{eq:harmonicdecompose}) to the integral over complex $J$ (\ref{eq:wsintegral}). In blue is the original contour. The first step is to deform it into one continuous contour (red) that wraps all poles, and finally to open it and make it run vertically close to the imaginary axis (orange). A small $\epsilon$ is needed to not hit the pole at $J=0$.}
\label{fig:WSint}
\end{figure}

The sum (\ref{eq:harmonicdecompose}) only converges when the two points are on the sphere. To analytically continue to all $\sigma\in\mathbb{C}\backslash[1,\infty)$, we can perform a Watson-Sommerfeld transform by introducing poles with unit residue at $J\in\mathbb{N}$ and then writing a contour integral that wraps them. The contour can then be opened to run along the $J=-\epsilon$ axis, where $\epsilon$ is a small positive number. This continuation introduces no further terms to the integral if we assume for simplicity that $[\mathcal{O}]_J$ has no poles in the right half complex $J$ plane, something which is true for all cases we studied. See Figure \ref{fig:WSint} for a representation of these contour manipulations. 
In equations,
\begin{equation}
    \langle\mathcal{O}(Y_1)\mathcal{O}(Y_2)\rangle=\int_{-\epsilon-i\mathbb{R}}\frac{dJ}{2i}\frac{1}{\sin(\pi J)}[\mathcal{O}]_Jf_J(-Y_1\cdot Y_2)\,,
    \label{eq:wsintegral}
\end{equation}
where we used the reflection formula for Gegenbauer polynomials
\begin{equation}
    C_J(z)=e^{i\pi J} C_J(-z)\,.
\end{equation}
and introduced a kernel with unit residue $k(J)=\frac{\pi}{\sin(\pi J)}e^{i\pi J}$. For later convenience, we define the shorthand notation
\begin{equation}
    \int_J(\ldots)\equiv\frac{\Gamma(\frac{d}{2})}{2\pi^{\frac{d}{2}}}\int_{-\epsilon-i\mathbb{R}}\frac{dJ}{2\pi i}\frac{J+\frac{d}{2}}{\sin(\pi J)}(\ldots)
    \label{eq:shortJ}
\end{equation}
This integral representation is intimately related to the K\"allénn-Lehmann decomposition. The two are connected by a simple analytic continuation where we let the orange contour in Figure \ref{fig:WSint} slide all the way to Re$J=-\frac{d}{2}$. If $[\mathcal{O}]_J$ has poles in the strip Re$J\in(-\frac{d}{2},0)$, residues on their positions must be taken into account. Because of unitarity, poles in that strip must appear on the real line, and be eventually related to complementary series irreps appearing in the K\"allén-Lehmann representation. Call this set of poles $\{J_C\}$. Then, under this procedure, we land on 
\begin{equation}
\begin{aligned}
    \langle\mathcal{O}(Y_1)\mathcal{O}(Y_2)\rangle=&\frac{\Gamma(\frac{d}{2})}{2\pi^{\frac{d}{2}}}\int_{-\frac{d}{2}-i\mathbb{R}}\frac{dJ}{2\pi i}\frac{J+\frac{d}{2}}{\sin(\pi J)}[\mathcal{O}]_JC_J(-Y_1\cdot Y_2)\\
    &-\frac{\Gamma(\frac{d}{2})}{2\pi^{\frac{d}{2}}}\sum_{J\in\{J_C\}}\frac{J+\frac{d}{2}}{\sin(\pi J)}\underset{J'=J}{\text{Res}}[\mathcal{O}]_{J'}C_{J}(-Y_1\cdot Y_2)\,.
    \label{eq:OJJ}
\end{aligned}
\end{equation}
Notice that, if we call $J=-\Delta$, we have the following identity
\begin{equation}
    \frac{\Gamma(\frac{d}{2})}{2\pi^{\frac{d}{2}}}\frac{1}{\sin(\pi\Delta)}C_{-\Delta}(-Y_1\cdot Y_2)=2G_\Delta(Y_1\cdot Y_2)\,,
    \label{eq:Gfromgegen}
\end{equation}
where $G_\Delta(\sigma)$ is the propagator of a free massive scalar with $m^2=\Delta(d-\Delta)$ in de Sitter, see (\ref{eq:freeprop}). We are very close to the K\"allén-Lehmann representation. We simply need to use the symmetries of the integral to write it as
\begin{equation}
    \langle\mathcal{O}(Y_1)\mathcal{O}(Y_2)\rangle=\int_{\frac{d}{2}+i\mathbb{R}}[d\Delta]\varrho^{\mathcal{P}}(\Delta)G_\Delta(Y_1\cdot Y_2)+\sum_{\Delta\in\{\Delta_C\}} \varrho_{\Delta}^{\mathcal{C}}G_{\Delta}(Y_1\cdot Y_2)\,.
\end{equation}
where $\{\Delta_C\}=\{-J_C\}$ and \cite{Hogervorst:2021uvp,Chakraborty:2023qbp}
\begin{equation}
    \varrho^{\mathcal{P}}(\Delta)=\frac{1}{2\pi}\left(\Delta-\frac{d}{2}\right)\left([\mathcal{O}]_{-\bar\Delta}-[\mathcal{O}]_{-\Delta}\right)\,, \qquad \varrho_{\Delta}^{\mathcal{C}}=2\left(\frac{d}{2}-\Delta\right)\underset{J=-\Delta}{\text{Res}}[\mathcal{O}]_J
    \label{eq:rhofromcharm}
\end{equation}
This relation between harmonic coefficients and spectral densities is very useful, and it can be inverted as follows \cite{Chakraborty:2025myb}
\begin{equation}
    [\mathcal{O}]_J=\int_{\frac{d}{2}+i\mathbb{R}}[d\Delta]\frac{\varrho^{\mathcal{P}}(\Delta)}{J(J+d)+\Delta\bar\Delta}+\sum_{\Delta\in\{\Delta_C\}}\frac{\varrho^{\mathcal{C}}_{\Delta}}{J(J+d)+\Delta\bar\Delta}\,,
    \label{eq:GJint}
\end{equation}
This equation is valid for $\text{Re}[J]>-\min\left(\{\Delta_C\},\frac{d}{2}\right)$, and the extension to the rest of the complex $J$ plane can be obtained by analytic continuation. 

\begin{figure}
\centering
\includegraphics[scale=1.4]{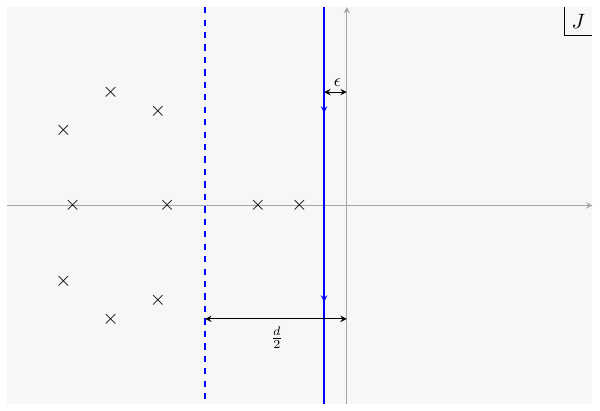}
\caption{The generic analytic structure of a ``momentum space" two-point function $[\mathcal{O}]_J$. Typically they are fully analytic on the right half of the complex plane. Poles appear in complex conjugate pairs and in the strip Re$J\in(-\frac{d}{2},0)$ they can only appear on the real axis due to unitarity. In blue we show the contour in the integral representation (\ref{eq:wsintegral}).}
\label{fig:WSint2}
\end{figure}

In practice, sometimes one needs to subtract the asymptotic behavior of the spectral density to get a finite expression from the integral in (\ref{eq:GJint}). Let us parametrize the integral over the principal series with $\Delta=\frac{d}{2}+i\mu$ and let us indicate the asymptotic behavior of the spectral density as $\varrho_{\mathcal{O}}^{\mathcal{P}}(\frac{d}{2}+i\mu)\sim c\mu^a$ for large real $\mu$. Then, assuming there are no subleading divergences, truncating the integral at some $\mu=\mu_{\text{max}}$ we can well approximate the harmonic coefficient as \cite{Chakraborty:2025myb}
\begin{equation}
    [\mathcal{O}]_J\sim 2\int_0^{\mu_{\text{max}}}d\mu\frac{\varrho_{\mathcal{O}}^{\mathcal{P}}(\frac{d}{2}+i\mu)}{J(J+d)+\frac{d^2}{4}+\mu^2}-2c\frac{\mu_{\text{max}}^{a-1}}{a-1}+\text{complementary}\,.
    \label{eq:numreg}
\end{equation}
\paragraph{Late time behavior from poles in $J$} There is another advantage to the representation (\ref{eq:wsintegral}). Typically the coefficient $[\mathcal{O}]_J$ only has poles on the left half of the complex $J$ plane. Moreover, because of unitarity, we established that poles in the strip Re$J\in(-\frac{d}{2},0)$ must lie on the real line (see Figure \ref{fig:WSint2} for a representation of the generic analytic structure of $[\mathcal{O}]_J$).
This simpler analytic structure makes it more immediate to extract the late time behavior compared to the K\"allén-Lehmann representation, especially if we adopt the further assumption that all complementary series contributions come from poles crossing the contour of integration over the principal series when continuing in some parameter of the theory. This is what happens in every known example, and it implies that
\begin{equation}
     \varrho^{\mathcal{C}}_{\mathcal{O},\Delta} =-4\pi\underset{\Delta'=\Delta}{\text{Res}}\left[\varrho^{\mathcal{P}}_{\mathcal{O}}(\Delta)\right]
     \label{eq:polesassumption}
\end{equation}
where $\Delta$ is the scaling dimension associated to the pole that crosses from right to left. Comparing this with (\ref{eq:rhofromcharm}), we deduce that this assumption implies that
\begin{equation}
    \underset{J=\Delta-d}{\text{Res}}[\mathcal{O}]_J=0\,, \qquad \forall\Delta\in\{\Delta_C\}\,.
\end{equation}
At the same time, as was derived in the main text in Section \ref{subsec:latetimes}, the K\"allén-Lehmann representation for scalar two-point functions can be recasted as \cite{Loparco:2023rug, Hogervorst:2021uvp}
\begin{equation}
\begin{aligned}
    \langle\mathcal{O}(Y_1)\mathcal{O}(Y_2)\rangle=&-4\pi\sum_{\Delta\in\{\Delta_P\}}N_{\Delta}\underset{\Delta'=\Delta}{\text{Res}}\left[\varrho^{\mathcal{P}}_{\mathcal{O}}(\Delta')\right]G_\Delta^{\text{AdS}}(\sigma)\\
    &+\sum_{\Delta\in\{\Delta_C\}}N_{\Delta}\varrho_{\mathcal{O},\Delta}^{\mathcal{C}}\left[G_\Delta^{\text{AdS}}(\sigma)-G_{d-\Delta}^{\text{AdS}}(\sigma)\right]
    \label{eq:latescalar}
\end{aligned}
\end{equation}
where $\{\Delta_P\}$ is the set of poles of the principal series spectral density which lie on the right of the Re$\Delta=\frac{d}{2}$ line. This representation is valid only for spacelike configurations with $\sigma<-1$.

If indeed (\ref{eq:polesassumption}) is valid, it must be that all the complementary series terms that go like $d-\Delta$ are canceled by some elements of the sum over $\{\Delta_P\}$. Overall, we would get
\begin{equation}
\begin{aligned}
    \langle\mathcal{O}(Y_1)\mathcal{O}(Y_2)\rangle=&-4\pi\sum_{\Delta\in\{\Delta_P'\}}N_{\Delta}\underset{\Delta'=\Delta}{\text{Res}}\left[\varrho^{\mathcal{P}}_{\mathcal{O}}(\Delta')\right]G_\Delta^{\text{AdS}}(\sigma)\\
    &+\sum_{\Delta\in\{\Delta_C\}}N_{\Delta}\varrho_{\mathcal{O},\Delta}^{\mathcal{C}}G_\Delta^{\text{AdS}}(\sigma)
\end{aligned}
\end{equation}
where $\{\Delta_P'\}$ is now the set of poles in the principal series spectral density with Re$\Delta>\frac{d}{2}$ and that are different than $d-\Delta$ with $\Delta\in\{\Delta_C\}$. Now, comparing with (\ref{eq:rhofromcharm}), we realize that this set of powers corresponds exactly to the set of poles of $[\mathcal{O}]_J$ with Re$J<0$, which we call $\{J_*\}$. Then, we can compactly write
\begin{mdframed}[backgroundcolor=shadecolor,linewidth=0pt]
\begin{equation}
    \langle\mathcal{O}(Y_1)\mathcal{O}(Y_2)\rangle=\sum_{J\in\{J_*\}}c_J\underset{J'=J}{\text{Res}}\left[\mathcal{O}\right]_{J'}\left[\left(\frac{\eta^2}{\mathbf{y}^2}\right)^{-J}+\ldots\right]
    \label{eq:finallatescalar}
\end{equation}
\end{mdframed}
\noindent where the dots stand for contributions coming from descendants, which are completely fixed by symmetry \cite{Hogervorst:2021uvp}, and 
\begin{equation}
    c_J=\frac{\Gamma (-J) \Gamma(\frac{d}{2}+J+1)}{2\pi^{\frac{d}{2}+1}}\,.
\end{equation}
\subsection{Decompositions in complete bases: antisymmetric two-point functions}
\label{subsec:bases2}
For two-point functions of antisymmetric tensors, the situation is totally analogous. The decomposition in Gegenbauer polynomials reads
\begin{equation}
    \langle B^{AB}(Y_1)B^{CD}(Y_2)\rangle=\frac{\Gamma(\frac{d}{2})}{2\pi^{\frac{d+2}{2}}}\sum_\pm\sum_{J=0}^\infty\left(J+\frac{d}{2}\right)[B]_{\pm,J}\Pi_{\pm}^{ABCD}C_J(Y_1\cdot Y_2)\,,
    \label{eq:sphereBB}
\end{equation}
where the differential operators $\Pi_{\pm}^{ABCD}$ are precisely the same as those that appear in the K\"allén-Lehmann representation (\ref{eq:dsproj}). 

Notice that we are on $S^{d+1}$ with $d$ arbitrary, while the decomposition of antisymmetric tensors in Section \ref{sec:kallanlehmann} was strictly derived in 4 dimensions. One should thus worry whether the basis changes drastically when $d+1\neq4$. While we do not have a proof, we observe empirically that when a two-point function of antisymmetric tensors is analytic in $d$, the decomposition (\ref{eq:sphereBB}) converges to the correct two-point function for any $d\in\mathbb{R}$ (we explicitly checked the case of the two-point function of massless scalars studied in \ref{subsec:masslessscalars} with coefficients (\ref{eq:masslessscalars6F5})). We will thus trust this decomposition in an arbitrary number of dimensions.

\noindent To make contact with the main text (e.g. eq. (\ref{eq:Binharms})), we have
\begin{equation}
    f_{\pm,J}^{AB,CD}(Y_1,Y_2)\equiv\frac{\Gamma(\frac{d}{2})(J+\frac{d}{2})}{2\pi^{\frac{d+2}{2}}}\Pi_\pm^{ABCD}C_J(Y_1\cdot Y_2)\,.
    \label{eq:fabcddef}
\end{equation}
As in the scalar case, the sum over Gegenbauer polynomials only converges on the sphere. Following the same exact steps as in the previous subsection, we can do a Watson-Sommerfeld transform and retrieve the following integral representation 
\begin{equation}
    \langle B^{AB}(Y_1)B^{CD}(Y_2)\rangle=\sum_\pm\int_J[B]_{\pm,J}\Pi^{ABCD}_\pm C_J(-Y_1\cdot Y_2)
    \label{eq:WSantisy}
\end{equation}
which converges for all $\sigma\in\mathbb{C}\backslash[1,\infty)$. Just like in the scalar case, this integral can be continued to the Re$J=-\frac{d}{2}$ line, up to some poles crossing the contour
\begin{align}
   \langle B^{AB}(Y_1)B^{CD}(Y_2)\rangle=&\frac{\Gamma(\frac{d}{2})}{2\pi^{\frac{d}{2}}}\sum_\pm\int_{-\frac{d}{2}-i\mathbb{R}}\frac{dJ}{2\pi i}\frac{J+\frac{d}{2}}{\sin(\pi J)}[B]_{\pm,J}\Pi^{ABCD}_\pm C_J(-Y_1\cdot Y_2)\label{eq:BJJJJ}\\
    &-\frac{\Gamma(\frac{d}{2})}{2\pi^{\frac{d}{2}}}\sum_\pm\sum_{J\in\{-1,J_C^\pm\}}\frac{J+\frac{d}{2}}{\sin(\pi J)}\underset{J'=J}{\text{Res}}[B]_{\pm,J'}\Pi^{ABCD}_\pm C_{J}(-Y_1\cdot Y_2)\,.\nonumber
\end{align}
where we indicated explicitly that the poles that can cross the contour are at $J=-1$, corresponding to the photon UIR, and $\{J_C^\pm\}$, the sets of poles of the $+$ and $-$ coefficients that can lie on the complementary series, $-\frac{d}{2}<J<-1$.

Comparing with the K\"allén-Lehmann representation using (\ref{eq:Gfromgegen}), we deduce the following relation between spectral densities and harmonic coefficients
\begin{equation}
\begin{aligned}
    \varrho^{\mathcal{P},\pm}(\Delta)=\frac{1}{2\pi}\left(\Delta-\frac{d}{2}\right)&\left([B]_{\pm,-\bar\Delta}-[B]_{\pm,-\Delta}\right)\,, \qquad \varrho^\gamma=\sum_\pm\underset{J=-1}{\text{Res}}\left[[B]_{\pm,J}\right]\\
    &\varrho_{\Delta}^{\mathcal{C},\pm}=2\left(\frac{d}{2}-\Delta\right)\underset{J=-\Delta}{\text{Res}}[B]_{\pm,J}\,.
    \label{eq:rhofromharms2}
\end{aligned}
\end{equation}
This equation can be inverted as in the scalar case, giving
\begin{equation}
    [B]_{\pm,J}=\int_{\frac{d}{2}+i\mathbb{R}}[d\Delta]\frac{\varrho^{\mathcal{P},\pm}(\Delta)}{(J+\Delta)(J+\bar\Delta)}+\sum_{\Delta\in\{\Delta_C^\pm\}}\frac{\varrho^{\mathcal{C},\pm}_{\Delta}}{J(J+d)+\Delta\bar\Delta}+\frac{\varrho^\gamma}{2(J+1)(J+2)}
    \label{eq:Bfromprinc}
\end{equation}
where the domain of validity is meant to be Re$J>-1$, and the function in the rest of the complex $J$ plane can be reconstructed by analytic continuation. Notice that we can distribute the $\varrho^\gamma$ contribution in $[B]_{+,J}$ and $[B]_{-,J}$ arbitrarily, given that the result of the integral over $J$ is the same for this term.  

The only aspect that requires some subtlety when translating results from the scalar case to the antisymmetric tensor case, is the late time behavior. In fact, the behavior at late times will depend on the specific components of the antisymmetric tensor which we are considering. In Section \ref{subsec:latetimes} we related the late time behavior of two-point functions of antisymmetric operators to the poles in the principal series densities, the complementary series terms, and the photon contribution. Here we report some further details of that derivation. First of all, we need to know how the projectors $\Pi_\pm$ act on $G^{\text{AdS}}_\Delta$ in conformally flat coordinates, something we used in the main text to derive (\ref{eq:lateB0i}) and (\ref{eq:lateBij}). From now on we focus on $d+1=4$.
\begin{equation}
\begin{aligned}
    \Pi_+^{\mu\nu\rho\sigma}G^{\text{AdS}}_\Delta(\sigma)=&\left(\partial_2^{\sigma]}\partial_1^{[\nu}\sigma\right)\partial_1^{\mu]}\partial_2^{[\rho}G^{\text{AdS}}_\Delta(\sigma)\,, \\
    \Pi_-^{\mu\nu\rho\sigma}G^{\text{AdS}}_\Delta(\sigma)=&f_1(\sigma)\left(\partial_2^{\sigma}\partial_1^{[\nu}\sigma\right)\left(\partial_1^{\mu]}\partial_2^{\rho}\sigma\right)+f_2(\sigma)\left(\partial_1^{[\nu}\sigma\right)\left(\partial_1^{\mu]}\partial_2^{[\sigma}\sigma\right)\left(\partial_1^{\rho]}\sigma\right)\,,
\end{aligned}
\end{equation}
where $\sigma$ is the chordal distance in the preferred coordinates, see for example (\ref{eq:sigmadef}), and\footnote{Here $\mathbf{F}$ is the regularized hypergeometric function
\begin{equation}
    \mathbf{F}(a,b,c,z)\equiv\frac{1}{\Gamma(c)}\ _2F_1(a,b,c,z)\,.
\end{equation}}
\begin{align}
        f_1(\sigma)=&\frac{\Gamma(\Delta-1)\Gamma(\Delta+1)}{2^{4-\Delta}\pi^2}\frac{(\Delta-3)(\sigma+1)\mathbf{F}\left(\begin{matrix}\Delta-1 & \Delta\\2(\Delta-1)& \end{matrix};\frac{2}{1+\sigma}\right)+2\sigma\mathbf{F}\left(\begin{matrix}\Delta-1 & \Delta+1\\ 2(\Delta-1) & \end{matrix};\frac{2}{1+\sigma}\right)}{(-1-\sigma)^{\Delta+1}}\nonumber\\
        f_2(\sigma)=&\frac{\Gamma(\Delta-1)\Gamma(\Delta)}{2^{4-\Delta}\pi^2}\frac{1}{(1-\sigma)(-1-\sigma)^{\Delta+2}}\Bigg(2(\Delta-1)(1+\sigma^2)\mathbf{F}\left(\begin{matrix}\Delta & \Delta\\2(\Delta-1)& \end{matrix};\frac{2}{1+\sigma}\right)\nonumber\\
        &+(2+\sigma(2\sigma+\Delta(\Delta-3)(1+\sigma)))\mathbf{F}\left(\begin{matrix}\Delta-1 & \Delta\\2(\Delta-1)& \end{matrix};\frac{2}{1+\sigma}\right)\Bigg)\,.
\end{align}
Focusing on specific components, going to conformally flat coordinates (\ref{eq:coordinates}) and at late (equal) times, we get
\begin{equation}
\begin{aligned}
    \Pi_{+,0i0j}G^\text{AdS}_\Delta(\sigma)&=c_\Delta\frac{\Delta-1}{4}\frac{(\eta^2)^{\Delta-2}}{(\mathbf{y}^2)^\Delta}\left[\delta_{ij}-2\frac{\mathbf{y}_{i}\mathbf{y}_{j}}{\mathbf{y}^2}\right]+\ldots\\
    \Pi_{+,ijkl}G^\text{AdS}_\Delta(\sigma)&=c_\Delta\frac{(\eta^2)^{\Delta-1}}{(\mathbf{y}^2)^{\Delta+1}}\left[\delta_{i[k}\delta_{l]j}+(\Delta+1)\frac{\mathbf{y}_{[i}\delta_{j][k}\mathbf{y}_{l]}}{\mathbf{y}^2}\right]+\ldots\\
    \Pi_{-,0i0j}G^\text{AdS}_\Delta(\sigma)&=-c_\Delta\frac{(\eta^2)^{\Delta-1}}{(\mathbf{y}^2)^{\Delta+1}}\left[\Delta\delta_{ij}-(\Delta+1)\frac{\mathbf{y}_{i}\mathbf{y}_{j}}{\mathbf{y}^2}\right]+\ldots\\
    \Pi_{-,ijkl}G^\text{AdS}_\Delta(\sigma)&=-c_\Delta\frac{\Delta-1}{4}\frac{(\eta^2)^{\Delta-2}}{(\mathbf{y}^2)^{\Delta}}\left[\delta_{i[k}\delta_{l]j}+2\frac{\mathbf{y}_{[i}\delta_{j][k}\mathbf{y}_{l]}}{\mathbf{y}^2}\right]+\ldots
    \label{eq:lateads}
\end{aligned}
\end{equation}
where the dots stand for subleading contributions at late times and
\begin{equation}
    c_\Delta\equiv\frac{2\Gamma(\Delta+1)}{\pi^\frac{3}{2}\Gamma(\Delta-\frac{1}{2})}\,.
\end{equation}
One can then obtain the electric and magnetic components in the locally inertial frame of a free falling observer in de Sitter through the definition (\ref{eq:EBdefinitions}) to derive (\ref{eq:lateB0i}) and (\ref{eq:lateBij})

\subsection{Details on the cubic scalar theory}
Here we report some details on the computations in the cubic scalar theory of Section \ref{sec:cubicscalar}. For example, it is useful to know the spectral density of the composite operator $:\chi^2:$ where $\chi$ is some fundamental free field of mass $m^2=\Delta(d-\Delta)=\frac{d^2}{4}+\mu^2_\chi$ \cite{Bros:2010rku,Akhmedov:2017ooy}
\begin{equation}
    \varrho_{\chi^2}(\frac{d}{2}+i\mu)=\frac{\mu\sinh\pi\mu}{2^5\pi^{3+\frac{d}{2}}\Gamma(\frac{d}{2}\pm i\mu)}\prod_{\pm,\pm,\pm}\Gamma\left(\frac{\frac{d}{2}\pm i\mu\pm i\mu_\chi\pm i \mu_\chi}{2}\right)
    \label{eq:rhochi2}
\end{equation}
Another quantity that is useful is the ``momentum space" two-point function of the operator $:\chi^2:$. This is maybe the only case for which such an object has been computed explicitly, apart from the two-point functions of elementary fields in free theories. It was first computed in \cite{Marolf:2010zp} and has this form:
\begin{equation}
\begin{aligned}
	&[\chi^2]_J=\frac{\Gamma(2-d)\Gamma(J+1)\Gamma(\frac{d+J}{2})\Gamma (\Delta ) \Gamma(\frac{J+2\Delta}{2}) \Gamma \left(\frac{2J+2\Delta-d +2}{4}\right) \Gamma \left(\frac{2J+2\Delta-d +4}{2}\right)}{8\pi^\frac{d-2}{2}\Gamma(\Delta+\frac{1}{2})\Gamma(\frac{1}{2}-\Delta)\sin(\frac{\pi}{2}(d-2\Delta))}\\
&\times\ _7\tilde{F}_6\left(\begin{matrix}\frac{2J+2\Delta-d +2}{2}, &\frac{J+\Delta-\frac{d}{2} +3}{2}  &\frac{2-d}{2},& J+1 ,&\Delta-d +1,& \frac{d+J}{2},& \frac{J+2\Delta}{2}\\ 
\frac{J+\Delta-\frac{d}{2} +1}{2},& J+\Delta +1,& \frac{d+2J+2}{2},& \frac{2\Delta-d+2}{2},& \frac{J+2\Delta-2d+4}{2},& \frac{J-d+4}{2}\end{matrix};1\right)\\
&\qquad\qquad +(\Delta\to d-\Delta)
\label{eq:gsqrd}
\end{aligned}
\end{equation}
where $\ _7\tilde F_6$ is a regularized generalized hypergeometric function.

    Even if it is not manifest from this expression, this quantity is finite in $d<3$. Close to $d=3$, it goes as
\begin{equation}
    [\chi^2]_J=-\frac{1}{8\pi^2}\frac{1}{d-3}+[\tilde\chi^2]_J\,.
\end{equation}
where $[\tilde\chi^2]_J$ is a complicated linear combination of derivatives of hypergeometric functions. We do not report it here but we checked that, for example, the sum over Gegenbauer polynomials with coefficients $[\tilde\chi^2]_J$ reproduces the two-point function of $\chi^2$.
\subsection{Details on the massless scalars}
\label{subsec:interactmasslessscalars}
In section \ref{subsec:interacphot} we considered the following theory of interacting massless scalars
\begin{equation}
    S=-\frac{1}{2}\int d^dx\sqrt{g}\left(\partial_\mu\phi_1\partial^\mu\phi_1+\partial_\mu\phi_2\partial^\mu\phi_2+g\partial_\mu\phi_1\partial_\nu\phi_2\partial^{[\mu}\phi_1\partial^{\nu]}\phi_2\right)
\end{equation}
and we focused on the two-point function of the composite operator
\begin{equation}
    B_{\mu\nu}\equiv\ :\partial_{[\mu}\phi_1\partial_{\nu]}\phi_2:
    \label{eq:BBBB}
\end{equation}
In Section \ref{subsec:masslessscalars} we found the spectral densities for this operator, which we report here for convenience
\begin{equation}
    \varrho^{\mathcal{P},+}_{\partial\phi_1\partial\phi_2}(\Delta)=\frac{(5-\Delta) (\Delta +2) (2 \Delta -3) \cot (\pi  \Delta )}{384 \pi ^2}\,, \qquad \varrho^\gamma_{\partial\phi_1\partial\phi_2}=\frac{1}{8\pi^2}\,.
    \label{eq:masslesssssss}
\end{equation}
The naive momentum-space two-point function of this operator diverges even in the free theory, as can be inferred from eq. (\ref{eq:Bfromprinc}). We thus need to regulate it. This can generically be done numerically following (\ref{eq:numreg}), but in this specific case we can extract an analytic form. To do so, we use dimensional regularization. In fact, we found an expression for the spectral density of (\ref{eq:BBBB}) that reproduces the two-point function of $B$ in any number of dimensions when the two masses of $\phi_1$ and $\phi_2$ are equal and in the principal series
\begin{equation}
    \varrho^{\mathcal{P},+}_{\partial\phi_1\partial\phi_2}(\frac{d}{2}+i\mu)=\frac{\mu\sinh\pi\mu}{16\pi^{\frac{d}{2}+3}\Gamma(\frac{d+2}{2})\Gamma(\frac{d}{2}+1\pm i\mu)}\prod_{\pm,\pm,\pm}\Gamma\left(\frac{\frac{d}{2}+1\pm i\mu\pm i\mu_\phi\pm i\mu_\phi}{2}\right)\,.
\end{equation}
where we are using $m_{\phi_1}^2=m_{\phi_2}^2=\frac{d^2}{4}+\mu_\phi^2$. In this regime, there is no photon contribution.
The expression (\ref{eq:masslesssssss}) is reproduced by continuing  $\mu_1\to i\frac{d}{2}$, $\mu_2\to i\frac{d}{2}$ and $d\to 3$. 

Then, notice the following fact:
\begin{equation}
    \varrho_{\partial\phi_1\partial\phi_2}^{\mathcal{P},+}(\frac{d}{2}+i\mu)=2\pi\varrho_{\phi^2}^{\mathcal{P}}(\frac{d}{2}+i\mu)\Big|_{d\to d+2}
\end{equation}
where $\varrho_{\phi^2}^{\mathcal{P}}(\Delta)$ is the spectral density of the composite operator $:\phi^2:$ where $\phi$ is a free scalar with mass $m_\phi^2$ (see eq. (\ref{eq:rhochi2})). At the same time, we have the following relation to derive the momentum space two-point function (\ref{eq:Bfromprinc})
\begin{equation}
    [\partial\phi_1\partial\phi_2]_{+,J}=\int_{\frac{d}{2}+i\mathbb{R}}[d\Delta]\frac{\varrho_{\partial\phi_1\partial\phi_2}^{\mathcal{P},+}(\Delta)}{(\Delta+J)(\bar\Delta+J)}
\end{equation}
Then, we perform the following manipulations
\begin{equation}
\begin{aligned}
    [\partial\phi_1\partial\phi_2]_{+,J}=&\int_{\mathbb{R}}d\mu \frac{\varrho_{\partial\phi_1\partial\phi_2}^{\mathcal{P},+}(\frac{d}{2}+i\mu)}{(\frac{d}{2}+i\mu+J)(\frac{d}{2}-i\mu+J)}\\
=&2\pi\int_\mathbb{R}d\mu\frac{\varrho_{\phi^2}^{\mathcal{P}}(\frac{d}{2}+i\mu)\Big|_{d\to d+2}}{(\frac{d}{2}+i\mu+J)(\frac{d}{2}-i\mu+J)}\\
=&2\pi\int_\mathbb{R}d\mu\frac{\varrho_{\phi^2}^{\mathcal{P}}(\frac{d}{2}+i\mu)}{(\frac{d}{2}+i\mu+J-1)(\frac{d}{2}-i\mu+J-1)}\Big|_{d\to d+2}\\
=&2\pi[\phi^2]_{J-1}\Big|_{d\to d+2}\,.
\label{eq:partphifromphi2}
\end{aligned}
\end{equation}
where $[\phi^2]_J$ is given in (\ref{eq:gsqrd}) (with the subtlety that one has to express $\Delta=\frac{d}{2}+i\mu$ before taking $d\to d+2$). 

The trick in (\ref{eq:partphifromphi2}) saves us a lot of work and allows us to have an explicit expression for $[\partial\phi_1\partial\phi_2]_{+,J}$. To get back to our massless scalars we continue back $\mu_\phi\to i\frac{d}{2}$ and obtain the following expression
\begin{equation}
\begin{aligned}
   & [\partial\phi_1\partial\phi_2]_{+,J}=\frac{\Gamma \left(\frac{d+J+1}{2}\right)}{4\pi^\frac{d}{2}\Gamma \left(\frac{J+1}{2}\right)} \Bigg(\frac{\Gamma (-\frac{d}{2}) \cot (\pi  d) \Gamma (J) \Gamma \left(\frac{2d+J+1}{2}\right)}{\Gamma \left(\frac{J+1-d}{2} \right) \Gamma (d+J+1)}-\frac{\Gamma (-d) \Gamma \left(\frac{J+1}{2}\right)^2 \Gamma \left(\frac{2J-d}{4}\right) \Gamma \left(\frac{J+1-2d}{2}\right)}{\pi  (d+2 J)} \\
    &\quad\quad\quad\times\Bigg(d \Gamma (\frac{d}{2}) \Gamma (J) \, _6\tilde{F}_5\left(\begin{matrix} \frac{d+J+1}{2} & \frac{J+1}{2} & \frac{2 J+4-d}{4} & -d & J  & \frac{2J-d}{2}\\
    \frac{J+1-2d}{2} & \frac{J+1-d}{2} & \frac{2 J-d}{4} & J+1 &\frac{d+2J+2}{2} &
\end{matrix};1\right)\\
&\qquad\qquad\qquad+\frac{2 \pi}{\sin(\frac{\pi  d}{2})} \, _6\tilde{F}_5\left(\begin{matrix}-d & -\frac{d}{2} &\frac{J+1}{2} &\frac{-d+2 J+4}{4} &\frac{d+J+1}{2} &\frac{2J-d}{2}\\
\frac{2-d}{2} &\frac{J+1-2d}{2}  &\frac{J+1-d}{2} &\frac{2 J-d}{4}&\frac{d+2J+2}{2}\end{matrix};1\right)\Bigg)\Bigg)
\label{eq:masslessscalars6F5}
\end{aligned}
\end{equation}
It can be checked that around $d=3$ it behaves as
\begin{equation}
    [\partial\phi_1\partial\phi_2]_{+,J}=\frac{1}{d-3}\frac{(J-2) (J+5)}{96 \pi ^2}+ [\widetilde{\partial\phi_1\partial\phi_2}]_{+,J}\,.
    \label{eq:divphiphi}
\end{equation}
where $[\widetilde{\partial\phi_1\partial\phi_2}]_{+,J}$ is finite, and is given by a sum of derivatives of hypergeometric functions, which we do not report here.

The momentum space representation of these two-point functions has ambiguities. For example, if we shift $[B]_{\pm,J}$ by adding $a_0+a_1J(J+3)$ for any $a_0$ and $a_1$, the result of (\ref{eq:Binharms}) does not change. That means we can remove the divergence in (\ref{eq:divphiphi}) and redefine the momentum space two-point function of this operator as simply $[\widetilde{\partial\phi_1\partial\phi_2}]_{+,J}$. We checked that, for example, the Watson-Sommerfeld integral representation (\ref{eq:WSantisy}) is valid for this operator.
\bibliography{bibliography}
\bibliographystyle{utphys}
\end{document}